\documentclass[acmsmall]{acmart}
\setcopyright{none}
\AtBeginDocument{%
  }

\authorsaddresses{}
\author{Naman Gupta}
\affiliation{
    \department{Computer Sciences}
    \institution{University of Wisconsin--Madison}
    \city{Madison}
    \state{Wisconsin}
    \country{USA}
\address{1205 University Ave, Madison, WI 53706, United States}
}
\email{n@cs.wisc.edu}

\author{Sophie Stephenson}
\affiliation{
    \department{Computer Sciences}
    \institution{University of Wisconsin--Madison}
    \city{Madison}
    \state{Wisconsin}
    \country{USA}
\address{1205 University Ave, Madison, WI 53706, United States}
}
\email{sophie.stephenson@cs.wisc.edu}

\author{Chung Chi Yeung}
\affiliation{
    \department{Computer Sciences}
    \institution{University of Wisconsin--Madison}
    \city{Madison}
    \state{Wisconsin}
    \country{USA}
\address{1205 University Ave, Madison, WI 53706, United States}
}
\email{cyeung6@wisc.edu}

\author{Wei Ting Wu}
\affiliation{
    \department{Computer Sciences}
    \institution{University of Wisconsin--Madison}
    \city{Madison}
    \state{Wisconsin}
    \country{USA}
\address{1205 University Ave, Madison, WI 53706, United States}
}
\email{wwu297@wisc.edu}

\author{Jeneile Luebke}
\affiliation{
    \department{Nursing}
    \institution{University of Wisconsin--Madison}
    \city{Madison}
    \state{Wisconsin}
    \country{USA}
\address{5111 Signe Skott Cooper Hall, Madison, WI, 53705, United States}
}
\email{jmluebke@wisc.edu}

\author{Kate Walsh}
\email{klwalsh2@wisc.edu}
\orcid{0000-0002-3996-2683}
\affiliation{%
    \department{Psychology and Gender \& Women Studies}
  \institution{University of Wisconsin-Madison}
  \city{Madison}
  \state{Wisconsin}
  \country{USA}
\address{1202 University Ave, Madison, WI 53706, United States}
}

\author{Rahul Chatterjee}
\affiliation{
    \department{Computer Sciences}
    \institution{University of Wisconsin--Madison}
    \city{Madison}
    \state{Wisconsin}
    \country{USA}
    \address{1205 University Ave, Madison, WI 53706, United States}
}
\email{rahul.chatterjee@wisc.edu}

\usepackage{booktabs}
\usepackage{setspace} 
\usepackage{amsmath}
\usepackage{xurl}
\usepackage[most]{tcolorbox}
\usepackage{xspace}
\usepackage{multicol}
\usepackage{multirow}
\usepackage{tabularx}
\usepackage{array}   

\usepackage[dvipsnames,table,xcdraw]{xcolor}
\usepackage{enumitem}
\usepackage{subcaption}
\usepackage{graphicx}
\usepackage{comment}
\usepackage{xparse}
\usepackage{mathtools}
\usepackage{ragged2e}
\usepackage{amsthm}
\usepackage{microtype}  
\usepackage{scalerel}
\usepackage{bm}
\usepackage{array}
\usepackage{amsfonts}
\usepackage{pifont}
\usepackage[normalem]{ulem}
\usepackage[strings]{underscore}
\usepackage{epigraph}
\usepackage{color,soul}
\usepackage{longtable}

\usepackage{csquotes}
\usepackage{tikz}
\usepackage{venndiagram}
\usepackage{calc}
\usepackage{fp}
\newcounter{z}
\usepackage{svg}
\usepackage{float}
\usepackage{fontawesome5}

\usepackage{listings}

\usepackage[all=normal,bibbreaks=tight,paragraphs=tight]{savetrees}

\renewcommand{\subparagraph}[1]{
\vspace{2mm}
\textit{#1}
}

\newcommand{\bnm}{\begin{newmath}}
\newcommand{\enm}{\end{newmath}}

\newcommand{\bea}{\begin{neweqnarrays}}%
\newcommand{\eea}{\end{neweqnarrays}}%

\newcommand{\bne}{\begin{newequation}}
\newcommand{\ene}{\end{newequation}}

\newcommand{\bal}{\begin{newalign}}
\newcommand{\eal}{\end{newalign}}

\newtcolorbox{cutegreenbox}[1]{colback=green!5,colframe=green!35!black,fonttitle=\bfseries,title={#1}}

\definecolor{DarkGreen}{RGB}{1,50,32}

\newcommand{\lgbt}{LGBTQ2S+\xspace}

\newcommand{\univ}{[U.S. University]\xspace}  
\newcommand{\wi}{[US. State]\xspace}  
\newcommand{\techclinic}{[Victim-Advocacy organization]\xspace} 
\newcommand{\hochunk}{[Tribe]\xspace}
\newcommand{\ldf}{[Tribe]\xspace}
\newcommand{\lco}{[Tribe]\xspace}
\newcommand{\badriver}{[Tribe]\xspace}
\newcommand{\teejop}{[City]\xspace}
\newcommand{\was}{[National-level Conference on Violence against Native peoples]\xspace}

\newcommand{\MMIW}{Missing and Murdered Indigenous Relatives\xspace}
\newcommand{\mmiw}{MMIR\xspace}

\newcommand{\victim}{relative\xspace}

\newcommand{\victims}{relatives\xspace}
\newcommand{\Victims}{Relatives\xspace}

\newlist{researchquestions}{enumerate}{1}
\setlist[researchquestions]{label*=\textbf{RQ\arabic*}}

\usetikzlibrary{shapes,positioning,calc}

\newcommand{\onlypercent}[2]{%
    \FPeval{\result}{round((#1 / #2) * 100, 2)}%
    \result\%%
}

\newcommand{\ntribes}{2,265\xspace}
\newcommand{\nqueries}{54,528}
\newcommand{\ntotalresults}{348,380}
\newcommand{\nuniquedomains}{28,800}
\newcommand{\nuniqueresultsprose}{123,029}

\newcommand{\nuniqueresults}{123029}

\newcommand{\ncodedpages}{140}
\newcommand{\morenews}{23}

\newcommand{\nbarriers}{five\xspace}
\newcommand{\nactions}{six\xspace}
\newcommand{\nrecommendations}{five\xspace}

\newcommand{\actortable}[2]{
\textbf{#1}\textsuperscript{\autoref{#2}}\xspace
}

\definecolor{headerDark}{HTML}{2C3E50}      
\definecolor{headerWarm}{HTML}{8B1A1A}      
\definecolor{rowAlt}{HTML}{F5F0EB}          
\definecolor{rowBase}{HTML}{FFFFFF}         
\definecolor{labelCol}{HTML}{1C1C1C}        
\definecolor{accentRule}{HTML}{8B1A1A}      

\newcommand{\barrier}[1]{%
  \tcbox[colback=headerDark!10,
         colupper=headerDark!50!black,
         colframe=headerDark!70,
         boxrule=0.3pt,
         arc=1pt,
         boxsep=0.5pt,
         left=2pt,
         right=2pt,
         top=0pt,
         bottom=0pt,
         fontupper=\itshape\bfseries,
         on line]{%
    \hyperref[tab:barriers]{\textcolor{headerDark}{{\footnotesize\faIcon{exclamation-circle}} \texttt{B#1}}}%
  }%
  \label{tab:barriers-B#1} 
}

\newcommand{\barrierref}[1]{%
  \tcbox[colback=headerDark!10,
         colupper=headerDark!50!black,
         colframe=headerDark!70,
         boxrule=0.3pt,
         arc=1pt,
         boxsep=0.5pt,
         left=2pt,
         right=2pt,
         top=0pt,
         bottom=0pt,
         fontupper=\itshape\bfseries,
         on line]{%
    \hyperref[tab:barriers-B#1]{\textcolor{headerDark}{{\footnotesize\faIcon{exclamation-circle}} \texttt{B#1}}}%
  }%
}

\newcommand{\action}[1]{%
  \tcbox[colback=headerWarm!10,
         colupper=headerWarm!50!black,
         colframe=headerWarm!70,
         boxrule=0.3pt,
         arc=1pt,
         boxsep=0.5pt,
         left=2pt,
         right=2pt,
         top=0pt,
         bottom=0pt,
         fontupper=\itshape\bfseries,
         on line]{%
    \hyperref[tab:actions]{\textcolor{headerWarm}{{\footnotesize\faIcon{heart}} \texttt{A#1}}}%
  }%
  \label{tab:actions-A#1} 
}

\newcommand{\actionref}[1]{%
  \tcbox[colback=headerWarm!10,
         colupper=headerWarm!50!black,
         colframe=headerWarm!70,
         boxrule=0.3pt,
         arc=1pt,
         boxsep=0.5pt,
         left=2pt,
         right=2pt,
         top=0pt,
         bottom=0pt,
         fontupper=\itshape\bfseries,
         on line]{%
    \hyperref[tab:actions-A#1]{\textcolor{headerWarm}{{\footnotesize\faIcon{heart}} \texttt{A#1}}}%
  }%
}

\newcommand{\recommendation}[1]{%
  \tcbox[colback=violet!15,
         colupper=violet!50!black,
         colframe=violet!40,
         boxrule=0.3pt,
         arc=1pt,
         boxsep=0.5pt,
         left=2pt,
         right=2pt,
         top=0pt,
         bottom=0pt,
         fontupper=\itshape\bfseries,
         on line]{%
    \hyperref[sec:recommendations]{{\footnotesize\faIcon{compass}} \texttt{R#1}}%
  }%
  \label{sec:recommendations-R#1} 
}

\newcommand{\recref}[1]{%
  \tcbox[colback=violet!15,
         colupper=violet!50!black,
         colframe=violet!40,
         boxrule=0.3pt,
         arc=1pt,
         boxsep=0.5pt,
         left=2pt,
         right=2pt,
         top=0pt,
         bottom=0pt,
         fontupper=\itshape\bfseries,
         on line]{%
    \hyperref[sec:recommendations-R#1]{{\footnotesize\faIcon{compass}} \texttt{R#1}}%
  }%
}

\newcommand{\mycaption}[2]{
    \caption[#1]{\textbf{#1} -- {\small\textmd{#2}}}
}

\newenvironment{myquote}[1]{%
  \def\quoteauthor{#1}
  \begin{list}{}{%
    \leftmargin0.3cm
    \rightmargin\leftmargin
  }%
  \item\relax
  \itshape ``%
}{
  ’’
  \hspace{0.2cm}\normalfont\small\hfill{--- \quoteauthor}%
  \end{list}%
}

\newenvironment{myquotel}[1]{%
  \def\quoteauthor{#1}
  \begin{list}{}{%
    \leftmargin0.3cm
    \rightmargin\leftmargin
  }%
  \item\relax
  \itshape``
}{
  ''
  \par\vspace{-0.5ex}
  \normalfont\small\hfill--- \quoteauthor%
  \end{list}%
}

\newcommand{\fixme}[1]{{\color{Red}{[\textbf{FIXME: #1}]}}}

\def\diff{0}

\newcommand{\removed}[1]{%
  \ifnum\diff=1{\color{magenta}\sout{#1}}\fi%
}

\newcommand{\changed}[1]{%
  \ifnum\diff=1%
    {\color{blue}\uline{#1}}%
  \else%
    #1%
  \fi%
}

\DeclareRobustCommand{\greybox}[1]{
\begin{tcolorbox}[boxrule=1pt, left=2pt, right=2pt, bottom=1pt, top=0pt]
    {\small \textit{#1}}
  \end{tcolorbox}
}

\def\authnote{1}
\newcounter{rcnote}[section]
\newcommand{\rcnote}[1]{\ifnum\authnote=1\refstepcounter{rcnote}{\bf \textcolor{magenta}{$\ll$RC: {\sf #1}$\gg$}}\fi}

\newcommand{\tabfontsize}{\footnotesize}

 \newcounter{tablerow}
\newcommand{\rowlabel}[1]{\refstepcounter{tablerow}\label{#1}\arabic{tablerow}}

\definecolor{acm-purple}{RGB}{90, 53, 158}
\definecolor{acm-green}{RGB}{0, 102, 0}
\definecolor{acm-darkblue}{RGB}{0, 51, 102}

\hypersetup{
  colorlinks   = true,
  linkcolor    = acm-purple,
  citecolor    = acm-green,
  urlcolor     = acm-darkblue,
  filecolor    = acm-darkblue
}

\newlength{\saveparindent}
\newlength{\saveparskip}
\newcounter{ctr}

\begin{document}

\newcommand{\mytitle}{``Lighting The Way For Those Not Here'': How Can Technology Researchers Help Resist the Missing and Murdered Indigenous Relatives (MMIR) Crisis?}
\newcommand{\mytitleshort}{How Can Tech Help Resist the MMIR Crisis?}
\title[\mytitleshort]{\mytitle}

\begin{abstract}
    Indigenous peoples across Turtle Island face disproportionate rates of disappearance and murder, a genocide rooted in settler-colonial violence and systemic erasure. 
    Technology plays a crucial role in the Missing and Murdered Indigenous Relatives (MMIR) crisis: it perpetuates systemic violence and impedes investigations, yet also enables sites of advocacy, healing, and resistance. For example, Native communities utilize AMBER alerts, digital news, sovereign crowdsourced databases, social media groups, and resistance movements to mobilize searches, amplify awareness, and honor missing relatives. Yet little research in HCI has critically examined the role of technology in shaping the MMIR crisis. Thus, we qualitatively analyze 140 webpages to identify sociotechnical barriers that hinder communities' efforts, while highlighting actions that foster healing, safety, and resilience. 
    We grounded our analysis in stories that resist epistemic erasure through relational accountability, critical humility, cultural sensitivity, and refusal. Finally, we provide recommendations for HCI to recognize self-determination and sovereignty of Indigenous technologies, direct action to support families, and honor Indigenous onto-epistemologies that cease epistemic violence.
\end{abstract}
\begin{teaserfigure}
     \centering
    \includegraphics[height=0.32\textwidth]{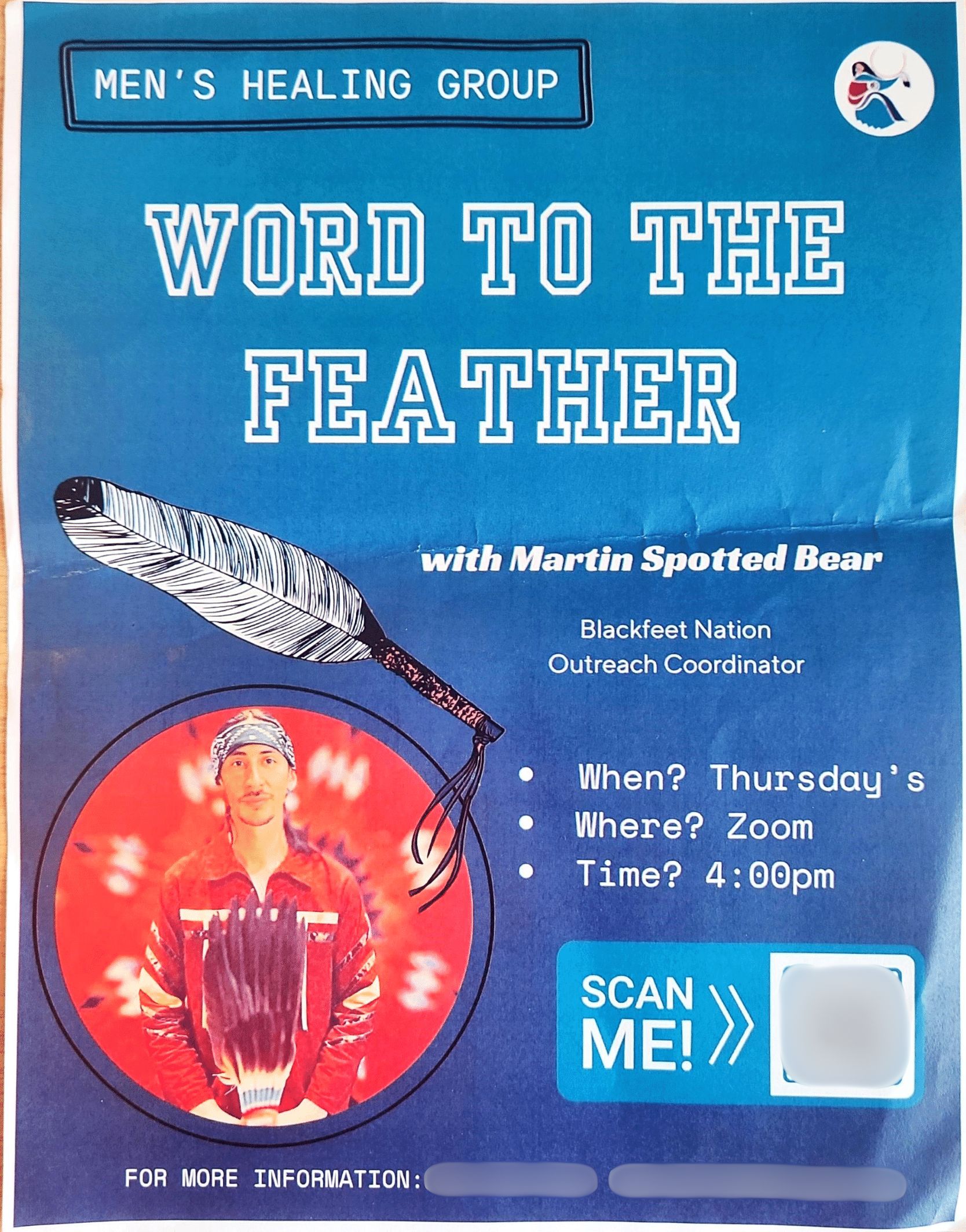}\hspace{1mm}
    \includegraphics[height=0.32\textwidth]{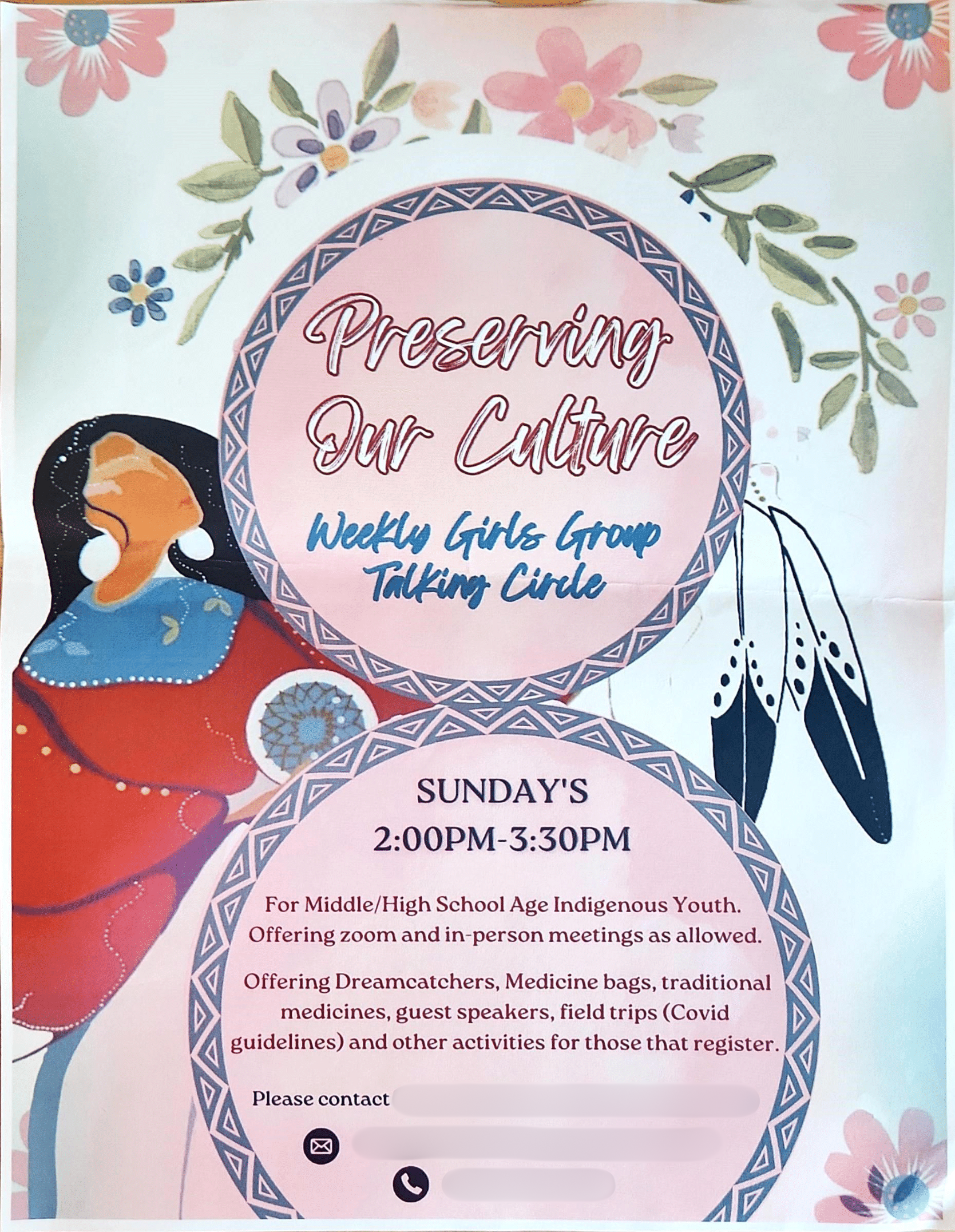}\hspace{1mm}
    \includegraphics[height=0.32\textwidth]{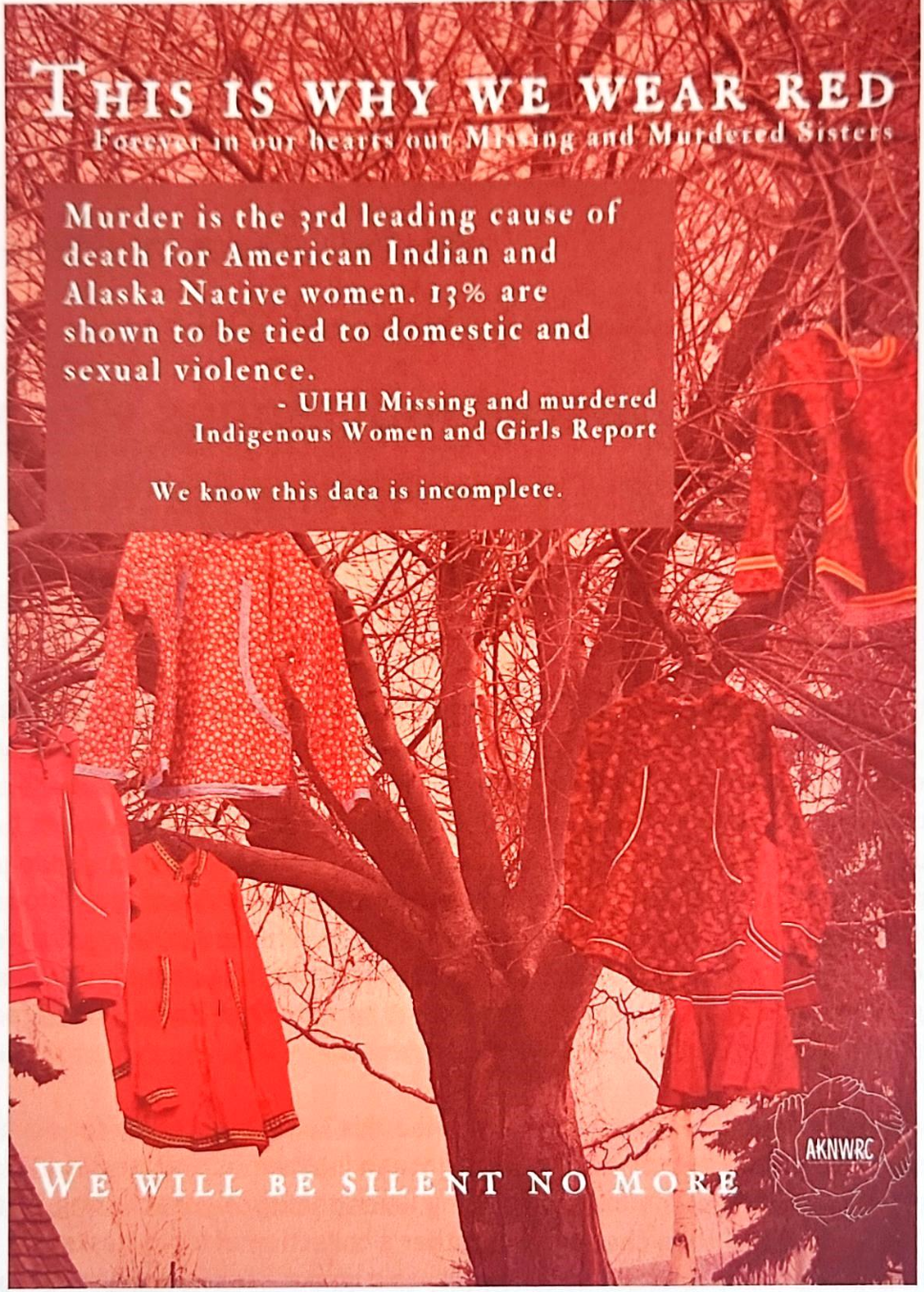}\hspace{1mm}
    \includegraphics[height=0.32\textwidth]{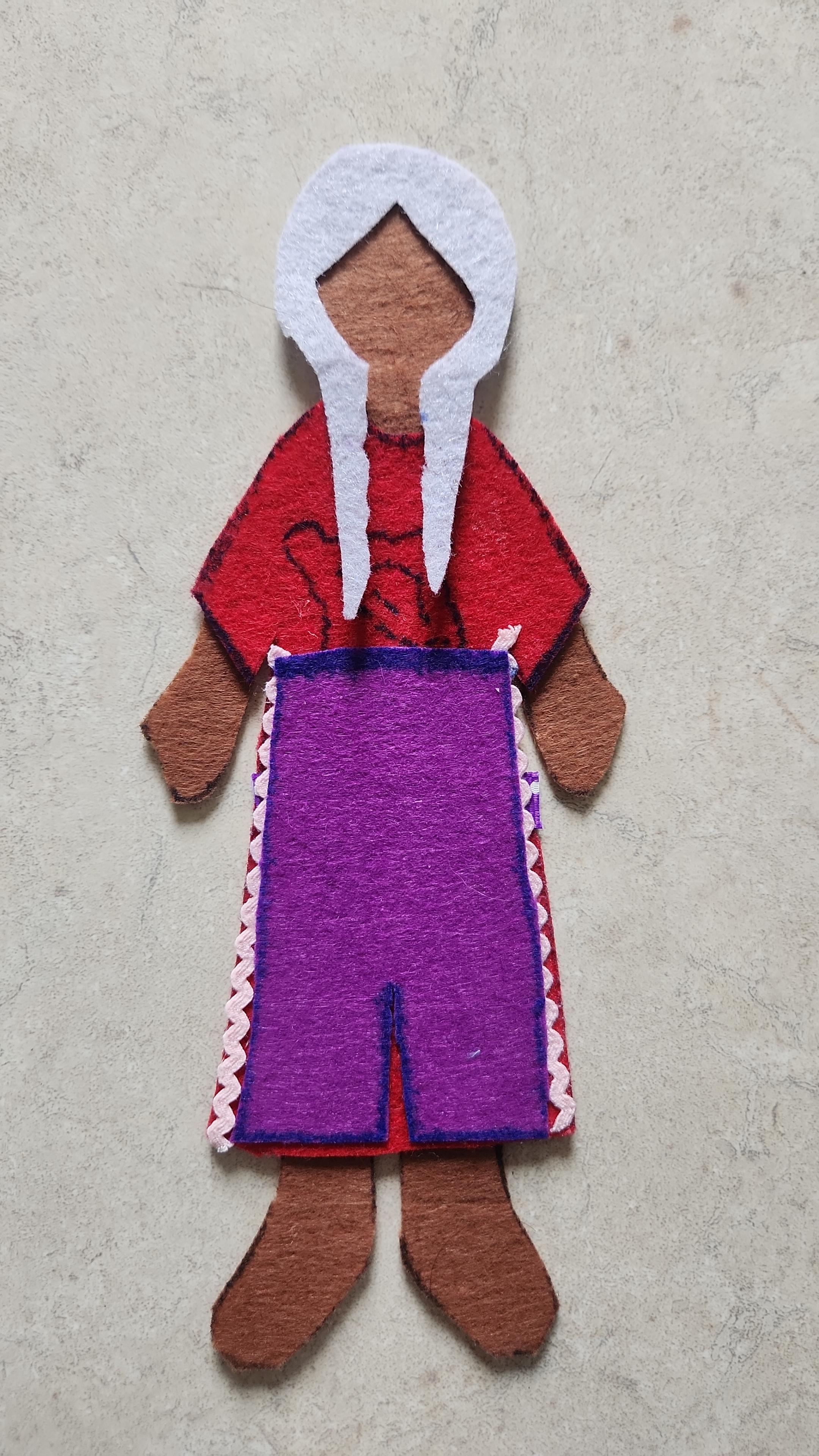}
    \mycaption{Artifacts}{Advocates and families shared noteworthy artifacts with us at \was: The (\textbf{left two}) flyers for talking circles run by Mother Nation, (\textbf{middle}) an action plan for families to find their relatives, created by Alaska Native Women's Resource Center~\cite{alaska_native_womens_resource_center_alaska_2025}, and  (d) (\textbf{right}) a doll we created to honor our lost relatives in the doll-making workshop. Talking circles and doll-making workshops are sacred traditional healing practices in some Indigenous communities (\actionref{3}).}
  \Description{The figure consists of four distinct sections, each featuring different elements. The first section on the left is a flyer for a "Men's Healing Group." It features a large feather illustration and an image of a man playing a drum next to turquoise text reading "Word to the Feather with Martin Spotted Bear." The background is a gradient of dark to light blue.
    The second section is a flyer with a floral border. It advertises "Preserving Our Culture Weekly Circle Group: Talking Circle," with images of traditional Native American design elements, including feathers and bright colors.
    The third section is a poster featuring red dresses and text about the violence against Native American and Alaska Native women. The background is predominantly red with a faded image of red dresses hanging in trees.
    The fourth section depicts a hand-crafted doll with long white braids, wearing a red cape and dark purple dress. The doll appears to be made from fabric and felt materials.
    }
    \label{fig:cultural-sensitivity}
\end{teaserfigure}

\maketitle

\greybox{\large \textbf{Content warning:} The paper may be disturbing for some readers. The paper contains stories of physical and sexual violence, genocide, trafficking, stalking, homicide, substance abuse, forced sterilization, profanity, harassment, and abduction. Please take care of yourself while reading the article.}

\epigraph{\em Think of a woman in your life with whom you are close. Consider how special she is to you, how much you care about her, and how she makes your life better. Now. . . What if she disappeared? The experience, for many, is that while you worry about her greatly, you trust authorities will find and bring her back safely. But for some of us, these authorities do not seem interested in helping. Pause and feel this. The authorities are not helping you find her. You do not know where else to turn. Weeks, months, and years pass, and yet you hear nothing about what happened to your loved one. No explanation from authorities. No one even seems to be talking about finding your loved one; in fact, no one has really acknowledged that she is missing. How would you feel? Sadness? Anger? Anguish? What if no one even validated your grief? How would you go on from day to day? What would you do? Where would you turn?}{\textit{Ficklin et al.~\cite{ficklin-fighting-2022}}}



\section{Introduction}



 \MMIW (\mmiw\footnote{While ``Missing and Murdered Indigenous People'' (MMIP) or Missing and Murdered Indigenous Women, Girls, and Two-Spirit people (MMIW or MMIWG2S) are also widely used terms, many advocates prefer ``\MMIW'' (\mmiw) to emphasize kinship, relationality, and global solidarity. 
 \actortable{Jodi Voice Yellowfish}{jodivoiceyellowfish}~\cite{yellowfish-missing-2023} verbalizes this strong solidarity within \mmiw movement  ``I haven't been [to] a space where everyone \ldots
 doing that work were so open to having so much care for the work, that everyone is called a ``relative'' in that space.''}) 
 is a human rights and public safety crisis impacting Indigenous people throughout Turtle Island (North America). 
 Indigenous peoples 
 across 
 genders 
 and ages 
 face high rates of violence and disappearances. 
Indigenous women are murdered at rates more than ten times the national average in the United States~\cite{bachman_violence_2008} and six times the national average in Canada~\cite{mahony_women_2017}. Native women are over-represented among domestic violence victims in Alaska by 250\%~\cite{indian-law-and-order-commission-roadmap-2015}. In the US, more than four in five Indigenous women (84.3\%) have experienced violence --- almost always by non-Indigenous perpetrators (96\% of cases)~\cite{rosay-violence-2016} --- and 40\% of sex trafficking victims in the US are Native women~\cite{national-congress-of-american-indians-policy-research-center-human-2016}. 
The 2019 National Inquiry into \mmiw, established by the Government of Canada in 2016, concluded that the systemic violence experienced by Indigenous peoples --- particularly Indigenous women and girls --- constitutes an ongoing genocide~\cite{national-inquiry-into-missing-and-murdered-indigenous-women-and-girls-reclaming-2019}.
In the US, federal reports and commissions have similarly recognized the \mmiw crisis as a systemic national issue shaped by jurisdictional failures, inadequate data systems, and the enduring impacts of settler colonialism, though without formally adopting the genocide framing~\cite{gao-mmiw-2021, not-invisible-act-2023}.


To address the \mmiw crisis, State and grassroots organizations are using technologies to find and protect Indigenous relatives~\cite{bailey-missing-2016, felt_mobilizing_2016, niwrc-mmiw-2024, moeke-pickering-understanding-2018}. For example, to help investigate and collect information about missing persons, governments use a centralized database~\cite{whyfederalgovernment, royal-canadian-mounted-police-working-2017}, DNA sequencing technology~\cite{bailey-missing-2016}, and alert systems~\cite{amberalertindian, the-amber-adovcate-amber-2025}. Families and tribal police frequently use social media platforms to raise awareness of the crisis~\cite{felt_mobilizing_2016, bleeker-efficacy-2023, moeke-pickering-understanding-2018, diehl-is-2019}. 
Despite these efforts, current technologies fail to make a real difference. We argue that this is because technologies perpetuate epistemic silencing through a lack of Indigenous voices and knowledge in their design and deployment~\cite{mbembe-necropolitics-2016, escobar-designs-2018}. These technologies embed deeply rooted settler-colonial and oppressive logics and perpetuate systemic erasure, rather than protecting Indigenous lives. Without centering community needs, colonial infrastructure renders the effort ineffective and actively harmful to the families of missing or murdered \victims.  \citeauthor{ficklin-fighting-2022} ~\cite{ficklin-fighting-2022} questions the ubiquity of modern technology and its ineffectiveness in addressing the crisis: \textit{``Technology is so embedded in our society that it seems impossible for anyone to maintain their privacy, much less go missing.''}
For example, despite the efforts of using technology to record \mmiw cases in US and Canada, comprehensive data on missing and murdered Indigenous survivors does not exist, leaving most cases uncounted, unsolved, or improperly investigated. 
\mmiw advocates at Urban Indian Health Institute (UIHI)~\cite{lucchesi-missing-2018} --- an Indigenous-led independent effort to document the missing or murdered victims ---  reported 5,712 cases,  yet only 116 (2\%) appeared in the US National Missing and Unidentified Persons System (NamUs) and National Crime Information Center (NCIC). In Canada, a 2014 RCMP (Royal Canadian Mounted Police)~\cite{royal-canadian-mounted-police-working-2017} recorded 1,181 cases, excluding suspected homicides, deaths deemed suspicious, and cases where the victim was missing for more than 30 days. Importantly, most grassroots advocacy groups believe this to be a severe undercount.

Technology has increasingly become central to address the \mmiw crisis.  The \mmiw resistance movement emerged from stories shared by grannies and aunties of K'emk'emeláy (now Vancouver, Canada) to make Indigenous loss  within dominant public discourse~\cite{audrey-huntley-beyond-2020}. Social media has become a critical infrastructure for sustaining the movement, enabling families, survivors, advocates, and Tribal communities to mobilize transnational solidarity~\cite{felt_mobilizing_2016, bleeker-efficacy-2023, moeke-pickering-understanding-2018, diehl-is-2019}.
However, technologies are not neutral infrastructures: they shape whose voices are amplified, which cases become visible, and how communities access healing and justice. The \mmiw crisis reveals how sociotechnical systems facilitate settler-colonial violence by side-stepping Indigenous agency and self-determination. This tension reveals an urgent need for HCI engagement with the \mmiw crisis. Yet, little is known about how technologies shape \mmiw response efforts in practice, including how Indigenous communities adapt, resist, and reconfigure them for search, advocacy, remembrance, and healing. 
To address this gap, we conduct a large-scale content analysis of articles available online related to \mmiw to examine how technologies support or hinder community-led \mmiw responses and how HCI researchers and practitioners can better support Indigenous communities through culturally grounded and sovereignty-centered design.  We ask the following research questions:

\begin{researchquestions} \it
    \item 
    How do technologies create barriers for Indigenous communities in searching missing or murdered relatives?
    \item How do communities reconfigure technologies to resist, heal, and advocate for the \mmiw movement?
    \item How can technology researchers and practitioners better support Indigenous communities to address the \mmiw crisis?
\end{researchquestions}


In this work, we show how settler-colonial and oppressive logics embedded in modern sociotechnical systems fuel violence, creating ``ecosystems of erasure'', instead of affording safety to the communities (see \autoref{tab:missing-persons-approaches}). Inaccurate data collection, opaque data sharing policies, extractive programs and lack of safe online spaces create barriers for communities to find safety.
Colonial institutions repeatedly misuse or appropriate Indigenous knowledge.
For example, the RCMP installed five 5G cell towers along the ``The Highway of Tears'' to close the gap in cell service. However, they used a narrow definition of “missing”/“murdered” and left
out many highways with spotty cell coverage.
Moreover, most Tribes lack the agency and access to the State criminal databases and are unable to release alerts, forcing dependence on colonial institutions.
%
At the same time, Indigenous communities mount a generational resistance against State technologies and reclaim them as spaces of advocacy, memory, and  care. 
Families and tribal police circulate missing-person posters, coordinate search efforts, document violence, and preserve collective memory.
Communities form ``ecosystem of care'' through social media campaigns, sovereign and crowd-sourced databases, and digital maps supporting three action pillars: (1) search and rescue for \victims (\autoref{actions:find}), (2)  providing support and healing to \victims and their families (\autoref{actions:support}), and (3) raising awareness of the \mmiw crisis (\autoref{actions:advoacy}).  
%

By centering Indigenous voices, we make the following four contributions across disciplines:
%


\begin{enumerate}[leftmargin=*]
    \item \textbf{Knowledge Contribution:} We outline \nbarriers technology-related barriers (denoted by \barrier{X}) addressing the MMIR crisis. 
    To fight systemic injustice, we highlight \nactions socio-technical actions (denoted by \action{X}) taken by Indigenous communities to find missing or murdered \victims (\autoref{actions:find}), seek safety, support, and healing (\autoref{actions:support}), and mobilize the \mmiw movement (\autoref{actions:advoacy}).
    \item \textbf{Methodological Contribution}: We 
    created a culturally-sensitive content analysis pipeline to identify and analyze \ncodedpages\ pages. We present our findings through storytelling, rooted in decolonial feminist onto-epistemologies, and employ reflexivity, reciprocity, and relational accountability; critical humility and cultural sensitivity; and refusal (\autoref{sec:positionality}). We demonstrate how empirical HCI methods and automation can be re-imagined to engage critically with settler-colonial systems by centering Indigenous meaning-making.
        
        \item \textbf{Data Contribution}: We create a dataset of web pages that resists epistemic erasure and transcends Western academic knowledge, holding sacred stories of missing or murdered \victims, families, advocates, and tribal police. The dataset includes news articles, reports by advocates and police agencies, podcasts, and court hearings. 
         This dataset will be open-sourced to support future HCI research and Indigenous advocacy.\footnote{The dataset is provided in the supplementary materials.}  
    \item \textbf{Design and Practice Recommendations}: {We invite future HCI scholars to design interventions that respect the sacred continuity and connectedness within and between three action pillars, rather than fragmented, parachuting ``solutions''. We provide provide \nrecommendations recommendations encouraging that scholars (a) recognize the self-determination and sovereignty of data and technologies (\autoref{recommendations:data_sovereignty}), (b) take direct action to support families, advocates and tribal police (\autoref{recommendations:tech}) and (c) honor Indigenous onto-epistemologies to cease epistemic violence (\autoref{recommendations:research}).
    }
\end{enumerate}

\section{Background}
\label{sec:bg}
{A historical grounding is crucial in enabling readers to resonate with intergenerational violence and trauma faced by missing or murdered \victims, their families, advocates, and tribal police~\cite{asher_blackdeer_decolonizing_2026, smith-decolonizing-2021}. Addressing the \mmiw crisis necessitates reckoning with these historical harms. We briefly outline the settler-colonial root causes of violence underlying the \mmiw crisis (see Dunbar-Ortiz~\cite{dunbar-ortiz-2014}, Blackhawk~\cite{blackhawk-2023}, and Stannard~\cite{stannard-1992} for a comprehensive historical account). Then, we outline 
limited prior HCI research that resists extraction and centered Indigenous meaning-making. Finally, we describe how we utilize decolonial feminist lenses to ground our research.
} 




\subsection{Historical Overview of the \MMIW Crisis}
\label{bg:root-cause}


\begin{myquote}{\actortable{Desi Small-Rodriguez Lonebear}{desismallrodriguez}~\cite{vice-news-indigenous-2020}}
     Why is it that we are more likely to be raped and murdered than go to college? Why is it that our young girls are just trying to survive?
\end{myquote}

Intersecting regimes of settler-colonial governance and coercively negotiated treaty frameworks exemplify the institutionalized normalization of violence against Indigenous peoples that underlies the \mmiw crisis. {Rooted in the `doctrine of discovery' or `manifest destiny', `White savior complex' emerged from white-supremacist ideologies that justified the theft of Indigenous land under the guise of ``saving'' Indigenous peoples~\cite{kherbaoui-bleeding-2021, mccurdy-privileged-2016, frey-white-2016, khan-white-2023, williams-like-2005}.}

\subsubsection{Structural Dispossession and Isolation}
\label{bg:dispossession}
The legacy of land dispossession, extractive industries, and residential schools has produced intergenerational trauma, perpetuating identity loss, cultural disconnection, poverty, and cycles of violence that continue to harm Indigenous communities today and fuel the crisis.
%
Colonial land policies such as forced relocation, land allotment, and the creation of reservations stripped access to lands that Indigenous communities have stewarded since time immemorial~\cite{wolfe_settler_2006}. In the U.S., the General Allotment Act of 1887 (Dawes Act) divided communally held Indigenous lands into individual parcels, sold to European settlers
~\cite{wilkins-uneven-2001, noauthor-dawes-2025}. The reservation system confined Indigenous peoples to small, often remote tracts of land with few economic opportunities~\cite{palmater-indigenous-2015, wilkins-uneven-2001}. 
Similarly, on the pretense of reparations for stolen land, federal recognition process fail to reflect the full diversity of Indigenous identities and governance systems, while reinforcing colonial control over who is considered ``legitimately'' Indigenous~\cite{whyfederalgovernment}. The US and Canada recognize 574 and 634 tribal nations, respectively, leaving unrecognized tribes vulnerable (e.g. Métis and Inuit First Nations)~\cite {wilkins-uneven-2001, palmater-beyond-2011}.
Recognition perpetuates settler-colonial dynamics; forcing Tribal Nations to depend on the State for economic resources, creating significant barriers to justice, health, and victim services such as limited access to protections, legal recourse, and culturally appropriate services~\cite{casselman-injustice-2016}. 
%
 Forced displacement into urban centers under the guise of employment, without adequate housing or social support, often left them isolated and vulnerable~\cite{fixico-termination-1986, echo-hawk-our-2020, luebke-barriers-2022}. 
Currently, over 70\% of Indigenous peoples reside in urban areas~\cite{urban-indian-health-institute-urban-nodate}.
However, \mmiw crisis continues in both rural and urban cities, shattering misconception that violence only happens in reservations. For example, as many as 94\% of Native women reported being raped or coerced in urban Seattle, US~\cite{echo-hawk-our-2020}.
%

%
Moreover, the expansion of extractive industries (oil-gas pipelines, mineral mining, cannabis, and recently, AI data centers) on reservation lands trespasses on sovereignty and endangers safety.
Industries establish \textit{man camps}, housing that brings thousands of transient labor (predominantly male), which have been linked to increased rates of domestic violence, sexual assault, stalking,  trafficking, and labor exploitation, adding stress on the tribal police's capacity~\cite{national-inquiry-into-missing-and-murdered-indigenous-women-and-girls-reclaming-2019, amnesty-international-canada-2016, siteadmin-indigenous-2013}.
For example, ``man camps'' in North Dakota have led to a 300 percent increase in sex crimes~\cite{siteadmin-indigenous-2013}. 
Modern extractive industries (e.g., AI data centers) replicate colonial dispossession and exploitation in remote areas (see our limitations in~\autoref{sec:limitations}).
Finally, coercive residential schools used law enforcement to forcibly remove Indigenous children from their families (1880s to 1970s in the US; till 1996 in Canada)~\cite{noauthor-indian-nodate,national-centre-for-truth-and-reconciliation-imagine-2021}. 
Children were prohibited from speaking their languages, practicing cultural traditions, or maintaining family ties. Many were sent to remote, inaccessible institutions far from home. Students suffered widespread physical, emotional, and sexual abuse, as well as neglect through malnutrition, overcrowding, and unsanitary conditions. Thousands died and were buried in unmarked graves near school sites~\cite{noauthor-indian-nodate}.  
The consequences remain profound. Indigenous children are still disproportionately represented in state child welfare systems~\cite{native-womens-association-of-canada-sisters-2010, blackstock-keeping-2004}. 
%
%

%



\subsubsection{Jurisdictional Maze and Law Enforcement Neglect}
\label{bg:jurisdictional-tension}
The persistence of police violence and jurisdiction tension reflects the broader extension of colonial control where Indigenous lives are routinely devalued. {In the US,} the 1978 Oliphant vs. Suquamish Indian Tribe~\cite{noauthor-oliphant-2025} ruling curtailed jurisdictional powers of Indigenous nations and banned tribal courts from prosecuting non-Native perpetrators for their crimes on reservations. {This split authority leaves tribal police under-resourced, and State agencies slow to intervene, leading to significant delays in investigations~\cite{deer-beginning-2015}, especially when 96\% of domestic violence perpetrators are non-Native~\cite{rosay-violence-2016} or the crime is committed on non-tribal lands~\cite{casselman-injustice-2016}}
On the contrary, the Federal Bureau of Investigation (FBI) can prosecute violent felonies on tribal lands; however, closed zero cases of sexual assault from non-Native perpetrators~\cite{us_department_of_justice_indian_2019}. The Public Law 280 (PL280)\footnote{Six are mandatory PL280 states--- Alaska (except the Metlakatla Reservation), California, Minnesota (except Red Lake), Nebraska, Oregon (except Warm Springs), and Wisconsin. Ten states assume full or partial jurisdiction --- Arizona, Florida, Idaho, Iowa, Montana, Nevada, North Dakota, South Dakota, Utah, and Washington.} gave some US states authority over criminal and certain civil matters on tribal lands (some tribes are exempt from PL280, most are not).
In Canada, the 1876 Indian Act governs the legal status of First Nations (except Metìs and Inuit Nations~\cite{palmater-beyond-2011}), but it does not recognize separate Indigenous criminal jurisdiction. Although tribal police exists, the criminal prosecutions on reserves are administered by provinces, with no equivalent to the tribal courts of the US. 
%
%
%
%
%
State law enforcement agencies fail to protect Indigenous peoples and investigate cases of disappearance or murder~\cite{national-inquiry-into-missing-and-murdered-indigenous-women-and-girls-reclaming-2019}.  Reports made by families are often dismissed, misclassified, or subjected to victim-blaming narratives that minimize the seriousness of the violence~\cite{razack-gendering-2016, beniuk_indigenous_2012}. Many cases go cold for decades without recourse for families~\cite{dateline-nbc-missing-2024, lucchesi-missing-2018}. Indifference from the police has historically accompanied harassment, physical violence, including fatal shootings, and even sexual assault~\cite{palmater-shining-2016, deer-beginning-2015, razack-gendering-2016}. 
This neglect and abuse of power reinforces mistrust of state institutions and discourages families from seeking help.

\subsubsection{Media Invisibility and Health Disparities}
\label{bg:invisibility}

Invisibility in Western media and the legacy of forced sterilization persists today, enduring racist narratives and lost trust in State institutions.
Capitalist and patriarchal belief systems (e.g., viewing women as less valuable, violence against women, ownership of women~\cite{ficklin-fighting-2022}) and hierarchical institutional structures eroded and replaced matrilineal societies where women often held central positions~\cite{deer-beginning-2015, anderson-recognition-2016, gilberg-addressing-2003, lafromboise-changing-1990}. 
Colonial narratives systematically devalue Indigenous peoples as ``less worthy'', irrational, and hypersexualized them; legitimizing their mistreatment and violence and reinforcing cycles of abuse and impunity~\cite{razack-race-2002, marubbio-killing-2006, razack-gendering-2016, jiwani-symbolic-2009, custalow_true_2007}. 
%
Mainstream media (News media, Hollywood, etc.) reproduce these violent narratives and stereotypes such as ``Indian Princess'' or the ``sexualized squ*w'', and patterns of invisibility through ``colonial amnesia''~\cite{sierp_eu_2020}.
Cases are often covered through a deficit-centered gaze, highlighting the vulnerability and brokenness of Native lives; hapless victims as opposed to survivors of targeted violence spanning generations~\cite{jiwani-symbolic-2009, gilchrist-newsworthy-2010, furey-missing-2023, stillman-missing-2007, harding-historical-2006, drache-what-2016, sommers-missing-2016, lucchesi-missing-2018, razack-race-2002, armstrong-media-2013}. 
%
%
UIHI\cite{lucchesi-missing-2018} found that news coverage comprised ``racism or misogyny or racial stereotyping, including references to drugs, alcohol, sex work, gang violence, victim criminal history, victim-blaming, making excuses for the perpetrator, 
and publishing images/video of the victim's death.'' 
In contrast, cases involving white women tend to garner far more extensive and sympathetic coverage, described as the ``missing white woman syndrome''~\cite{gilchrist-newsworthy-2010, armstrong-media-2013, harding-historical-2006}. 
This lack of visibility reflects and reinforces systemic racism, minimizing the perceived urgency of the violence and reducing public and political pressure for accountability. 
Health disparities among communities cause high rates of lifetime substance abuse, suicide, homicide, incarceration, and anxiety/affective disorders~\cite{czyzewski-colonialism-2011, evans-campbell-indian-2012, ehlers-measuring-2013, walters-bodies-2011, gone-american-2012, kirmayer-mental-2000, brave-heart-historical-2011, brave-heart-american-1998}. 
%
%
In rural communities, health facilities may be underfunded and understaffed, forcing women to travel long distances to receive care~\cite{bachman_violence_2008, grenier_failure_2007, luebke-barriers-2022, klingspohn_importance_2018}. In urban cities, \victims and families often encounter racism and discrimination in ``mainstream'' healthcare and social services, actively violating and discouraging them from seeking assistance~\cite{allan-first-2015, luebke-barriers-2022, palmater-beyond-2011}. 
The absence of accessible shelters, underfunded Indian Health Services (IHS) facilities, and culturally-grounded and trauma-informed healing programs 
disenfranchises families' grief, leaving many without protection or resources~\cite{brave-heart-historical-2011, doka-disenfranchised-1989, luebke-barriers-2022, brave-heart-american-1998}. Finally, under the guise of public health, eugenics, and population control, the US and Canada implemented forced sterilization programs to exercise control over Indigenous women’s bodies, devaluing their lives and treating them as expendable. 
These programs sterilized thousands of Indigenous women (1960s and 1970s, till 2000s in Canada~\cite{stote-act-2015}), often without proper consent or through coercive practices such as withholding medical care~\cite{lawrence-indian-2000, rutecki_forced_2011, stote-act-2015}. 

 \subsubsection{Inadequate {Legislation and} Policy Efforts}
\label{bg:policy}
Advocacy led by Indigenous women guided a series of legal policies and reforms to address the crises, yet the enforcement remains uneven and inadequate. 
{Unlike policy reforms that seek to make colonial systems more inclusive (e.g., diversity, equity, and inclusion policies), advocates have fought for the restoration of Indigenous sovereignty, recognizing Indigenous jurisdiction over justice systems and reclaiming control of land and resources.
After years of deliberate ignorance and inaction, the US finally addressed the crisis through the Tribal Law and Order Act (2010) and the Violence Against Women Act (VAWA) (reauthorized in 2013\footnote{The 2013 Reauthorization of the Violence Against Women Act (VAWA) reaffirmed tribes' inherent power to exercise Special Domestic Violence Criminal Jurisdiction (SDVCJ)  ``though it does not cover all forms of domestic violence.''~\cite{national_congress_of_american_indians_ncai_research_2021}. Till 2021, only 28 tribes implemented SDVCJ, leading to 128 prosecutions of perpetrators~\cite{national_congress_of_american_indians_ncai_vawa_2018}.} and 2022), aimed to improve coordination across justice systems~\cite{reporter-council-2018, deer-beginning-2015, house-fact-2022}. 
The 2019 ``Operation Lady Justice'' (aka  U.S. Presidential Task Force)  established state task forces to improve coordination among State agencies.\footnote{Departments of Justice (DoJ), Interior (DOI), Health and Human Services (HHS), Federal Bureau of Investigation (FBI), Bureau of Indian Affairs (BIA), along with state and tribal law enforcement, organize listening sessions with Native communities.} However, the initiative faced criticism for a lack of transparency and tribal involvement. 
More recently, Ashlynne Mike AMBER Alert In Indian Country Act (2018)\footnote{Ashlynne Mike, an 11-year-old Navajo girl, was abducted and killed. The lack of a sovereign alert system on the Navajo Nation and jurisdictional confusion among law enforcement agencies. Named after her passing, the law provides funding to help tribes establish their own emergency alert systems.}, Savanna’s Act (2020),\footnote{Named after Savanna LaFontaine-Greywind, a Spirit Lake Tribe member, had dreamed of becoming a registered nurse and was 8 month pregnant when she was brutally murdered.} and the Not Invisible Act (2020)~\cite{the-not-invisible-act-commission-not-2023} were passed to improve alerting systems, data collection and inter-agency task forces (35 states have a MMIR taskforce)~\cite{native_americans_today_mmiw_2025}. 
In Canada, the 2019 national inquiry~\cite{national-inquiry-into-missing-and-murdered-indigenous-women-and-girls-reclaming-2019} forced the government to make specific recommendations to improve collaboration among federal, provincial, and Tribal partners. 
Beyond recognition, implementation has been slow and inconsistent, with critics noting a lack of Native representation, adequate funding, oversight, and concrete action; some calling it performative liberal politics to garner Native votes~\cite{felt_mobilizing_2016}. Similarly, the National Action Plan on MMIWG (2021) exists largely as an aspirational document without enforceable accountability mechanisms~\cite{royal-canadian-mounted-police-working-2017}.

\subsection{Decolonial Academic Research to Address MMIR}
\label{bg:research}
\paragraph{Coloniality in Academic Research} {Coloniality has long silenced Indigenous peoples' voices, languages, histories, and experiences as inferior, thereby justifying dispossession, assimilation, and neglect, rendering them ``invisible''} {~\cite{quijano-coloniality-2007, mignolo-decoloniality-2018, maldonado-torres-coloniality-2007, phyak-epistemicide-2021, mignolo-epistemic-2009, grosfoguel-epistemic-2007, fanon-black-2008, fanon-wretched-2002, said-orientalism-1979, wynter-unsettling-2003, mignolo-introduction-2007, escobar-worlds-2007, mignolo-darker-2011, mbembe-out-2021}.} 
Following the ``colonial impulse''~\cite{ali-brief-2016, dourish-ubicomps-2012}, HCI often reproduces such colonial and hegemonic power structures through Eurocentric epistemologies, universalizing design methods, parachuting, and extractive methodologies~\cite{andrus-data-2025, lin-techniques-2021, strohmayer-chiversity-2018, irani-postcolonial-2010, bidwell-moving-2016, alcoff_extractivist_2022}, deficit-oriented design~\cite{tuck-suspending-2009, to-flourishing-2023, wong-villacres-reflections-2021}, and lack of citations on methods from Majority World~\cite{connell-southern-2021, kumar-braving-2021, kumar-hci-2019, kumar-hci-2020, cannanure-hci-2023, kumar-hci-2018}.
``Epistemicide'' is not just the systematic destruction of Indigenous knowledge systems, but also the absence in state records, as demonstrated by the vast discrepancies in the data on missing persons in government databases~\cite{lucchesi-missing-2018}. Unfortunately, much of the scholarship has been produced through extractive empiricist frameworks. 
Computing research extracts knowledge from marginalized communities without reciprocity and tangible benefits~\cite{volzer_finite_2025, satariano-global-2023, tacheva_ai_2023, crawford_atlas_2021, mcelroy_digital_2019, kwet_digital_2019, hao_empire_2025}; sidelining Indigenous perspectives on justice and safety. We discuss how the extractive logics embedded within State technologies perpetuate epistemic violence by failing to address the \mmiw crisis.

\paragraph{Decolonial Academic Research} Decolonial {research} challenges the dominance of Western epistemologies, instead emphasizing relationality, pluraliversality, and the co-creation of knowledge with marginalized communities, re-imagining design and technology as sites of care, resistance, and possibility~\cite{escobar-designs-2018}. 
`Decoloniality' in computing resists narratives of white-hetero-patriarchal construction of technology and ``progress'' as inevitable~\cite{bidwell-moving-2016, akama_speculative_2016, abdelnour-nocera-reframing-2013, awori-transnationalism-2015, bidwell_landscapes_2008, winschiers-theophilus-determining-2010, allard-tremblay-rationalism-2021, erete_method_2023, petterson_co-designing_2024, sondergaardfeministvoicesecological2022, schlesinger_intersectional_2017, hankerson_does_2016}. 
%
{HCI scholars have worked with Indigenous communities through participatory design practices across the globe in Africa~\cite{bidwell-moving-2016, awori-transnationalism-2015, winschiers-theophilus-determining-2010, kotut-clash-2020, kotut-trail-2021, kotut-winds-2022, farao-transformative-2024, adamu-remembering-2024, tran-oleary-who-2019, stichel_namibian_2019, warrick_social_2016}, Latin America~\cite{escobar-designs-2018, guerrero-millan-cosmovision-2024, wong-villacres-lessons-2021, pinto_re-learning_2024, lehuede_alternative_2024}, Arab world~\cite{lazem-challenges-2022}, Polynesia~\cite{noe-where-2024, barcham_collaborative_2025}, Australia~\cite{peters-participation-2018, woodlock_living_2023, shaw_mobile_2014, brereton_beyond_2014}, and Asia~\cite{winschiers-theophilus-determining-2010, peter_navigating_2024, sheh-hamidulfuad-collaborative-2024, arias-indigenous-2018}. Many of these partnerships have resulted in innovative design technologies such as broadcasting technologies~\cite{guerrero-millan-cosmovision-2024}, language revitalization applications~\cite{leeanne-how-2015, hopwood-decolonizing-2023, simon-fraser-university-researchers-2021, garcia-how-nodate, roche-articulating-2019}, and online groups to preserve stories and Indigenous knowledge~\cite{kotut-winds-2022, kotut-clash-2020, kotut-trail-2021}. } 
%
{We use Indigenous decolonial feminist lenses to acknowledge the impacts of historical and ongoing injustices on Indigenous peoples. 
Indigenous feminist lenses call for a transformational shift; depict stories of resilience and healing rooted in Indigenous onto-epistemologies, and not just stories of pain and suffering through a Western empirical lens~\cite{verges-decolonial-2021, dignazio_counting_2024, byrd-being-2023, blackdeer-unsettling-2023, cortez_decolonial_2021, dorries-beyond-2020, mack-our-2019, deer-beginning-2015, arvin_decolonizing_2013, smith-decolonizing-2021}. Prior work in HCI has utilized transformational justice lenses through asset-based HCI~\cite{hardy_turn_2024, wong-villacres-reflections-2021, wong-villacres-lessons-2021, coughlin-strength-2020} and transformative justice or restorative justice~\cite{chordia-social-2024} in a variety of contexts such as child sexual abuse~\cite{sultana_shishushurokkha_2022}, domestic abuse~\cite{rabaan_exploring_2021, rabaan_survivor-centered_2023, rabaan_healing_2024}, youth trust in social media~\cite{schoenebeck-youth-2021} and criminal justice~\cite{musgrave-techno-mediated-2025}.}
{We are inspired by Anzaldúa’s ``boundary work''~\cite{anzaldua-borderlands-2021, anzaldua-light-2015}, Mbembe's necropolitics~\cite{mbembe-necropolitics-2016, mbembe-out-2021, de-finney-rekinning-2022}, and Smith's decolonial research~\cite{smith-decolonizing-2021} as guiding lenses for capturing this transformation and reconfiguration. Inspired by Anzaldúa’s theorization of nepantla/mestiza, we embrace the liminal space between our identities as colonized peoples from the Majority World.
We become co-participants in this continuously reshaping liminal space that allows us to center this research work through reflexivity (\autoref{position:reflexivity}), reciprocity and relational accountability (\autoref{position:reciprocity}), critical humility and cultural sensitivity (\autoref{position:sensitivity}), while respecting communities' right of refusal (\autoref{position:refusal}).
}

\paragraph{Research to Address MMIR}
{Academic research on the \mmiw crisis has played a dual role; it has illuminated the scale of the violence and, at times, ceased epistemic violence and brought interventions to investigation agencies. 
%
\citeauthor{ficklin-fighting-2022}~\cite{ficklin-fighting-2022} represent stories of \mmiw victims highlighting the art-based and academic decolonial actions taken by Indigenous scholars in Psychology. 
%
\citeauthor{bailey-missing-2016}~\cite{bailey-missing-2016} call technology a modern tool of colonial oppression; how it enables stalking, domestic violence, online harassment, and trafficking to harm Indigenous women. \citeauthor{bailey-missing-2016} contextualize how the RCMP has historically used DNA technologies to take away agency in the guise of ``effective investigations''.
In HCI, scholars traced Native political tweets such as \#MMIW, \#NativeLivesMatter 
and \#pipeline 
during elections in Canada~\cite{felt_mobilizing_2016} and U.S~\cite{vigil-hayes-indigenous-2017, vigil-hayes-complex-2019}.
%
~\citeauthor{bleeker-efficacy-2023}~\cite{bleeker-efficacy-2023}, ~\citeauthor{diehl-is-2019}~\cite{diehl-is-2019}, and ~\citeauthor{moeke-pickering-understanding-2018}~\cite{moeke-pickering-understanding-2018}, show how Indigenous advocates ``reframe'' a racialized violent discourse on social media as a display of sovereignty, colloquially known as ``Indigenous Data Sovereignty''~\cite{ricaurte-data-nodate, lehuede_alternative_2024, kukutai_indigenous_2016, yanchapaxi_indigenous_2025, tsosie_models_2020, carroll_care_2020}.
%
{Continuing with its ethos, cartographers like  
Lucchesi~\cite{lucchesi-mapping-2019, lucchesi-indians-2018, lucchesi-spatial-2020, lucchesi-mapping-2022} and ~\citeauthor{kidd-extra-activism-2019}~\cite{kidd-extra-activism-2019} create digital maps to trace the connections between land-based violence and Indigenous bodies.
~\citeauthor{miner-informatic-2022}~\cite{miner-informatic-2022} argue that these are how visual mapping technologies (e.g., Native Land Map\footnote{Native Land Map http://native-land.ca} and Kitikmeot Atlas Project\footnote{Kitikmeot Atlas Project https://atlas.kitikmeotheritage.ca}) disrupt settler-cartographic practice to decolonize maps and borders. 
%
%
Finally, Waking Women Healing Institute created a resource map to help families quickly locate grassroots agencies~\cite{noauthor_waking_nodate}. In the absence of data, ~\citeauthor{ricaurte-data-nodate}~\cite{ricaurte-data-nodate} argued that grassroots data activism practices around feminicide acts as a form of ``epistemic disobedience'' to the colonial extraction prevalent in Western data science. D'Ignazio~\cite{dignazio_counting_2024, dignazio_feminicide_2022} extends Ricaurte's framework to generate counterdata that can combat this absence of data. 
Moreover, ~\citeauthor{suresh_towards_2022}~\cite{suresh_towards_2022} designed an ML model that supports Sovereign Bodies Institute (SBI) efforts to collect and monitor \mmiw data effectively. However, their work centers only theoretically engage with feminist theories. However, their work centers on Western empiricism and falls short of providing an actionable tool that meets activists' needs.
%

\paragraph{Research Gap.}
Limited HCI research has honored the experiences of murdered or missing \victims, families, advocates and tribal police. The lack of grassroots voices at the forefront of the \#MMIR movement has long been overlooked in technology design practices in HCI, perpetuating cycles of violence and invisibility.} 
Thus, in this research, we investigate the socio-technical barriers faced and grassroots strategies to locate missing or murdered relatives, seek safety and support, and raise awareness of the \mmiw movement. 
\section{Situating Ourselves}
\label{sec:positionality}


\begin{myquote}{\actortable{Cutcha Risling Baldy}{cutcharislingbaldy}~\cite{mueller-nevada-2020}}
    We repeat statistics about our sexual assault and violence and the attempts to take our futures from us and the stories about our deaths, not our lives, not our futures. That's not what I want to do ... this is a story about us and if we recenter our resurgence, our fights, our resistance and the fact that we will do it.
\end{myquote}

\subsection{Reflexivity}
\label{position:reflexivity}
We acknowledge our research training in Western epistemologies with a strong sensitivity to Decolonial and Indigenous onto-epistemologies, creating reflexive distance to the Indigenous peoples of Turtle Island~\cite{patel-beyond-2006, paris-humanizing-2014, gani-positionality-2024, pillow-confession-2003, pillow-epistemic-2019, pillow_dangerous_2010}.
We are an interdisciplinary team that holds academic positions in a Predominantly White \univ. We embrace the hybridity of Western and Indigenous ways of being, and we weave the empirical data statistics with grassroots stories (\autoref{methods:data_analysis}). 
%
%
All but two authors provide direct support services to survivors of domestic violence. Three authors have provided trauma-informed support for 150+ survivors through \techclinic, a non-profit organization that works in collaboration with 23 violence shelters in \wi. One author is a nurse with expertise in Indigenous healthcare, an enrolled member of \badriver, and serves in the data subcommittee of \wi Department of Justice's \mmiw Taskforce.  
%
Finally, a few authors are abuse survivors, and their experiences have been instrumental in embodying trauma-informed lenses (see our approach to qualitative analysis~\autoref{methods:data_analysis}).
%
Reading and writing about the traumatic stories is emotionally daunting. The authors felt the vicarious trauma from engaging with this work and took intentional steps to care and support each other (e.g., taking breaks, checking in with each other)~\cite{coles-qualitative-2014, bellini2023sok}. {Finally, we obtained approval from the \univ Institutional Review Board (IRB) and the \univ Tribal Liaison Office. The project was determined to be non-human-subjects research.} 

%
%


\subsection{Reciprocity and Relational Accountability}
\label{position:reciprocity}
We led this study from a place of knowing and heart, with a greater sense of purpose to reckon with and give back to the land that we live on and generations of Indigenous peoples that it has supported. 
Three authors were born and raised in the US. Four authors immigrated from Majority World countries; we share a deep connection with the colonized Indigenous communities across the globe. All authors deeply respect the knowledge embedded in the \hochunk custodianship of \teejop and recognize their continuing connection to land, water, and community in \teejop. \univ occupies \hochunk land and has the largest grouping of Indian burial mounds on a university campus; anywhere in the world.
Our university could not have been established or sustained were it not for state and federally sponsored settler colonialism that dispossessed and displaced American Indian nations and communities. Through this work, we \removed{must now} confront the outcomes of unjust land treaties and the harm caused by \univ's complicity in an ongoing genocide as we seek reconciliation with Indigenous communities of \wi.
%
Through the use of desire-centered framing, we turn the research process itself into a generative gift~\cite{jimmy-towards-2019, 
kimmerer-braiding-2013, marcel-mauss-gift-2015} 
-- an act of reciprocal resistance to bring medicine and healing to the Indigenous communities. 
Finally, we provide \mmiw advocates and future researchers with an editable spreadsheet of web pages during publication. This will allow the advocates to crowd-source stories residing in news articles and reports. {We have held conversations with NIWRC on how their advocates can utilize this data going forward.}  
For reviews, we have uploaded a CSV file (supplementary materials) containing a list of coded pages, domain categories, technology categories, LLM-generated summaries, and direct quotes from the pages. 

\subsection{Critical Humility and Cultural Sensitivity}
\label{position:sensitivity}
To personify utmost cultural sensitivity and respect for Indigenous ways of being and knowing, we followed several practices to develop ``thread sensitivities''~\cite{jimmy-towards-2019, yang-weaving-2023}. 
From May 2024 to August 2025, we made connections with visiting Elder scholars in \univ, and advocacy agencies including the National Indigenous Women's Resource Center (NIWRC)~\cite{niwrc-mmiw-2024}, Urban Indian Health Institute (UIHI)~\cite{urban-indian-health-institute-urban-nodate}, Navajo Missing \& Murdered Diné Relatives (MMDR)~\cite{missing--murdered-dine-relatives-mmdr-missing-nodate}, MMIWhoismissing~\cite{mmiwhoismissing-mmiwhoismissing-nodate}, Healing Intergenerational Roots (HIR) Wellness~\cite{hir-wellness-home-nodate}, and Our Native Daughters~\cite{not-our-native-daughters-not-nodate} in developing a reciprocal relationship. We were invited to present our research plan to the \wi \mmiw taskforce and \lco \& \ldf reservations, where members provided crucial feedback.
Moreover, NIWRC invited us to conduct two trainings (one virtual, one in-person) on technology-facilitated abuse to empower advocates to build agency over technology and abuse that stems from it. We absorbed Indigenous teachings, wisdom, and creation stories from families, advocates, and Elders. 
We participated in spiritual ceremonies, including Powwows, a doll-making workshop, and smudging to cleanse our spirits and heal from vicarious trauma (see \autoref{fig:cultural-sensitivity}). 
%
We witnessed stories of trauma, pain, and suffering from family members involved in support and advocacy work for decades~\cite{pillow-epistemic-2019}. The stories also highlighted the coordinated efforts to resist and heal, which further strengthened our resolve to center Indigenous analytical frameworks (see \autoref{methods:data_analysis}). Finally, we shared a paper draft with Indigenous scholars \& advocates for feedback.

\subsection{Right of Refusal}
\label{position:refusal}
We hope to bring forth Indigenous activism and knowledge systems to the attention of the Western academic world through ``citational justice''~\cite{itchuaqiyaq-citation-2020, absolon-kaandossiwin-2011}. At the same time, right of refusal means some forms of knowledge are ``sacred'' and should not be shared or scrutinized by academia, as doing so risks commodifying or misusing culturally significant information~\cite{tuck-r-words-nodate, simpson-ethnographic-2007, weiss-refusal-2016, garcia-no-2022, gottschalk-druschke-access-2022, tuck-unbecoming-2014}. Scholars argue for a protective boundary that respects knowledge as community-held and not necessarily subject to external validation or control. Not every issue within a community needs academic intervention. Sometimes, research fails to address the actual needs or wishes of a community. Therefore, we use the right of refusal as an intentional lens to make choices on how, why, and what to represent in this paper. We consulted with the \wi Department of Justice's \mmiw Taskforce, which led to careful considerations about how to represent the crisis in academia to drive meaningful action. Therefore, instead of centering pain, we make a careful effort to respect refusal by not analyzing missing posters in search and rescue groups, pages with Indigenous languages, and ensuring that traditional spiritual practices are not appropriated.
\begin{figure*}[t]
    \centering
\begin{tikzpicture}[
    font=\small,
    data/.style={
        draw,
        rounded corners=2mm,
        fill=yellow!20,
        minimum width=2.0cm,
        minimum height=1.1cm,
        align=center,
        line width=0.8pt
    },
    stepp/.style={
        draw,
        rounded corners=2mm,
        fill=blue!20,
        minimum width=2.0cm,
        minimum height=0.65cm,
        align=center,
        line width=0.8pt
    },
    output/.style={
        draw,
        rounded corners=2mm,
        fill=green!25,
        minimum width=3.0cm,
        minimum height=1.1cm,
        align=center,
        line width=0.8pt
    },
    arrow/.style={->, line width=0.8pt}
]

\node[data] (queries) at (0,0) {54k\\[0.25em] Queries};

\node[stepp] (search) at (3.2,1.55) {Google Search};
\node[data] (pages) at (3.2,0) {123k\\[0.25em] Web Pages};

\node[stepp] (downsample) at (6.4,1.55) {Downsample};
\node[data] (sample) at (6.4,0) {140\\[0.25em] Web Pages};

\node[stepp] (coding) at (9.6,1.55) {Coding};
\node[output] (codebook) at (9  .6,0) {Codebook};

\draw[arrow] (queries.east) -- ++(0.45,0) |- (search.west);
\draw[arrow] (search.south) -- (pages.north);

\draw[arrow] (pages.east) -- ++(0.45,0) |- (downsample.west);
\draw[arrow] (downsample.south) -- (sample.north);

\draw[arrow] (sample.east) -- ++(0.45,0) |- (coding.west);
\draw[arrow] (coding.south) -- (codebook.north);

\end{tikzpicture}
    \mycaption{Data Processing Pipeline}{shows the data collection, 
    downsampling,  and qualitative analysis  of web pages on technology use in the \mmiw crisis.}
    \label{fig:pipeline}
    \Description{The figure is a flowchart illustrating a data processing pipeline. It consists of three main rectangular blocks connected sequentially with arrows. The first block is tan and labeled "54k Queries," receiving input from a smaller blue rectangle labeled "Google Search." An arrow leads to the next tan rectangle labeled "123k Web Pages," indicating data addition. A blue rectangle labeled "Downsample" is above, directing to the "123k Web Pages" block, illustrating a transition to the next tan block labeled "140 Web Pages." Finally, a blue rectangle labeled "Coding" connects to a green rectangle titled "Codebook." The diagram visually depicts the stages of data collection, downsampling, and qualitative analysis.}
\end{figure*}

\section{Methods}
\label{sec:methods}

To answer the research questions, we collect web pages that cover facets of the \mmiw crisis instead of interviewing advocates and families to avoid retraumatization through an extractive inquiry. Our goal is not to collect new testimonies but rather to surface and represent existing knowledge that is already publicly available and not usually represented in academia. 
We conducted a preliminary search to examine the kinds of {resources} available and help ascertain that the web pages can be a viable data source.
We searched for MMIR-related articles popular platforms such as Google, Facebook, and Twitter/X.
We found news articles covering specific cases,  blog posts and reports that contextualize the broader crisis and social media posts with missing persons posters to help locate missing relatives.
These content were created by former survivors, families, and advocates, including comments expressing solidarity from relatives worldwide.
We see this as an initial, respectful methodology within HCI research that can later inform whether and how more direct, participatory forms of inquiry (e.g., interviews, workshops) might be needed or ethically pursued.
Next, we describe our data pipeline (see \autoref{fig:pipeline}) through data collection (\autoref{methods:data_collection}) and analysis  (\autoref{methods:data_analysis}). Finally, we acknowledge the limitations of our study, especially in using automation tools such as LLMs in a decolonial project and the epistemic and ecological harms they may entail (\autoref{sec:limitations}). 





\subsection{Data Collection}
\label{methods:data_collection}
\paragraph{Collecting Pages.}
We gather public pages indexed on the most widely used search engine (Google)\footnote{We recognize that not all information about the \mmiw crisis is likely to be indexed by Google (see our limitations noted in (\autoref{sec:limitations})).}.
%
%
We collected \ntribes tribe names that have inhabited Turtle Island (now North America) from 5 independent websites~\cite{nativeamericantribes, waldmanencyclopedianativeamerican2006, firstnationscanadaa, firstnationscanada, firstnationscanada2025}.  This approach allowed us to avoid scoping only federally-recognized tribes across the US-Canada colonial borders 
(see  \autoref{bg:jurisdictional-tension}). 
Next, we generated \nqueries\ seed queries by using four templates (`Murdered and Missing OR Missing ``[\textit{T}]'' [\textit{A}]',
    `Violence against ``[\textit{T}]'' [\textit{A}]', `Sex trafficking against ``[\textit{T}]'' [\textit{A}]', `Human trafficking against ``[\textit{T}]'' [\textit{A}]'). 
Here, \textit{T} can be Native, Indigenous, Native American, or a specific Tribe. \textit{A} can be women, girls, sisters, peoples, communities, or Nations. 
An example query is `Missing and Murdered ``Shawano'' women.'\footnote{We tried small variants of these queries, like replacing ``against'' with an ``AND'', and did not observe any meaningful difference in the results beyond minor reordering.}
Next, we 
queried Google Search\footnote{To respect Google's fair use and terms of service and throttling, we made fewer than two queries per second, which should have a negligible effect on Google’s regular operations.} and retrieved the first 20 results.
%
We collected data from December 2024 to February 2025.
%
In total, we collected \ntotalresults\ results. On average, we received 7 results per query. 
Next, we removed duplicates to create a set of \nuniqueresultsprose\ unique web pages hosted on \nuniquedomains\ domains. 
\begin{table*}[t]
\rowcolors{2}{rowAlt}{rowBase}
  \centering
  \mycaption{Overview of the Data}{Categories of pages' domains in our (a) full dataset and (b) sample we used for qualitative coding.}
  \Description{The figure is a table with the title "Table 1. Overview of the Data – Categories of pages’ domains in our (a) full dataset and (b) sample we used for qualitative coding." The table is divided into two main columns titled "Full Dataset (N=123,029)" and "Stratified Sample (N=140)." Each column has three sub-columns: "Category," "\# Pages," and "\% Pages." There are nine rows listing various categories: Education, Non Profit Organization, News, Social Media, Government, E-commerce platform, Blog article, International Organization, Law enforcement, and Unknown. The table provides the number and percentage of pages for each category in both the full dataset and the stratified sample.}
    \label{fig:domain-cats}
{%
    \begin{tabular}{lrrcrr}
    \toprule
    &\multicolumn{2}{c}{\textbf{Full Dataset (N=\nuniqueresultsprose)}} & & \multicolumn{2}{c}{\textbf{Stratified Sample (N=\ncodedpages)}}\\
    \cmidrule{2-3}
    \cmidrule{5-6}
    {Category} & {\# Pages} & {\% Pages} &  & {\# Pages} & {\% Pages} \\ 
    \hline \hline
        Education & 30,749 & \onlypercent{30749}{\nuniqueresults} &  & 25 & \onlypercent{25}{\ncodedpages} \\
        Non Profit Organization & 26,932  & \onlypercent{26932}{\nuniqueresults}  & & 16 & \onlypercent{16}{\ncodedpages} \\
        News & 19,224 & \onlypercent{19224}{\nuniqueresults}  & & 47 & \onlypercent{47}{\ncodedpages} \\
        Social Media & 17,328  & \onlypercent{17328}{\nuniqueresults}  & & 9 & \onlypercent{9}{\ncodedpages} \\
        Government & 14,587 & \onlypercent{14587}{\nuniqueresults}  & & 13 & \onlypercent{13}{\ncodedpages}\\
        E-commerce platform & 9,397 & \onlypercent{9397}{\nuniqueresults}  & & 6 & \onlypercent{6}{\ncodedpages}\\
       Blog article & 8,437 &  \onlypercent{8437}{\nuniqueresults} & & 21 & \onlypercent{21}{\ncodedpages} \\
        International Organization & 5,623 & \onlypercent{5623}{\nuniqueresults}  & & 6 & \onlypercent{6}{\ncodedpages}\\
        Law enforcement & 1,228 &  \onlypercent{1228}{\nuniqueresults}  & & 17 & \onlypercent{17}{\ncodedpages} \\
        Unknown &  531 & \onlypercent{531}{\nuniqueresults} & & 9 & \onlypercent{9}{\ncodedpages}\\
    \bottomrule
    \end{tabular}
}\\
\end{table*}

\paragraph{Overview of the Data}
\label{methods:data_categorization}
We categorize \nuniquedomains\ domains into 10 categories of pages (see \autoref{fig:domain-cats}). To do this, we use a llama-3.1:8b LLM model (aka Domain Categorizer (DC-LLM)) and prompt (see~\autoref{fig:prompt1}) with each domain's title and description meta HTML tag. We measured the accuracy of the DC-LLM (97\% accuracy; 3\% misclassifications) by manually verifying with 100 most popular domains. 
Educational pages, including academic articles, academic library websites, and Wikipedia articles ---were the most prevalent, making up 22\% of the dataset. 20\% of the pages were created by nonprofit organizations. News and social media websites accounted for 14\% and 13\% of the pages, respectively. 
%
Next, we show the top ten most popular domains in~\autoref{fig:top-domains}. 
Four of the ten most common domains were social media websites (Facebook, Instagram, Reddit, and LinkedIn). These four domains alone contributed 13,420 pages to our dataset (10.9\%). The popularity of the social media websites can be owed to the grassroots activism (\actionref{5}).
%
Other top domains related to national (cbc.ca) and Indigenous news (ictnews.org), education and research (en.wikipedia.org, researchgate.net, jstor.org), showing the initiative by Native journalists to represent families' stories through independent news websites.
%
%
Finally, the government pages (justice.gov) contained reports written by advocacy organizations, task forces, tribal police departments, and government agencies. 
The law enforcement and government pages focused on legislative policies, bills, and laws to improve investigations and reconciliation efforts with Tribal Nations (\actionref{6}). 

\begin{table*}[t]
\rowcolors{2}{rowAlt}{rowBase}
  \centering 
  \mycaption{Top 10 domains}{represented in our dataset. * denotes that the domain is created by Indigenous advocates.}
  \Description{The figure shows a table titled "Table 2. Top 10 domains – represented in our dataset." A note indicates that an asterisk () denotes domains created by Indigenous advocates. The table is split into two sections: "Full Dataset (N=123,029)" and "Stratified Sample (N=140)". Each section includes columns labeled "Domain," "\# Pages," and "\% Pages." The table lists various domains with corresponding numbers and percentages of pages. In the full dataset section, "facebook.com" has the highest number of pages at 7,776, representing 6.32\% of the pages. Other domains such as "instagram.com," "cbc.ca," and "reddit.com" follow. In the stratified sample section, "en.wikipedia.org" and "indianz.com" each have 4 pages, each making up 2.86\% of pages. The list includes diverse domains, with some marked by an asterisk to signify Indigenous advocacy creation.
}
    \label{fig:top-domains}
{%
    \begin{tabular}{lrrclrr}
    \toprule
    & \multicolumn{2}{c}{\textbf{Full Dataset (N=\nuniqueresultsprose)}} & & \multicolumn{3}{c}{\textbf{Stratified Sample (N=\ncodedpages)}}\\
    \cmidrule{2-3}
    \cmidrule{5-7}
    {Domain} & {\# Pages} & {\% Pages} & & {Domain} & {\# Pages} & {\% Pages}    \\ 
    \hline \hline
        facebook.com      &  7,776 & \onlypercent{7776}{\nuniqueresults} & & en.wikipedia.org & 4 & \onlypercent{4}{\ncodedpages} \\
        instagram.com     &  3,627 & \onlypercent{3627}{\nuniqueresults} & & indianz.com*  & 4 & \onlypercent{4}{\ncodedpages} \\
        cbc.ca            &  1,387 & \onlypercent{1387}{\nuniqueresults} & & facebook.com & 3 & \onlypercent{3}{\ncodedpages} \\
        reddit.com        &  1,175 & \onlypercent{1175}{\nuniqueresults} & & nativenewsonline.net* & 3 &  \onlypercent{3}{\ncodedpages}\\
        en.wikipedia.org  &  1,029 & \onlypercent{1029}{\nuniqueresults} & & doj.state.wi.us   &3 &  \onlypercent{3}{\ncodedpages}\\
        ictnews.org*       &   927 & \onlypercent{927}{\nuniqueresults} & & legendsofamerica.com* & 2& \onlypercent{2}{\ncodedpages} \\
        researchgate.net  &  868  & \onlypercent{868}{\nuniqueresults} & & t3ps.ca* & 2& \onlypercent{2}{\ncodedpages}\\
        linkedin.com      &   842 & \onlypercent{842}{\nuniqueresults} & & rcmp-grc.gc.ca & 2& \onlypercent{2}{\ncodedpages}\\
        jstor.org         &   825 & \onlypercent{825}{\nuniqueresults} & & gsps.ca*  & 2& \onlypercent{2}{\ncodedpages}\\
        justice.gov       &   771 & \onlypercent{771}{\nuniqueresults} & & tsuutinapolice.com* &2 & \onlypercent{2}{\ncodedpages}\\
    \bottomrule
    \end{tabular}
}\\
\end{table*}

\paragraph{Data Downsampling}
\label{methods:data_filtering}

Not all of \nuniqueresultsprose\  web pages discuss technologies used by the communities.
Manually filtering relevant pages from all \nuniqueresultsprose\ pages was not feasible.
Therefore, we employed an LLM-assisted filtering method to create a set of relevant pages that reference ``technology use''. To ensure that pages from all domain categories sampled roughly equally, from each domain category, we use a frequency-based stratified sampling method~\cite{neyman-two-1992}. (Note: pages may have multiple domain categories.) We use the domain category as the strata for the sampling method.
We start with the 500 most popular web pages (50 pages per domain category, 50 times 10 \= 500 pages). 
We prompted the Qwen2.5 14b 1M\footnote{We use a 1 million token context window size to adapt to the content size of the web pages.} LLM model (aka Content Categorizer (CC-LLM)) to categorize the page into one or more of the technology categories (see~\autoref{fig:prompt2}). 
%
CC-LLM categorized 335 pages into one or more technology-use categories (\autoref{fig:prompt2}). 
We excluded 166 pages --- 5 pages were PDFs which had parsing errors while loading into LLM, 8 resulted in timeout (limited LLM runtime to 10 minutes), 96 were too large (more than 25000 words), and 56 were too small (fewer than 20 words). 
We manually verified and adjusted the categorizations to ensure accuracy. For this, two authors read 335 pages collectively and reduced our sample to 116 pages, excluding 219 pages that were deemed not relevant to technology use. 
Since news articles contained heartfelt stories with direct quotes from survivors, families, advocates, and tribal police, we oversampled \morenews\ relevant news articles for qualitative analysis.  We ensured that Native-led websites were included in the final analysis to highlight grassroots voices. The final total that we analyzed was \ncodedpages\ pages.

\subsection{Qualitative Analysis}
\label{methods:data_analysis}

\paragraph{Inductive Coding.} 
Our coding process was rooted in Indigenous onto-epistemologies of oral traditions, relational accountability, and community-based ethics, where survivors', families', and advocates' stories embody both data and theory~\cite{kovach-indigenous-2021, absolon-kaandossiwin-2011, anderson-recognition-2016}.
%
To create an initial codebook, the first author coded the 50 most popular pages that contain ``technology'' in their URL to create the top-level codes.
%
Two authors independently coded \ncodedpages\ web pages and then collaboratively generated a initial codebook. Critically guided by one Native author to ensure cultural grounding, all authors discussed emerging themes, resolved disagreements, and iteratively refined the codebook until reaching thematic saturation. We carefully read the stories by inviting senses: emotional, spiritual, intellectual, and physical. We engaged with our own embodied memories, dreams, and experiences of violence, loss, and resilience. 
This interpretive stance allowed us to witness harm but also resilience, reclamation, and resistance~\cite{pillow-epistemic-2019}. Rather than siloed constructs, we viewed themes of barriers, actions, and recommendations as interconnected with personal and cultural loss, identity, and community well-being~\cite{absolon-kaandossiwin-2011}\footnote{\citeauthor{absolon-kaandossiwin-2011}~\cite{absolon-kaandossiwin-2011} posits that Indigenous research must be conducted ``with spirit'' (\emph{Kaandossiwin}), emphasizing that knowledge emerges through relationships, ceremony, and lived experiences rather than a detached objective and mechanical way.}.
 



\paragraph{Storytelling Representation.} 
Storytelling is both a methodological and political act: it validates the lived realities of Indigenous women and families, creates collective memory, and resists the epistemic violence that has reduced Indigenous lives to statistics. 
Indigenous epistemologies embed identity, womanhood, resistance, and healing through heartfelt stories~\cite{yang-weaving-2023, rowat-walking-2019, absolon-kaandossiwin-2011, kovach-indigenous-2021, anderson-recognition-2016, ficklin-fighting-2022, corntassel-indigenous-2009}. Many reports contained stories and ancestral teachings to guide their actions in seeking healing and justice. 
We ask the readers to emotionally connect to the stories and witness the grief and resilience of generations of Indigenous peoples. 
The names of the pages mirror this balance between grief and strength. Some signify the stories of intergenerational trauma felt from the violence (e.g. ``A mother's worst nightmare''~\cite{mabie-mothers-2022}, ``When Your Loved Ones Go Missing and Authorities Don’t Care''~\cite{gable-when-2023}, ``They Trespass Her Body Like They Trespass This Land''~\cite{swanson-they-2020}
) and resilience (e.g. ``Our Bodies, Our Stories''~\cite{echo-hawk-our-2020}, 
``Lighting the way for those not here''~\cite{german-lighting-2008}, ``Looking Ahead to Build the Spirit of Our Women''~\cite{greater-sudbury-police-service-looking-2017}, ``Every Number is a Person''~\cite{baldy-every-2020}
). We incorporate storytelling into the design of our own paper, allowing the title, section headings, and direct quotes to carry narrative weight. 
We ask the readers to emotionally connect to the stories and witness the grief and resilience of generations of Indigenous peoples. Finally, we attribute the quotes from MMIR advocates and their role in the \mmiw crisis (\autoref{tab:actors}).

\subsection{Limitations}
\label{sec:limitations}


Our work is not without limitations. First, although our web crawling collected 123K unique web pages, we manually coded only \ncodedpages\ pages. 
Moreover, we rely on Google's search index to create a comprehensive dataset of web pages. Search engines are known for localizing their search results to a geographical area, which may contribute to the epistemic exclusion of Indigenous onto-epistemologies~\cite{vaughan-search-2004, noble-algorithms-2018, novin-making-2017, goldman-search-2008}. 
This limits our ability to fully capture the landscape of how the \mmiw crisis and technology intersect. 
%
%
However, despite excluding pages, we ensured that our qualitative analysis reached thematic saturation.
To counter exclusion, we oversampled \morenews\ more news articles to amplify the grassroots stories of families. 

Second, we only analyzed English-language pages (with occasional Native language text). English, as a colonizing language, cannot fully capture the deeper meaning behind Indigenous voices, yet we avoid translating the texts into Indigenous languages to prevent misrepresentation and respect their cultural significance. 
We trust that authors deliberately represented stories in English without losing relationality {and kinship}~\cite{kimmerer-braiding-2013}. 

Third, we used LLMs to assist with preliminary analysis. We limited LLM use to summarization with manual filtering, manually verified all categorizations, and used a hosted LLM instance to protect the privacy. {Big Tech} LLMs are largely trained on dominant internet sources, reinforcing {epistemological erasure through} Western epistemologies and definitions of technology~\cite{zheng_epistemological_2023, mohamed-decolonial-2020, thais-misrepresented-2024, mollema_decolonial_2024, labelle_decolonial_2022, pang_understanding_2025}. {LLMs disproportionately extract from economically vulnerable, disenfranchised, and historically marginalized populations with reduced political power. The proliferation of AI data centers in Indigenous lands reinforces colonial logics of extraction through displacements, ecological damage, and ``man camps''~\cite{volzer_finite_2025, satariano-global-2023, tacheva_ai_2023, crawford_atlas_2021, mcelroy_digital_2019, kwet_digital_2019, hao_empire_2025}.}
Moreover, LLMs perform poorly in violence-related research due to built-in censorship guardrails~\cite{abishethvarman_xguard_2025}. Recognizing these harms, many communities refuse to use, train, or allow AI data-centers on their land~\cite{honor_the_earth_joint_2025}. We extend ethical invitations for future scholars to refuse AI or use AI sustainably with the communities (\recref{5}).

\begin{table*}[t]
    \rowcolors{2}{rowAlt}{rowBase}
    \centering
    \tabfontsize
    \renewcommand{\arraystretch}{1.4}
    \mycaption{Summary of Socio-Technical Barriers}{Faced by \victims, families, advocates, and tribal police.}
    \begin{tabular}{>{\centering\arraybackslash}
                    m{0.05\textwidth}
                    m{0.2\textwidth}
                    m{0.05\textwidth}
                    m{0.4\textwidth}
                    m{0.15\textwidth}}
\rowcolor{headerDark}
\multicolumn{1}{c}{\color{white}\textbf{B\#}} 
& 
\multicolumn{1}{c}{\color{white}\textbf{Barrier}}
& \multicolumn{1}{c}{\color{white}\textbf{Pages}} & \multicolumn{1}{c}{\color{white}\textbf{Summary}} & \multicolumn{1}{c}{\color{white}\textbf{Recommendations}}
\\
          \barrierref{1} & Inaccurate Data Collection  & 47 &
        Law enforcement does not collect the data on missing persons correctly.
        &  \recref{1} 
        \\
 \barrierref{2} & Opaque Data Sharing  Policies  & 47 &
        Law enforcement does not share the data on missing persons correctly.
        &         \recref{1} 
        \\

       \barrierref{3} & Extractive Programs & 10 &
        Colonial-extractive use of technologies and policies that silence Native voices.
        &   \recref{1}
\recref{4}
\recref{5}
 \\
      \barrierref{4} & {Lack of Safe Online Spaces} 
            & 10 &  
              Racial and colonial violence from non-Native perpetrators that transcends online spaces. 
        & 
         \recref{2} \recref{3}
         \recref{5}
        \\
  \arrayrulecolor{accentRule}\bottomrule
    \end{tabular}
    \Description{The figure is a detailed table titled "Summary of Socio-Technical Barriers." Each row lists specific barriers, with corresponding data on the number of pages, summaries, and recommendations.}
    \label{tab:barriers}
\end{table*}

\section{RQ1: Barriers Faced by Indigenous Communities: \textit{Safety is a Broken Promise}}
\label{sec:barriers}

%
%
%
%

We show how the extractive use of state and digital technologies perpetuates the epistemic exclusion of Indigenous peoples, fueling the \mmiw crisis (see \autoref{tab:barriers}).


\begin{myquotel}{\actortable{Jordan Marie Brings Three White Horses Daniel}{jordanmarie}~\cite{swanson-they-2020}}
They go missing in life, they go missing in the media, and they go missing in the data.
\end{myquotel}



\subsubsection*{\barrier{1} Inaccurate Data Collection}
%
State agencies fail to keep an accurate track of data related to \mmiw, rendering the missing or murdered \victims ``invisible'', ``both on reservations and in urban areas, at high rates.''~\cite{swanson-they-2020}. 
State datasets frequently contain incorrect names, gender, race, ethnicity, tribe affiliation{/citizenship}, location of the incident, and even the type of incident. Refer to \autoref{tab:missing-persons-approaches} for contrasting approaches to data collection between State and Indigenous communities.

\paragraph{Name:} {Traditional names, nicknames, and tribe affiliation/citizenship are an ``important kinship marker that often are not legalized.''~\cite{apok-we-2021}}
Failure to record the {Name}, correctly renders the \victims ``nameless''.
\actortable{Lela Mailman}{lelamailman}~\cite{monroe-is-2024} recalls her experience ``When the police finally opened an inquiry, it felt perfunctory. One report referred to Melanie [her daughter who went missing] as `Melissa' [Melanie's sister]; \ldots Melanie's name wasn't entered into NamUs \ldots until three years later.''
\removed{Further, traditional names or nicknames are often not recorded, which are an ``important kinship marker that often are not legalized.''~\cite{apok-we-2021}} 
%
Moreover, rooted in racial profiling, police conflate Indigenous names with Hispanic, African-American or Indian-American names \removed{(e.g., Singh)}~\cite{lucchesi-missing-2018}. 

\begin{myquotel}{Police Representatives from Santa Fe and Seattle~\cite{lucchesi-missing-2018}}
[Many] Native Americans adopted Hispanic names back during colonial times...\ldots
``N'' was being used in the 60s up through the late 70s and early 80s – meant Negro not Native American. 
\ldots Our crime systems are not flexible enough to pick out Native Americans from others in the system...it would be impossible to compile any statistically relevant information.
\end{myquotel}

\paragraph{Race \& Ethnicity.} 
When most crime is perpetrated by non-Native peoples (96\%~\cite{rosay-violence-2016}), the classification of race of both perpetrator and victim becomes crucial. A misclassification of race ``generally favors the larger race, so while [Indigenous peoples] are often misclassified as white, the reverse of that is rare.''~\cite{lucchesi-missing-2018}.
For example, following a ``bias-free policing policy'', the RCMP did not collect or disclose statistics on the perpetrators' race and rejected calls for internal scrutiny about their practice~\cite{wikipedia-missing-2025}. 
Similarly, Wasilla Police Department's missing persons reporting form collapses races into ``Other'' category~\cite{apok-we-2021} (see \autoref{fig:dropdown}).

\begin{myquote}{\actortable{Charlene Aqpik Apok et al.}{charleneaqpikapok}~\cite{apok-we-2021}}
The standard four-box race and ethnicity options of White, Black, Asian, Indian originating with the US Census have been a colonial tool that works to eliminate the existence of Indigenous peoples, instead of truly enumerating us.
\end{myquote}

%


\paragraph{Location and Cause of Death.}
The jurisdiction tension (\autoref{bg:jurisdictional-tension}) complicates the case when the location of the incident isn't accurately collected. Many times, the location is reported through a larger nearby hub in cities, which may obscure cases happening in small rural towns~\cite{apok-we-2021}. 
Further, cases are closed with false causes: ``unknown'' or death: ``undetermined'' due to lack of evidence.
\actortable{Rep. Tawna Sanchez}{tawnasanchez}
~\cite{reyna-secrecy-2024} recalls ``\removed{[The police] heard exactly what we told them they would hear,} suicides are not often investigated, because they’re assumed to be suicides rather than, possibly murders.''. 
\begin{myquote}{\actortable{Jamie Day}{jamieday}~\cite{day-missing-2022}}
Rhonda did not overdose and jump naked headfirst into a trash can. Kristin was not likely hiding in this TV cabinet when she died\ldots the people who found Megan used the word ``beaten'' to describe how she looked. It just doesn't make sense, and the undetermined status feels hurtful to those who are still seeking answers.
\end{myquote}

\begin{figure}
    \centering    \includegraphics[width=0.5\linewidth]{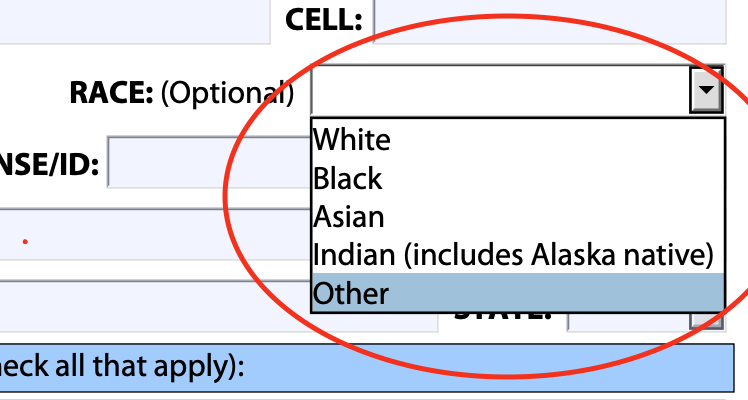}
    \mycaption{Online Reporting Form}{An online reporting form by the City of Wasilla Police Department.}
    \Description{The image shows a portion of an online form used by the City of Wasilla Police Department. The section displayed includes a dropdown menu for selecting race, marked as optional. The dropdown is expanded and lists five options: White, Black, Asian, Indian (includes Alaska native), and Other. The dropdown is highlighted with a red oval. The form also contains fields for "CELL:" and "LICENSE/ID:". Below the fields, there is a note about selecting all that apply.}
    \label{fig:dropdown}
\end{figure}
\footnotetext{https://www.cityofwasilla.gov/DocumentCenter/View/623/Citizens-Report-Form-PDF?bidId=}

\subsubsection*{\barrier{2} Opaque Data Sharing Policies}
Even if the data is collected, there is a lack of a standard policy to share data across tribal and State law enforcement, and no centralized system to correct and verify data. 
When tribes ask, agencies often provide incomplete or unreliable data; partial information, or recalled cases from memory, highlighting that ``institutional memory is not a reliable or accurate data source''~\cite{lucchesi-missing-2018}.
Moreover, getting access to data is crucial given the urgency of missing persons cases where time to report can be a decider between life or death. Unfortunately, 
``families have to wait a certain number of hours to file a missing persons report.''~\cite{apok-we-2021}. 

\begin{myquote}{\actortable{Charlene Aqpik Apok et al.}{charleneaqpikapok} ~\cite{apok-we-2021}}
The lack of centralized systems leads to missing data but also means in many cases reports cannot even be generated at a community or agency level. \ldots
the data collected is not even digitized to be used at the system level.
\end{myquote}

%






\newcolumntype{L}[1]{>{\RaggedRight\arraybackslash}p{#1}}   
\newcolumntype{B}{>{\bfseries\RaggedRight\arraybackslash}p{2.8cm}} 

\begin{table*}[t]
\rowcolors{2}{rowAlt}{rowBase}
  \centering

  \mycaption{State Databases vs. Indigenous Data Sovereignty}{Contrast in fundamental value-sensitivities in how State and Indigenous advocates collect missing persons data in databases.}
  \label{tab:missing-persons-approaches}
 
  \noindent
  \begin{tabularx}{\linewidth}{B m{5.0cm}|m{5.2cm}}

    & \cellcolor{headerDark}
    \color{white}\textbf{State Databases$^\dagger$} &
    \cellcolor{headerWarm}
    \color{white}\textbf{Indigenous Data Sovereignty} \\
 
 
    Primary Purpose &
    Procedural crime tracking and case closure. &
    Relational accountability, mapping violence patterns, securing justice. \\
 
 
    Definition of ``Missing'' &
    Narrow, rigid criteria (e.g., must happen exactly on a specific highway route). &
    Broad, holistic tracking including boarding school victims and urban disappearances. \\
 
 
    Categorization &
    Broad racial buckets, frequently misclassified; ignores tribal citizenship. &
    Specific tribal affiliations, traditional names, and intersectional identity markers. \\
 
 
    Access \& Control &
    Controlled by federal agencies; heavily delayed public access. &
    Governed by community; access granted only to those proving ethical intent. \\
 
 
    View of the Victim &
    A case number, a checkbox, a data point. &
    A sacred relative, a loved one, a spirit to be brought home. \\
 
    \arrayrulecolor{accentRule}\bottomrule
 
  \end{tabularx}
 
  \smallskip
  \footnotetext{$^\dagger$ State Databases include NamUs = National Missing and Unidentified Persons System;
  NCMPUR = National Crime Information Center Missing Person and Unidentified Remains.}
 
\end{table*}

\subsubsection*{\barrier{3} Extractive Programs.}

Institutions repeatedly misuse or misappropriate Indigenous knowledge, including community-generated data, testimony, and research findings. Some agencies reproduced community reports without attribution, distorted facts, or treated community participation as a procedural requirement rather than meaningful collaboration~\cite{wilbur-protect-2021}. Moreover, law enforcement often victim-blames and ``skeptical, uncaring or minimizing their loved one's disappearance''~\cite{the-not-invisible-act-commission-not-2023}. These practices
reinforce mistrust, particularly when institutions adopt digital systems or technologies without Indigenous governance.

\paragraph{Lack of Agency} 
State programs seldom include
``few Native American voices in their efforts.''~\cite{henshaw-4-year-old-2024}
%
For example, in 2019,  the National Inquiry  in Canada ended up extracting stories from families~\cite{national-inquiry-into-missing-and-murdered-indigenous-women-and-girls-reclaming-2019}.
Many refused to share their stories, disturbed by the lack of trauma{-informed} care offered by the commission as an act of defiance. 

\begin{myquote}{\actortable{Maggie Cywink}{maggiecywink}~\cite{brown-indigenous-2018}}
The [Canadian] Liberal government made the national inquiry a campaign promise. \ldots
It went from something that was personal, that was grassroots, that was family, to something that became a political thing.\ldots
 The national inquiry has bulldozed through our communities and with an extension will continue to exacerbate the emotional and psychological burden on the very people it is intended to solace.
\end{myquote} 

Most tribes lack access to the State criminal databases or the ability to release alerts. Advocates and families reported that alerts are not released or cases are never registered.
Some US states and Canada provinces have created missing Indigenous persons alerts that can be requested by families or tribes,\footnote{MMIR alerts in US states: California (Feather Alert), Colorado (Missing Indigenous Persons Alert), Washington (Missing Indigenous Persons), New Mexico and Arizona (Turquoise Alert or Emily's Law) and Canada provinces: Manitoba (Red Dress Alert).} however, they are not used properly and often detrimental. For example, 
Feather Alerts (California) are broadcast through ``broadcasters, social media, digital billboards and other outlets''~\cite{dadigan_feather_2024} are  denied even when missing relatives are facing a serious risk.
Since Jan 1, 2023, about 60\% of Feather Alert requests have been rejected by California Highway Patrol (CHP)~\cite{minkler-north-2024}\footnote{This statistic only includes numbers from CHP and does not include numbers rejected by local law enforcement.}. The alert system requires local law enforcement to determine whether the person is missing or seriously endangered enough to merit an alert. The request is sent to CHP, who do another evaluation before releasing the alert. This criterion is prone to fail due to a biased definition of who qualifies as ``missing'' and who it excludes. 
Recent amendments 
proposes to drop the criteria
``reducing the likelihood that bias or ignorance of Tribal cultures would affect the decisions.''~\cite{dadigan_feather_2024} 
It will also require them to respond within 48 hours with a written notice to tribes and families if an alert is denied~\cite{hall-missing-2024}. 
 


\begin{myquotel}{\actortable{Morning Star Gali}{morningstargali}~\cite{dadigan_feather_2024}}
A refusal to issue the alert for anyone who has a history of mental health or substance issues shows an incredible lack of understanding for the history of California Indian peoples, and why there are so many struggles. It is also absolutely a violation of Tribal sovereignty.
\end{myquotel}

Rural and remote tribal reservations may lack access to transportation, cell service, scarce support infrastructure or internet access, making investigations challenging (\autoref{bg:dispossession}). Geographic isolation and economic inequity
force many domestic violence survivors to hitchhike alone to nearby cities for economic independence, many falling into cycles of promiscuous sex work and substance abuse (e.g. Highway of Tears in British Columbia, Canada~\cite{culhane-their-2003, wikipedia-highway-2025}). 
In December 2024, RCMP addressed the cases by ``install[ing] five 5G cell towers along the highway to close a gap in cell service 
''~\cite{wikipedia-highway-2025}. RCMP~\cite{royal-canadian-mounted-police-working-2017} partnered with drivers to provide GPS devices to log time, date and coordinates when they see a hitchhiker.  
However, through a narrow definition of ``missing''/``murdered'', the RCMP left out many highways with spotty cell coverage. 
Strictly, RCMP 
requires  ``the crime to have happened within a mile of Highway 16, 97, or 5; their count rejects all cases that take place elsewhere along the route''~\cite{wikipedia-highway-2025}. These policies show dissonance, failing to meaningfully consult with tribes and understand their lived realities.

%

\paragraph{Silencing voices.} Even if families and advocates are invited, they are actively silenced and not allowed to critique the institutions that have harmed them for centuries.
For example, ``Operation Lady Justice'' held virtual sessions to allow the families to voice their opinions. Despite spotty cellular coverage, the online conference calls were moved to phone for the final session. Moreover, many families were deliberately muted and not allowed to speak after the 3-minute mark~\cite{agoyo-enough-2022}.  Even if Native advocates are invited to speak, they are questioned ---

\begin{myquote}{\actortable{Abigail Echo-Hawk}{abigailechohawk}~\cite{wilbur-protect-2021}}
I remember sitting there and \ldots
and every single one of them spoke directly to me. 
It was like little arrows trying to penetrate me as they directed the rest of their comments directly against what I had shared.
\end{myquote}

Families are more willing to share information and stories when those efforts are led by trusted members rather than external agencies, highlighting their desire for sovereignty~\cite{brown-indigenous-2018}.

\subsubsection*{\barrier{4} Lack of Safe Online Spaces}

Online platforms such as social media play a critical role for families and advocates in searching for relatives and raising awareness. However, these digital spaces are also sites of frequent harassment and racist commentary, creating harmful environments that re-traumatize families and hinder their advocacy efforts.

\paragraph{Normalized violence.}
Hate crimes, hate speech, and harassment are so normalized that they often transcend into online spaces. Racist online spaces threaten digital, physical, mental, emotional, and spiritual safety, continuing cycles of violence. 


\begin{myquote}{\actortable{Cutcha Risling Baldy}{cutcharislingbaldy}~\cite{baldy-every-2020}}
Think about everyday violence like comments on news stories or postings on social media that justify violence against us, history lessons and books that try to water down the attempted genocide of us.
\end{myquote}

Advocates draw connection between online harassment and criminal intent in offline spaces. For example, {court documents show the racist intent of perpetrators who often}\removed{``Price killed Keehner because she was Native American \ldots
according to a court report. Another record shows that they} ``expressed white supremacist beliefs online through Facebook''~\cite{bleir-murdered-2018}. 
As a reflection of hypersexualization and fetishization in Hollywood (\autoref{bg:invisibility}), many Native users on Reddit pointed out the apathy around murders of Native men and women among non-Native users, adding to epistemic exclusion.

\paragraph{Online Privacy Challenges.}
Families, advocates, and allies share missing posters online to boost investigations and visibility (\actionref{1}), sometimes with or without the family's consent. In the course of investigation, they face heavy financial challenges and emotional burnout and vicarious trauma: creating and posting content across platforms, monitoring responses, and removing outdated posts once cases are resolved or \victims are found. However, such noble actions may end up in unintentional harm. 
It may be hard for the missing poster to be taken down from platforms, leading to financial fraud and perpetrators preying on grieving families.

\begin{myquotel}{\actortable{Annita Lucchesi}{annitalucchesi} \& \actortable{Abigail Echo-Hawk}{abigailechohawk}~\cite{lucchesi-missing-2018}}
The average community member does not have thousands of dollars and unlimited time to continue to follow up for the data [on missing relative]. 
\end{myquotel}

Due to the sensitive nature of the posters, some families have to weigh a choice between privacy and providing accessible contact information. 

 
\begin{myquote}{\actortable{Jodi Voice Yellowfish}{jodivoiceyellowfish}~\cite{yellowfish-missing-2023}}
When you're in crisis mode,  [families] blast their cell phone number all over social media, all over a flyer,  
 and that is the most unsafe thing to do.  I have yet to see a case that doesn't have fake ransom  
 or fake tips.  \ldots
 They [scammers] exploit that family vulnerability and call them and give them a glimmer of hope that their kid  
 or their relative or whoever is alive.  \ldots
``Your so-and-so needs their medication, \ldots
 Send this much to this Cash app 
 and you can talk to them tonight at six o'clock.''  
\end{myquote}

%
%

%
%
To create a safe space for traumatized families, moderators of search-\&-rescue Facebook groups added safety norms and advisories to combat harassment.
Advocates and tribal police provide helpful privacy tips for sharing posts on social media~\cite{yellowfish-missing-2023, greater-sudbury-police-service-looking-2017}. For example, the Navajo Nation provides its tribal police's contact information to prevent fraud (see~\autoref{fig:social-media-actions}).

\begin{figure*}[t]
    \centering
    \includegraphics[width=1\textwidth]{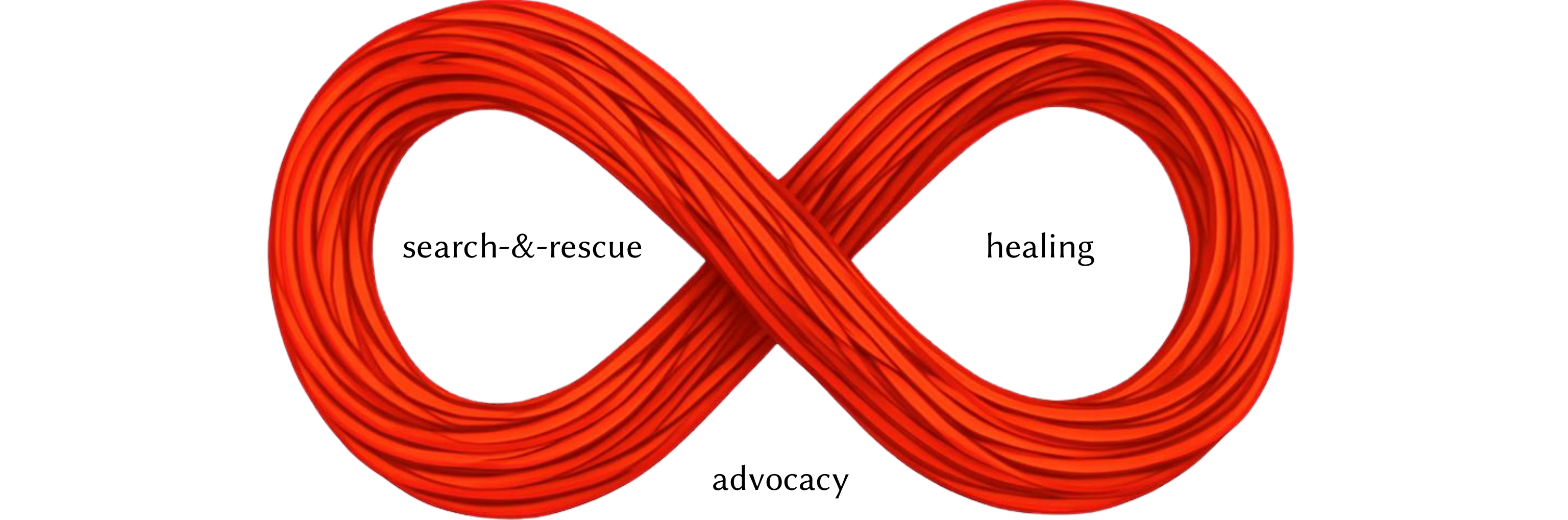}
    \mycaption{Interconnected Infrastructure of Care}{Indigenous communities have established a closed-loop socio-technical ecosystem supporting three action pillars: (a) search and rescue missing or murdered \victims (\autoref{actions:find}), (b)  provide support and healing to \victims and their families (\autoref{actions:support}), and  (c) raise awareness of the \mmiw crisis (\autoref{actions:advoacy}).  
    }
    \Description{}
    \label{fig:closed-loop}
\end{figure*}

\section{RQ2: Actions Taken by Indigenous Communities: \textit{``We Built Our Own Path''}}
\label{sec:actions}

Indigenous peoples transform their pain to showcase intergenerational resilience to resist 
systemic oppression, ``in spite of the police.''~\cite{hayes-indigenous-2022}. 
%
Communities reclaim and re-configure modern technologies to empower and extend  acts of resistance through relationality and care for ``sisters, daughters, sons, uncles, and cousins''.
We witness this crucial transformation by acknowledging that data and technologies are not limited to western concepts; ``instead, \ldots
[data are]
the stories of precious lives\ldots
data are people, loved ones, gifts from ancestors, who are each deeply missed,''~\cite{apok-we-2021}.
We expand the definition of ``socio-technical'' to include traditional spiritual practices that provide collective healing to the communities. 
In this section, we describe the socio-technical actions (see \autoref{tab:actions}) supporting three action pillars: (a) search and rescue missing or murdered \victims (\autoref{actions:find}), (b)  provide support and healing to \victims and their families (\autoref{actions:support}), and  (c) raise awareness of the \mmiw crisis (\autoref{actions:advoacy}).

\begin{myquotel}{\actortable{Morning Star Gali}{morningstargali}~\cite{hayes-indigenous-2022}}
MMIW movement is one of not just dismantling systems, but that we're also building. We're building a movement of healing [and] \ldots
    supporting our relatives through these traumatic events and through this crisis. \ldots
    \removed{What's the shift for me is where} 
    centering the voices and uplifting the stories and sharing those in a very human way. \ldots
    It's shifting that [deficit] narrative to say like, no, every individual deserves to be treated with respect and humanely and to have their story told in a way that is larger than all of us to recognize the systemic injustices 
    \ldots
    and the way that these systems have failed our relatives all around.
\end{myquotel}

\subsection{Actions to Search \& Rescue Missing and Murdered \Victims
: \textit{``We Carry the Search in Our Bones''}}
\label{actions:find}

\subsubsection*{\action{1} Investigation by Families and Advocates}

{Fed up with delayed investigation by law enforcement agencies, the families take matters into their own hands to search for their missing or murdered \victims.}
Recognizing their desire to investigate, many tribes have launched informal policing programs and search-\&-rescue parties.

\begin{myquote}{\actortable{Malinda Harris}{malindaharris}~\cite{monroe-is-2024}}
We took on the role of being the investigating police officer. I always think that was the police’s job, but when a mother loses her daughter, it’s natural instinct to do whatever you have to do to find her. I really didn’t know I was doing the police’s job; \textbf{I was just looking for my daughter.}
\end{myquote}

\paragraph{Collect evidence.}
\label{actions:evidence}
Families collect evidence from their community or the places their \victim last went missing. 
%
%
%
Many collected digital evidence from their \victims' accounts, posts on social media platforms, or security-camera footage from their last known location~\cite{monroe-is-2024, mabie-may-2022, ayer-first-2018}. {E.g.}, \actortable{Marilene James}{marilenejames}~\cite{monroe-is-2024} shared a Facebook post to collect tips and shared them with the police
``I had all these people calling and messaging me, giving me an idea of what happened, telling me about seeing her [Yazzie] there, or that their boyfriend had something to do with it''. 
%

\paragraph{Distribute case information online.}
\label{actions:distribute}

Despite harmful racist abuse and challenges in sharing information (\barrierref{4}), social media platforms have been instrumental for the communities. 
E.g., Selena used social media to broadcast her whereabouts and let her family know about her abduction.

\begin{myquote}{\actortable{Cheryl Horn}{cherylhorn}~\cite{mabie-mothers-2022}}
After her [Selena] car broke down on a desolate road, left a scribbled note on the back of an envelope. It provided her name, a description of what she wore and the time she'd left on foot to seek help `If I do not make it there and you do not hear from me idk (I don't know) someone probs took me'
\end{myquote}
Many Facebook search-\&-rescue groups and pages are run by survivors, families, activists, and tribal police to search for \victims (see \autoref{fig:social-media-actions}).

\begin{myquotel}{\actortable{Charlene Aqpik Apok et al.}{charleneaqpikapok}~\cite{apok-we-2021}}
Social media  is a huge repository of information that serves as a tool in gathering data. These days, \ldots
Facebook and Instagram are go-to communication outlets for family members trying to connect. It also is the first stream of communication to spread word when there is a concern for safety. 
 If a report is made, details are often shared first and widely on social media before any other information is provided to community members.
\end{myquotel}

\begin{figure}[t]
    \framebox{\includegraphics[height=0.3\textwidth]{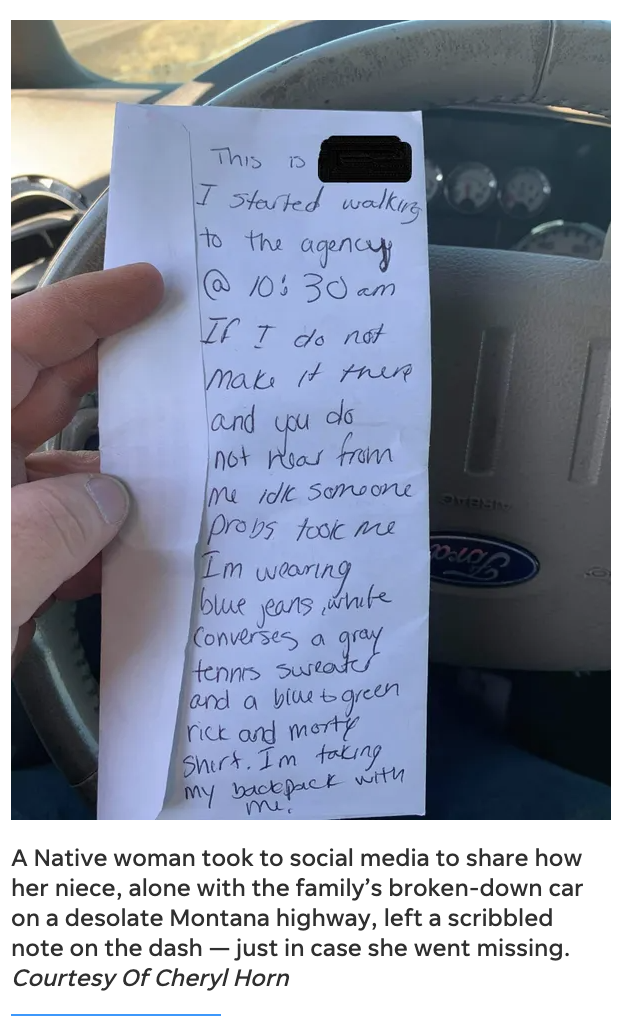}}
    \framebox{\includegraphics[height=0.3\textwidth]{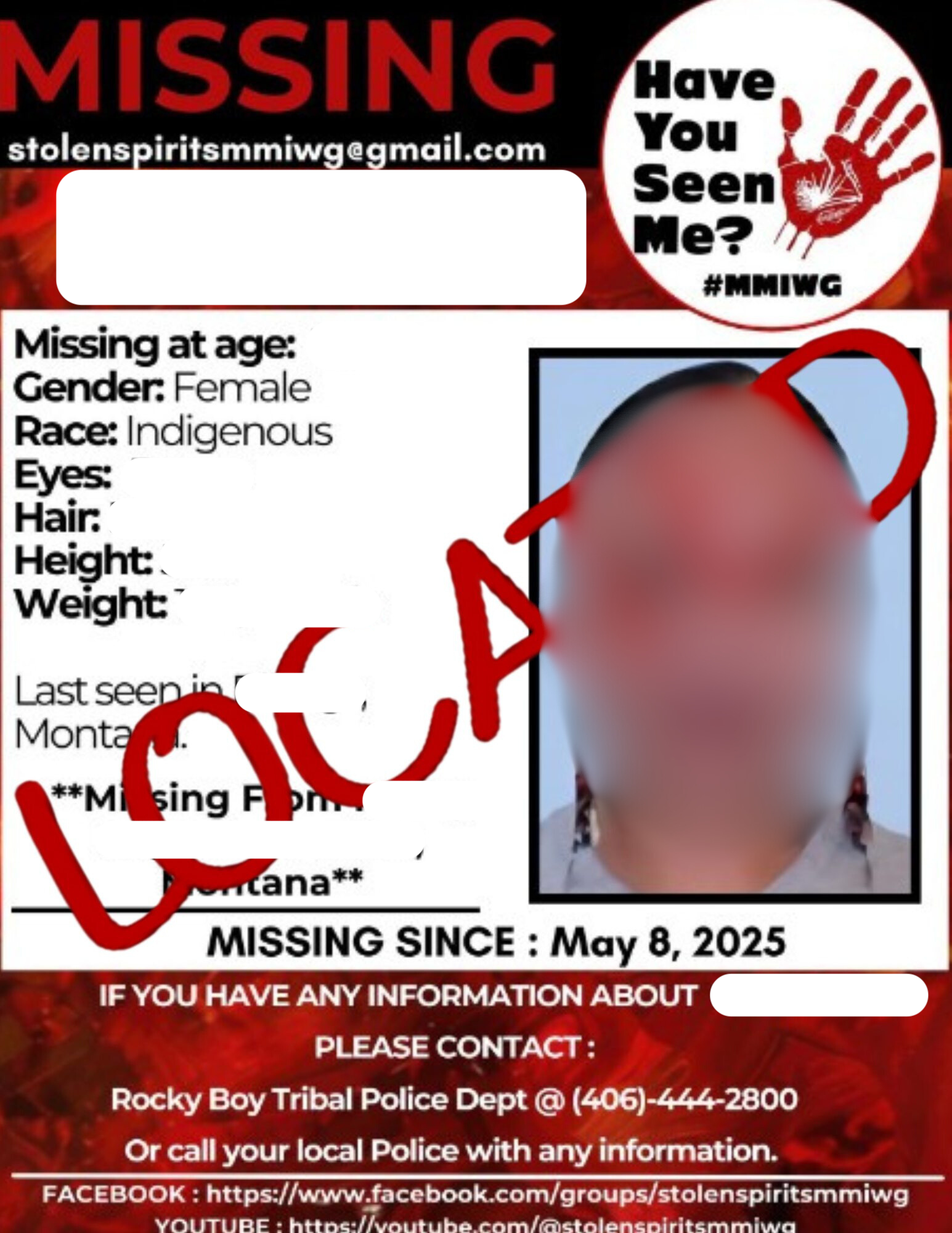}}
    \framebox{\includegraphics[height=0.3\textwidth]{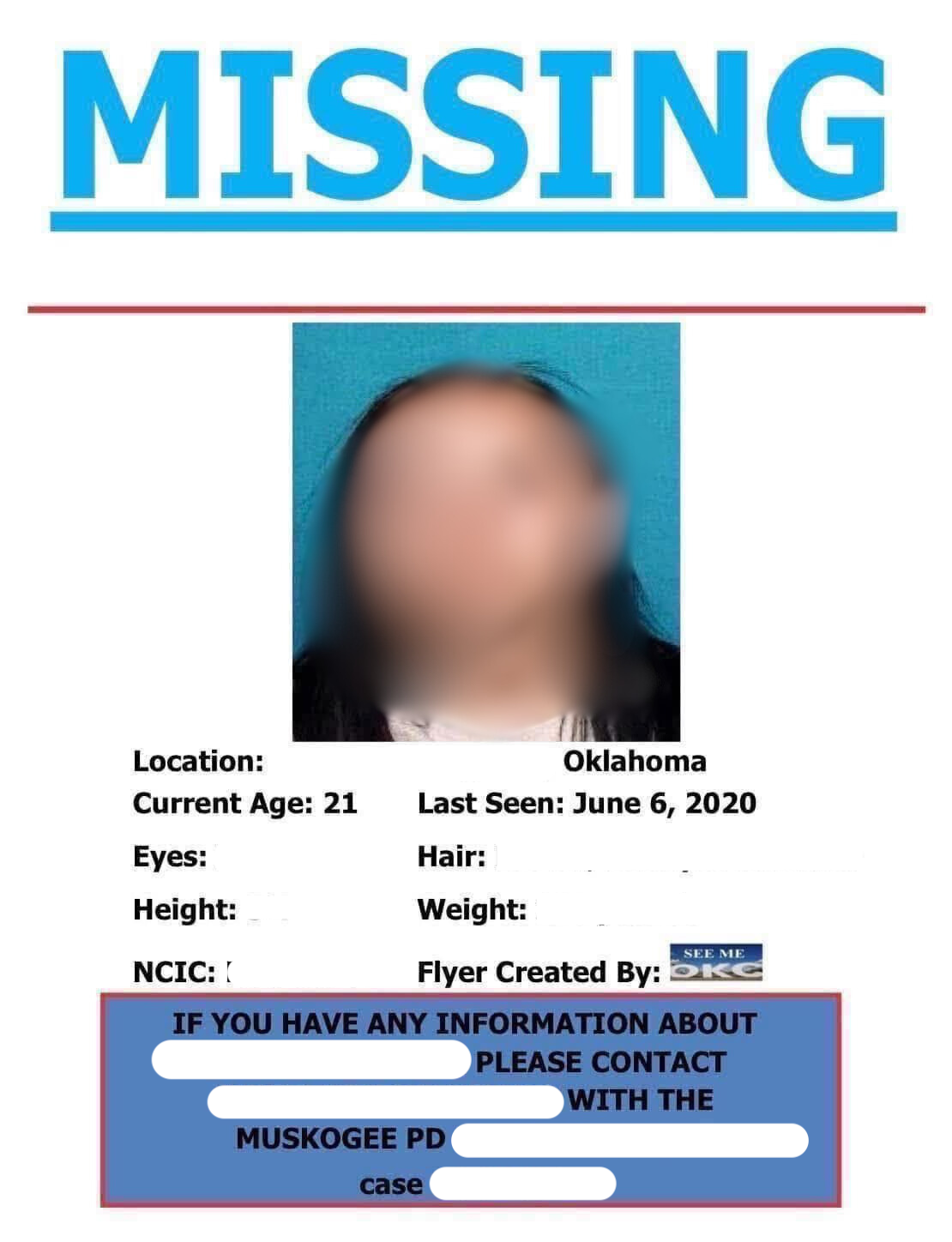}}
    \framebox{\includegraphics[height=0.3\textwidth]{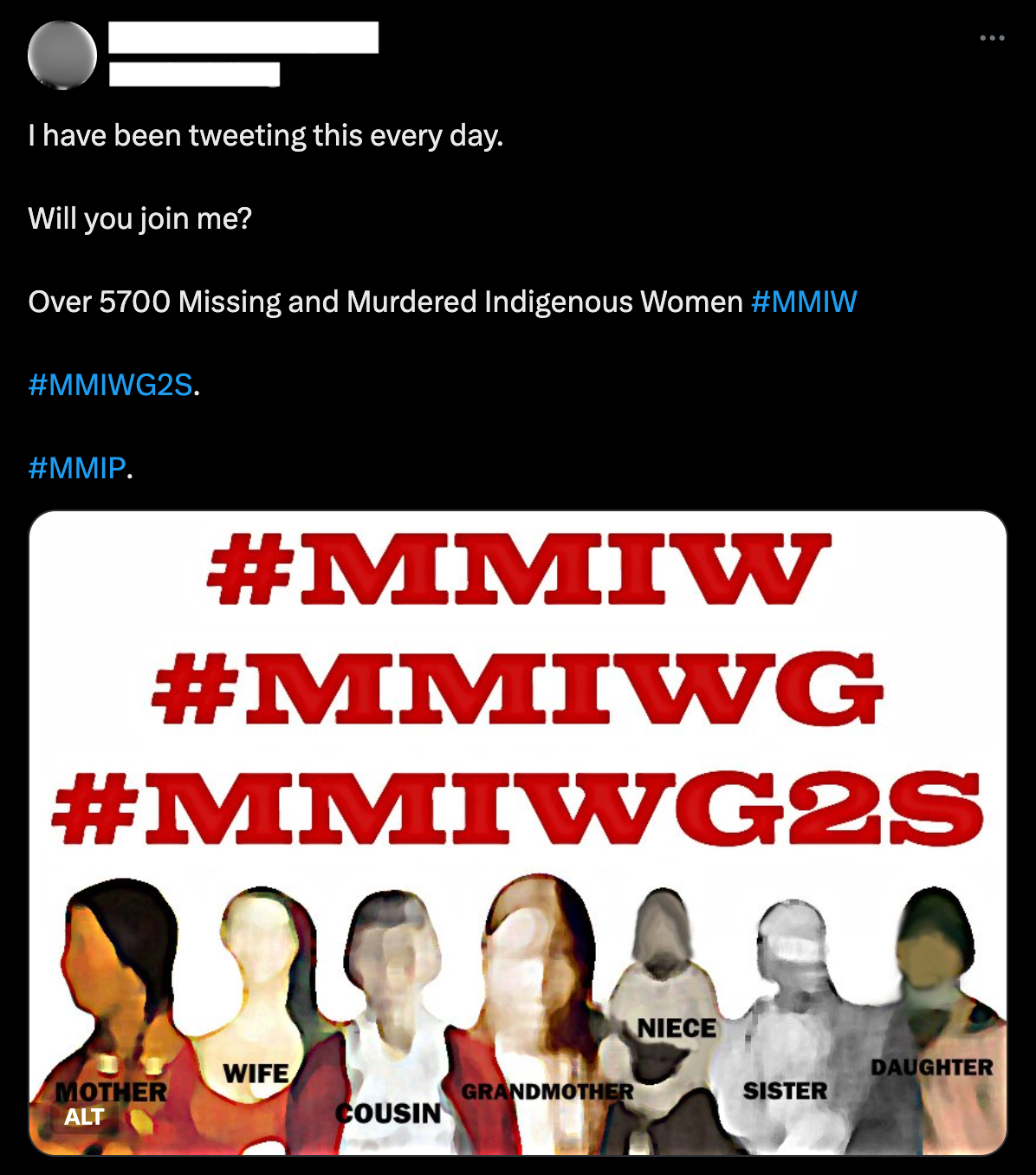}}
    \mycaption{Social Media Posts}{The posts illustrate actions taken by families, advocates, and tribal police to find missing relatives and raise awareness. (\textbf{left}) A note courtesy Cheryl Horn~\cite{mabie-mothers-2022} that her daughter left when she went missing. The note provided her name, a description of what she wore and the time she'd left on foot to seek help. The note says ``If I do not make it there and you do not hear from me idk (I don't know) someone probs took me''~\cite{mabie-may-2022}. The (\textbf{middle two}) posters found on social media for finding missing \victim  (\actionref{1}), and (\textbf{right}) image is a social media post to raise awareness about the \mmiw crisis.}
    \Description{The image features three distinct sections, each depicting efforts to raise awareness on missing persons. On the left, a photo shows a handwritten note being held by a person in a vehicle. The note details an individual's information and instructions in case of their disappearance. In the center, a red-hued missing person poster provides details such as gender, race, age, and a contact email, with a "LOCATED" stamp across it. On the right, a black social media post highlights a campaign for missing and murdered Indigenous women with hashtags like \#MMIW and includes an array of human silhouettes labeled with familial titles. Below the image, a caption elaborates on the images, mentioning their context in social media awareness campaigns.}
    \label{fig:social-media-actions}
\end{figure}


\subsubsection*{\action{2} Investigative Tools by Advocates and Tribal Police}
Advocacy organizations and tribal police provide crucial technical support infrastructure and tools to aid families in investigations, improve communications, and assert Indigenous data sovereignty so that the data stays within the community.

\begin{myquotel}{\actortable{Annita Lucchesi}{annitalucchesi} \& \actortable{Abigail Echo-Hawk}{abigailechohawk}~\cite{lucchesi-missing-2018}}
We will not let the lives of Native women be a checkbox that meets minimum requirements.
\end{myquotel}

%


\paragraph{Create maps and databases.}
\label{actions:database}


Indigenous cartographers created GIS maps and visualization tools~\cite{miner-informatic-2022, buffalo-mit-2023, lucchesi-mapping-2019, lucchesi-indians-2018, noauthor-mapping-nodate, ollivierre-participatory-2021, hetoevehotohkee-lucchesi-spatial-2020, rose-redwood-decolonizing-2020}, sovereign maps and databases~\cite{lucchesi-sovereign-2024, buffalo-mit-2023, missinginorg-alaska-2025, the-amber-adovcate-amber-2025, wikipedia-missing-2025, culbert-bcs-2023, apok-we-2021, kays-silent-2022, lucchesi-missing-2018, ray-vanished-2023} to support investigations and to trace violence in their communities. 
Witnessing the challenges with state databases (\barrierref{1},\barrierref{2}), advocates create their own sovereign databases by creatively cross-referencing various informal sources\footnote{\citeauthor{lucchesi-missing-2018}~\cite{lucchesi-missing-2018} used informal sources such as ``Freedom of Information Act (FOIA) requests to law enforcement agencies, state and national missing persons databases, searches of local and regional news media online archives, public social media posts, and direct contact with family and community members''.}. Many databases utilized advances in DNA sequencing technologies to enable identification, especially in ``cases long gone cold''~\cite{the-amber-adovcate-amber-2025}. 

\begin{myquote}{\actortable{Sasha Reid}{sasha-chelsea}~\cite{culbert-bcs-2023}}
The original intent of the database was not to solve crimes, \ldots [but to] identify actions that community activists, politicians and police could take to increase safety. \ldots but as the names and details grow, \ldots it could now be used to pinpoint geographic clusters or similar patterns of how victims disappear or are killed.
\end{myquote}

However, 
``no central database that is routinely updated, spans beyond colonial borders, and thoroughly logs important aspects of the data.''
~\cite{assistant-chief-jack-austin-jr-guides-2024}.
Addressing this concern, \actortable{Annita Lucchesi}{annitalucchesi} set up a centralized database through the ``Sovereign Bodies Institute (SBI)''
\footnote{Unfortunately, due to capacity constraints and dependence on philanthropy for funds, SBI shut down operations in December 2024~\cite{lucchesi-sovereign-2024}.}.
%
Moreover, the Yukon territory has created a complete dataset and policies to cross-reference data with the RCMP~\cite{buffalo-mit-2023}. 






\paragraph{Websites with guidance for families.}
Tribal organizations provided toolkits with resources to support families and tribal police. The toolkits share concrete actions for families 
including missing persons forms, checklists for investigative evidence, requests for AMBER alerts, additions to bulletin or police website, templates for social media posts and missing posters, guidelines to increase the reach and maximum visibility~\cite{native-hope-missing-2024, niwrc-mmiw-2024}.
%
%
For example, NIWRC~\cite{niwrc-mmiw-2024} created toolkits for families on steps to do within the first 72 hours of a \victim going missing and contact information of law enforcement specialists through the ``Jurisdiction Assessment Tool'' .
\paragraph{Purchase technical equipment.}
Tribal police purchase advanced technologies to aid in forensic investigations, such as GPS-enabled devices, body-worn cameras, digital evidence management systems, drones, infrared-enabled all-terrain vehicles, mobile terminals, and apps such as LiveScan that coordinates with State crime databases for criminal history checks through digital fingerprints~\cite{uccm-anishnaabe-police-service-annual-2023}.

\begin{table*}[t]
    \rowcolors{2}{rowAlt}{rowBase}
    \centering
    \tabfontsize
    \renewcommand{\arraystretch}{1.4}
    \mycaption{Summary of Socio-Technical Actions}{Taken by \victims, families, advocates, and tribal police.}
        \begin{tabular}{>{\centering\arraybackslash}
                    m{0.02\textwidth}
                    m{0.05\textwidth}
                    m{0.2\textwidth}
                    m{0.05\textwidth}
                    m{0.4\textwidth}
                    m{0.15\textwidth}}

\rowcolor{headerWarm}
\multicolumn{1}{c}{} 
& 
\multicolumn{1}{c}{\color{white}\textbf{A\#}}
&
\multicolumn{1}{c}{\color{white}\textbf{Action}}
& \multicolumn{1}{c}{\color{white}\textbf{Pages}} & \multicolumn{1}{c}{\color{white}\textbf{Summary}} & \multicolumn{1}{c}{\color{white}\textbf{Recommendations}}
\\
        \cellcolor{headerWarm} &  
       \actionref{1} & Investigation by Families and Advocates 
            & 54 &  
            Families and advocates lead their own investigation to find the missing \victim, collect evidence, and distribute information about the case online.
            & 
            \recref{1} \recref{2} \recref{4}
              \\
        \cellcolor{headerWarm} \multirow{-3}{*}{\rotatebox[origin=c]{90}{\color{white}Search \& Rescue}}  & \actionref{2} & Investigative Tools by Advocates and Tribal Police   
        & 18 &  
        Advocates have created their own databases and resource websites to aid the investigation. Tribal police purchase technical equipment and use tools to improve communication with families. 
        & 
         \recref{1} 
         \recref{4} 
           \recref{5} 
         \\        
    \arrayrulecolor{accentRule}\midrule
         \cellcolor{headerWarm} &  \actionref{3} & Traditional Storytelling and Spiritual Healing & 31 &
        Advocates and families heal trauma by sharing stories and practicing spiritual ceremonies through online spaces.
        & \recref{5} 
        \\
    
        \cellcolor{headerWarm} \multirow{-2.5}{*}{\rotatebox[origin=c]{90}{\color{white}Healing}} & \actionref{4} & Support Material Needs  & 7 &
        Advocates and tribal police support \victims and families with their material needs. 
        & 
       \recref{3}
       \\    
        
        \arrayrulecolor{accentRule}\midrule
         \cellcolor{headerWarm} & \actionref{5} & Advocacy Movements and Campaigns  & 81 &
        Advocates use online spaces to advocate for \mmiw and other anti-violence movements to represent accurate information, organize protests, vigils, prayers, and walks, and creative media. & 
        \recref{3}
        \\
        
        \cellcolor{headerWarm} \multirow{-2.5}{*}{\rotatebox[origin=c]{90}{\color{white}Advocacy}}  & \actionref{6} & Education and Reconciliation Programs  & 23 &
        Advocates and tribal police conduct safety awareness trainings in the community and collaborate with state and federal police to form advisory groups as a part of reconciliation. &
        \recref{5} 
        \\
     \arrayrulecolor{accentRule}\bottomrule
    \end{tabular}
    \Description{The figure is a table titled "Table 4. Summary of Socio-Technical Actions – Faced by relatives, families, advocates, and tribal police." The table is divided into columns labeled "Action," "\# of Pages," "Summary," and "Recommendations." There are three main sections: "Search \& Rescue," "Safety, Healing, \& Support," and "Advocacy." Each section contains some actions, identified with codes such as A1, A2, etc.  Each row lists specific actions, with corresponding data on the number of pages, summaries, and recommendations.
    }
    \label{tab:actions}
\end{table*}

\subsection{Actions to Seek Safety, Healing, and Support
: \textit{``Coming Back Through Ceremony''}}
\label{actions:support}

\removed{Even with the lack of resource inequity, advocates have made an instrumental ``transformation''.} 
Many \mmiw advocates have personal experiences of trafficking, violence, and {loss}. Using their own experiences, advocates provide crucial safety, comfort, healing, and validation to the families. 


\subsubsection*{\action{3} Traditional Storytelling and Spiritual Healing}

%
Indigenous spirituality is closely tied to the natural world, where ``land and community have the highest possible meaning and are places to honor and commune with spirits dedicated to helping one another''~\cite{native_education_healingsharing_2023}. 
In many tribes, ``Sacred Circle'' (\autoref{fig:spiritual}) recognizes the holistic connection between healing all senses: physical, mental, spiritual, ecological, and emotional. The ceremonies of smudging, sweat lodges, sharing circles, and traditional teachings offer a sacred space for personal reflection, healing, and cultural connection~\cite{treaty-three-police-service-annual-2023}. 
%
 %
Since the COVID-19 lockdown, Grandmothers operate online sweat lodges, sharing circles, and smudging to heal sexual violence and trafficking survivors and their families. 

\begin{myquote}{\actortable{Abigail Echo-Hawk}{abigailechohawk}~\cite{wilbur-protect-2021}}
I've actually been zooming in to the Pawnee nation's support group \ldots
[with] other Pawnee women \ldots
together. 
\ldots
And as a result of the pandemic, for the very first time, they've done that virtually. 
I've never been able to participate in that before but I'm accessing my tribal services, even though I'm living in Seattle, Washington.
\end{myquote}

\paragraph{Passing down stories.}
Healers, medicine keepers, caretakers, storytellers have provided ``sacred spaces'' for the communities to heal through generations. Communities are creatively adopting modern technology to pass down stories of resilience and traditional teachings. 
Advocates use Tribal Nations' Facebook pages to warn residents of imminent dangers.
\begin{myquote}{\actortable{Kandi Mossett}{kandimossett}~\cite{bleir-murdered-2018}}
You'll see alerts on there about a van that tried to grab three Native kids driving by the elementary school, I've seen four of those now, supposedly white offenders.
\end{myquote}

Although sharing stories can be re-traumatizing, that can ``reopen those wounds for them.''~\cite{wilbur-protect-2021}, families that have personal experience of loss, extend {their} support to {community members} to share grief, mourn, and heal.

 \begin{myquotel}{\actortable{Trisha Etringer}{trishaetringer}~\cite{butz-special-2021}}
From that healing, they're able to stand up and do something about it. It may be art. It may be song. It may be prayer. It may be actually going onto the front lines and advocating for equality.
 \end{myquotel}

\paragraph{Creative technological platforms.} Activists use online platforms to create podcasts~\cite{yellowfish-missing-2023, nanook-diversity--action-center-episode-2021, assistant-chief-jack-austin-jr-guides-2024, knoepp-murdered-2024, dateline-nbc-missing-2024, kays-silent-2022, wilbur-protect-2021}, plays, films, documentaries and TV shows~\cite{arts__film_2023, assistant-chief-jack-austin-jr-guides-2024, paige-showtime-2023, berg-2024-2024, gerian-yellowstone-2020, canada-who-2012, government-of-canada-royal-2021}, and artwork~\cite{native-hope-missing-2024, bull-as-2021, coit-marching-2023, day-missing-2022, hanson-songs-2018} to center stories ``as ethically as possible, and also with honor, to ensure that the families were able to speak their truth.''~\cite{horton-families-2023}. 
For example, 
The ``Giving Voice Initiative''~\cite{reconciliation_indigenous_2025} breaks the silence by centering survivors to ``safely express what they've been through, gain knowledge, and begin healing themselves and their communities.''

\begin{myquotel}{\actortable{Littledave, Jackson \& Stephens}{sheyashe-maggie-ahlisha}~\cite{knoepp-murdered-2024}}
[Families] were in charge of letting us know what they wanted to share with us. \ldots
%
We saw a need to tell these stories from an Indigenous perspective because these stories aren't told. We want to bring that to light because this is a silent epidemic here and across the country.
\end{myquotel}


%
%
 

\removed{Moreover, advocates strike a balance between humanizing the victims and presenting empirical statistics on stolen lives.}

Unfortunately, many non-Native advocates use  ``extractive storytelling'' to deem these stories as ``true-crime podcasts''. As \actortable{Luella Brien}{luellabrien}~\cite{paige-showtime-2023} adds ``These are not true-crime stories to us. These cases are our relatives.''. 
{Therefore, it is incredibly important to center silenced Indigenous voices.}


\begin{figure}[t]
     \centering
    \includegraphics[height=0.3\textwidth]{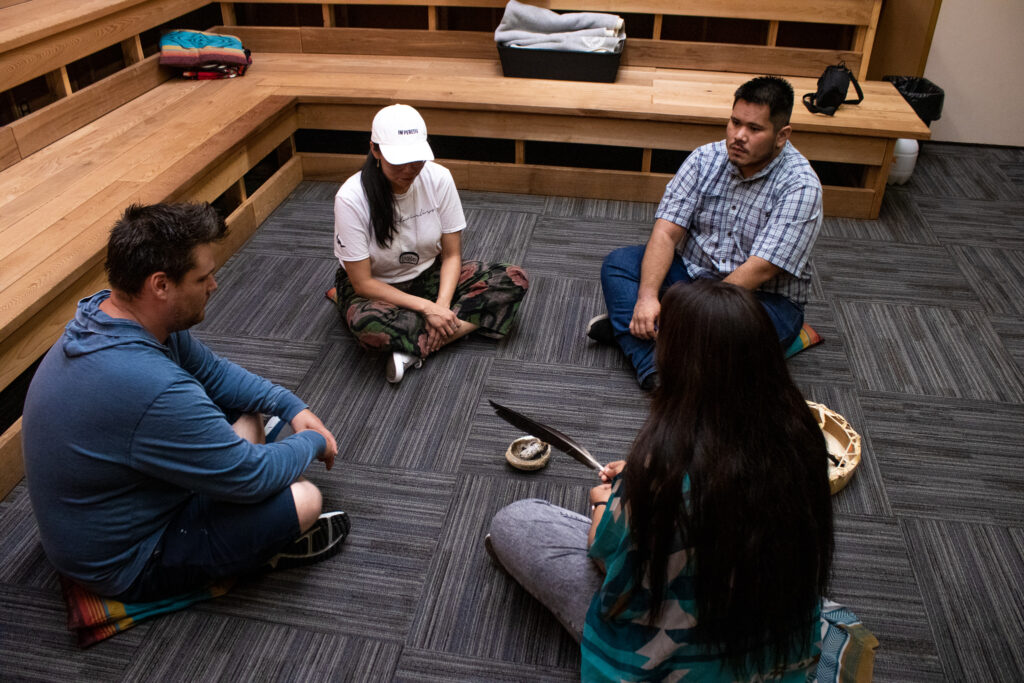}
    \includegraphics[height=0.3\textwidth]{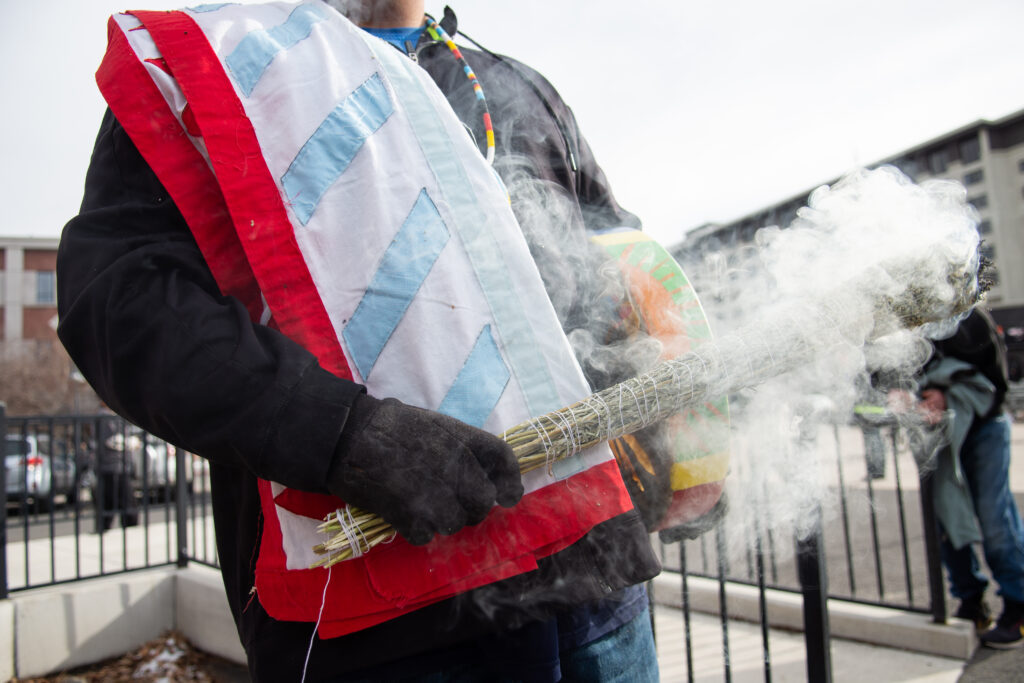}\\[1mm]
    \includegraphics[height=0.33\textwidth]{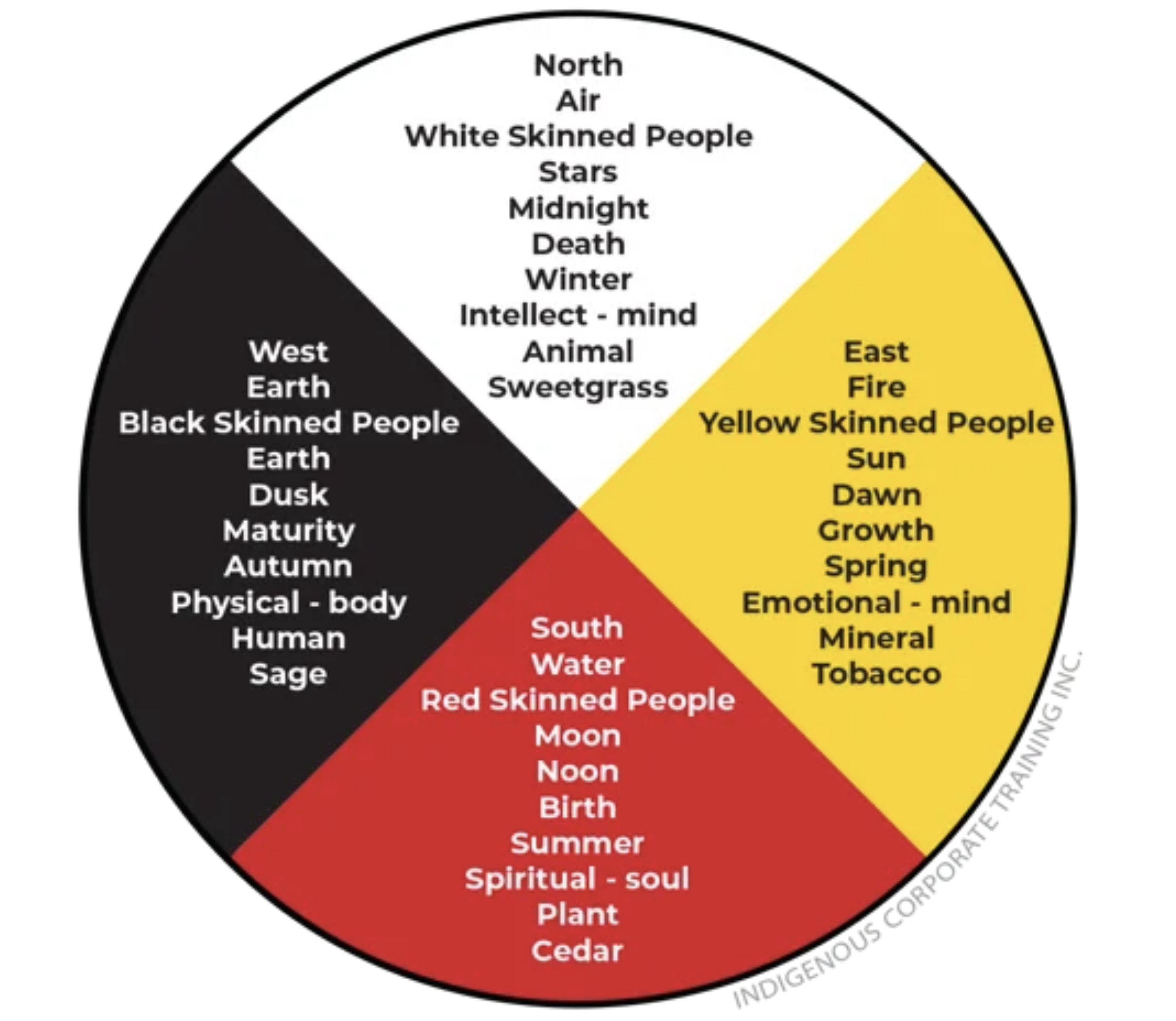}
    \includegraphics[height=0.3\textwidth]{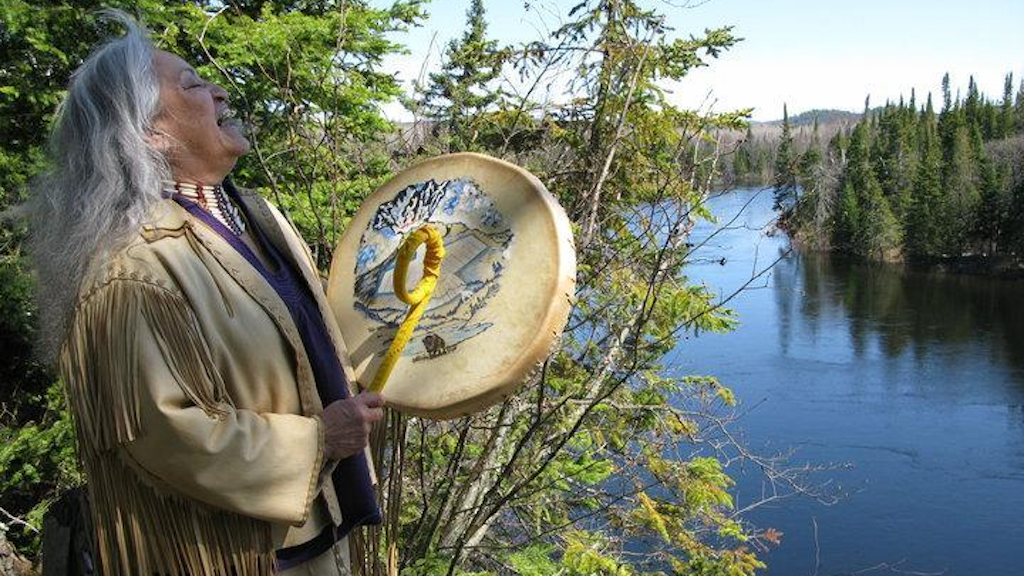}\par
    \mycaption{Traditional Healing Practices}{Many Indigenous peoples have been using traditional practices to heal and honor the spirits since time immemorial, such as (\textbf{top-left}) sharing circles help create a community of healing~\cite{native_education_healingsharing_2023}, (\textbf{top-right}) smudging, drumming, singing, food offerings and prayers are held by Elders to cleanse spirits~\cite{mueller-nevada-2020},  (\textbf{bottom-left}) Medicine wheel ``represent the alignment and continuous interaction of the physical, emotional, mental, and spiritual realities''~\cite{indigenous_corporate_training_inc_what_nodate}, and (\textbf{bottom-right}) sweat lodge ceremonies led by Elders and Grandmothers~\cite{auger-force-2023}.}
    \Description{The figure collage consists of four sections related to sacred traditional healing practices. In the top-left, four people are seated on the floor in a circle, engaging in what appears to be a communal or healing activity, with traditional objects placed at the center. The top-right image shows a person wearing a colorful blanket over their shoulders, holding a smoking stick, possibly involved in a smudging ceremony. The bottom-left section features a circular diagram divided into four quadrants, each colored differently with text describing characteristics like directions, elements, and times of day. The bottom-right image shows a person with long hair outdoors, holding a painted drum and looking upwards, set against a backdrop of trees and a body of water.}
    \label{fig:spiritual}
\end{figure}

\subsubsection*{\action{4} Support Material Needs}
Advocates support families and \victims with housing, finances, and economic support, and help them file a missing persons report and testify in court. 
%

\begin{myquotel}{\actortable{Kelly Hayes}{kellyhayes}~\cite{hayes-indigenous-2022}}
Supporting families in search of justice sometimes means showing up for people who are engaging with the criminal legal system, while knowing it will most likely disappoint them.
\end{myquotel}

\paragraph{Helplines.}
Agencies have set up culturally-sensitive helplines that ``understand the unique barriers to safety and justice that Native peoples face''~\cite{nimiipuu-tribal-tribune-missing-2022}. For example, the TsuuT\'ina  Nation Police{~\cite{tsuutina_nation_police_service_annual_2022, tsuutina_nation_police_service_annual_2021}} distributes radio advertisements to raise awareness of support services and information for the missing persons helpline on their website. {On a national scale, }
\textbf{StrongHearts Native helpline}, a safe and confidential helpline for suicide prevention, domestic and family violence, and substance abuse, has supported 13,000 survivors through calls, text, and online chat~\cite{niwrc-2025-nodate}. {Advocates urge communities to utilize these services.}
%

\begin{myquotel}{\actortable{Abigail Echo-Hawk}{abigailechohawk}~\cite{wilbur-protect-2021}}
You can call them, you can text them, you can spend time in conversation with people talking through. because 
there are going to be times when nobody is going to believe you. 
\removed{And that's what rape culture looks like for [Native] women and our LGBTQ+ community nationwide. Nobody believes and these systems have been set up to benefit men and allows for this violence to continue.}
\end{myquotel}



\paragraph{Apps.}
Tribal police have built apps to ``build resilience, reduce stigma, and promote mental wellbeing''~\cite{treaty-three-police-service-annual-2023}. 
For example, \actortable{Treaty Three Police}{ttp}~\cite{treaty-three-police-service-annual-2023, treaty-three-police-service-annual-2024} introduced the ``Peer Connect App [that] offers another avenue for members to access peer support, providing an easily accessible, user-friendly tool to connect with trained peers.'' 
Moreover, apps counter hate speech on \mmiw social media groups eg. \actortable{Kenora Makwa Patrol}{kenoramakwapatrol}~\cite{williams-ontarios-2023}
``address hate-motivated speech in social media and counteract it with true stories and an understanding of the effects of systemic and structural racism''. 

\subsection{Actions for Advocacy and Raising Awareness
: \textit{``Our Voices Won’t Be Silenced''}}
\label{actions:advoacy}

%
\begin{figure}[t]
     \centering
    \includegraphics[height=0.35\textwidth]{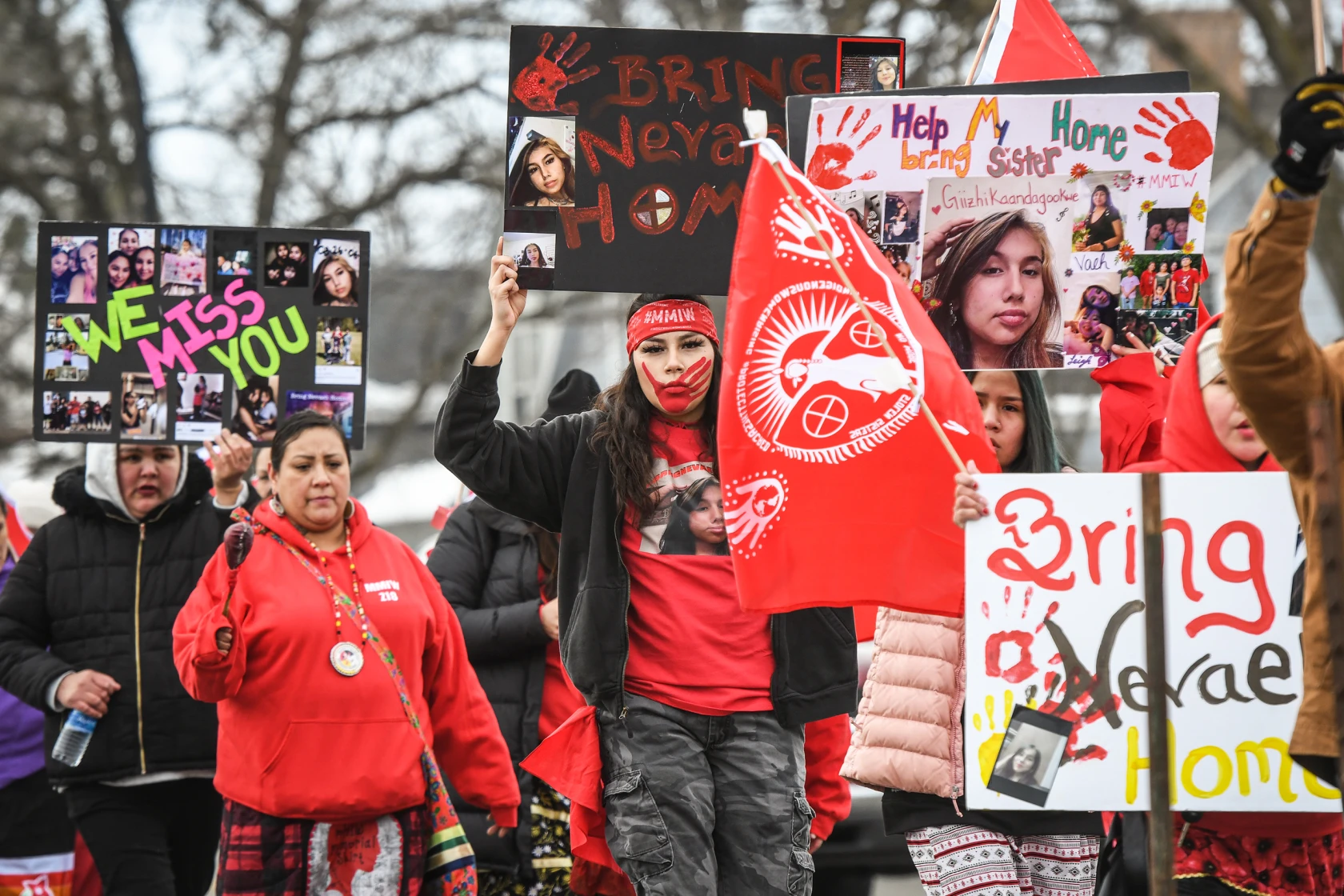} \hspace{1mm}
    \includegraphics[height=0.35\textwidth]{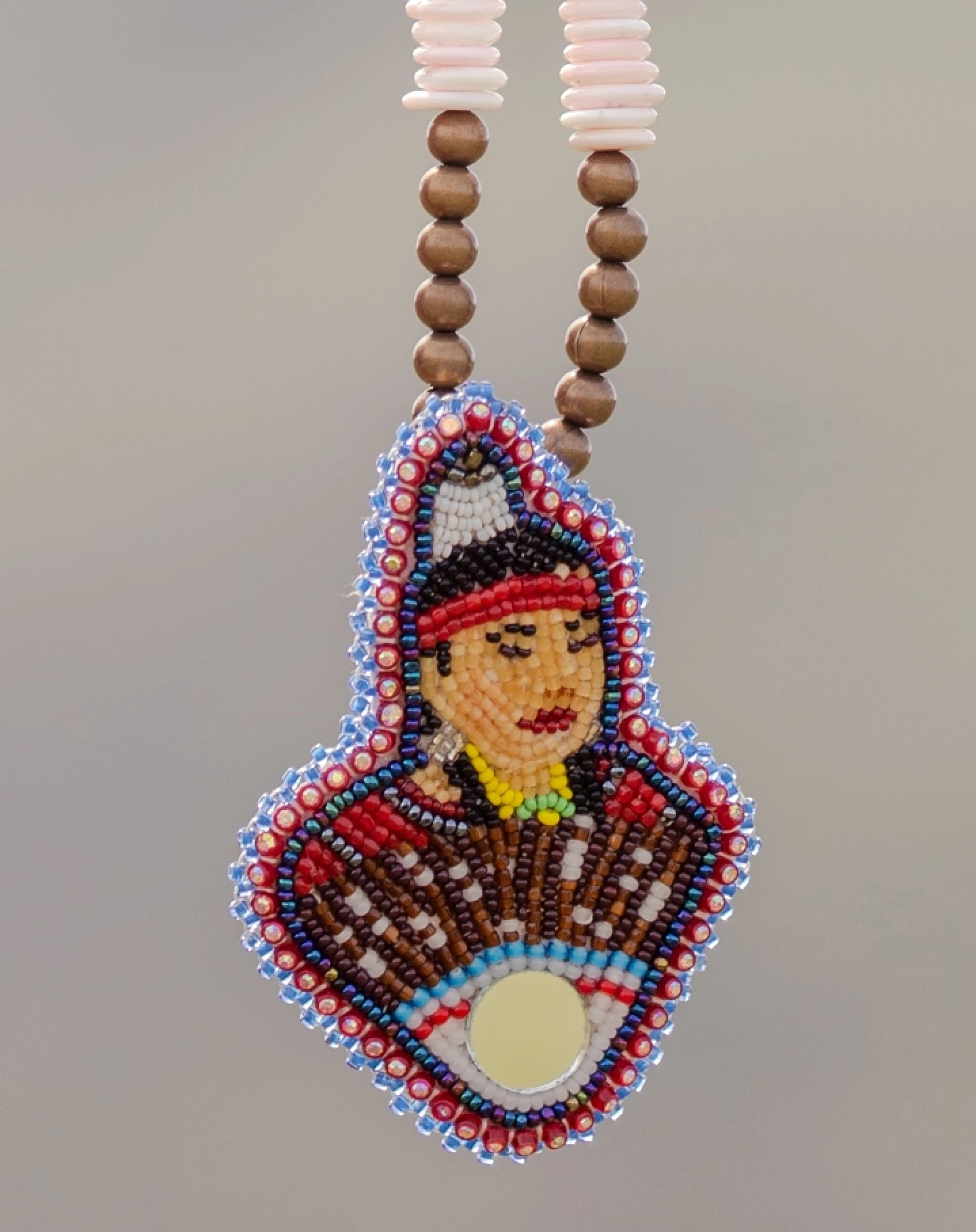}\\[0.8mm]
    {\setlength{\fboxsep}{0pt}\framebox{\includegraphics[height=0.344\textwidth]{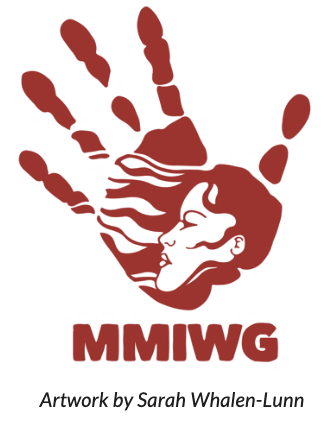}}}\hspace{5mm}
    \includegraphics[height=0.35\textwidth]{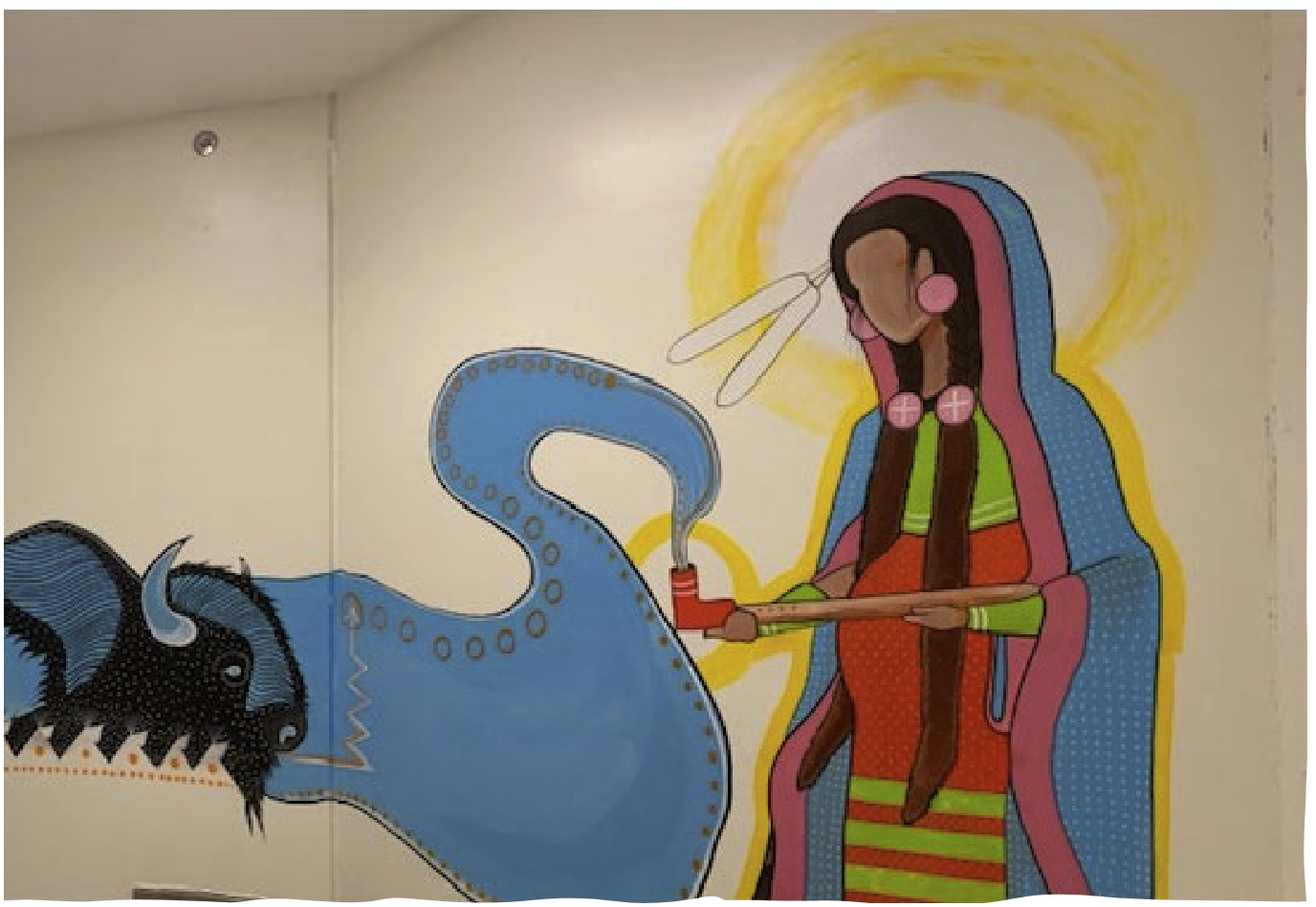}
    \mycaption{Art}{Indigenous activists symbolize the resilience of Indigenous peoples through the \#\mmiw movement. The artworks symbolize (\textbf{top-left}) communities wearing red colors at prayers, protests, candle vigils, marches, and marathons to honor the relatives and raise awareness about the crisis~\cite{haasken_bring_2023}, (\textbf{top-right}) The beadwork portrait of Mavis Kirk-Greeley created by her sister Merle Kirk symbolize of the [\mmiw] movement in Oregon~\cite{bull-as-2021}, (\textbf{bottom-left}) the red hand symbol embodying the \mmiw movement honoring the spirits of the ones who are lost.~\cite{apok-we-2021}, and (\textbf{bottom-right}) TsuuT’ina Cell Block: cellblocks dorn cultural revitalization art and literature depicting trauma-informed and restorative justice values~\cite{tsuutina_nation_police_service_annual_2022}}
    \Description{The figure is a collage showcasing various elements of Indigenous art and activism related to the \#MMIR movement. In the top-left, a group of individuals, dressed warmly for cold weather, hold signs with messages like "We miss you" and "Bring Nevae home," advocating for missing Indigenous people. The top-right image is a detailed beadwork portrait of a woman wearing a hat, surrounded by colorful beads. The bottom-left features a red handprint with a stylized image of a woman's face overlaid, labeled "MMIWG," which stands for Missing and Murdered Indigenous Women and Girls. The bottom-right is a mural depicting a person in traditional attire with long hair, interacting with a blue dragon-like creature, against a yellow circular backdrop, conveying themes of cultural revitalization.}
    \label{fig:art}
\end{figure}

Frustrated by the lack of State's urgency, activists and families have been steadfast in advocating for their rights, raising awareness, and sweeping legislation and policy change (\autoref{bg:policy}).
%
    \removed{\actortable{Maggie Jackson}{sheyashe-maggie-ahlisha}~\cite{knoepp-murdered-2024} lauds the instrumental efforts by the community ``It was honestly a community effort to just raise awareness to something that was haunting our people and we all knew that it needed to be solved. I don't know what the catalyst was but I feel like it was multiple different things that just really brought her story to life.''}
    %

\begin{myquote}{\actortable{Abigail Echo-Hawk}{abigailechohawk}~\cite{echo-hawk-mmiwg-2020}}
Decades of advocacy and activism fell on deaf ears, while more and more of our women went missing and were murdered. Legislators, government agencies, and media have been forced to pay attention because of the relentless work by the families of MMIWG victims, grassroots activists, tribes, and Native organizations across the country.
\end{myquote}

\subsubsection*{\action{5} Advocacy Movements and Campaigns}

%



The \mmiw movement relies on breaking silences and using social media platforms to amplify suppressed voices and mobilize reforms~\cite{mueller-nevada-2020}. The \#\mmiw movement emerged when stories of loss passed down across generations from mothers to their daughters; found their way on social media platforms with \#MMIW, \#MMIP, \#MMIWR, \#NoMoreStolenDaughters, \#NoMoreStolenSisters, \#NoMoreStolenRelatives.
In 2012, \actortable{Sheila North Wilson}{sheilanorth}~\cite{monroe-is-2024} started using \#mmiw on Twitter ``It surprised me at how fast the hashtag picked up, and how far it went.'' 


\begin{myquote}{\actortable{Alleen Brown}{alleenbrown}~\cite{brown-indigenous-2018}}
The growing movement \ldots
didn't arise out of data — it came from the fact that so many Indigenous women know someone who has died violently or disappeared. One of the hallmarks of the movement is that it does not center around how the woman was murdered or who killed her. It identifies the generations-long elimination of thousands of women from Indigenous communities as a direct result of government['s] attempts to eliminate Indigenous cultures.
\end{myquote}

\paragraph{Intersectionality with anti-violence movements.}
\label{actions:intersectionality}
%
%
    %
    %
{Honoring a belief system that treats land as a relative, communities often link violence against their people to violence against the land. As \actortable{Patina Park}{patinapark}~\cite{swanson-they-2020} posits ``We can't be surprised that people who would rape our land are also raping our people''.
%
The \mmiw movement have {always been deeply intersectional, aligning with anti-violence and feminist movements\footnote{\#MMIR movement intersects with \#LandBack, \#NoDAPipeLine, \#ResidentialSchoolSurvivors,\#NotOneDropForData, \#DataSovereignty, \#IndigenousEnvironmentalJustice, \#IndigenousFoodSovereignty, \#MeToo, \#TwoSpirit, \#Indiqueer, \#BlackLivesMatter, \#Sogorea Te`'LandTrust, and \#SeedingSovereignty.}. This stance challenges systemic harms that disproportionately affect marginalized Black and Brown communities. Similarly, rematriating the land and knowledge is inseparable from healing intergenerational trauma deeply felt by \mmiw families.
Rematriation aims to return land, culture, and governance under the leadership of matriarchs.
\paragraph{Organize protests, vigils, prayers, walks.}
 The online movement also mobilized people in offline spaces (see \autoref{fig:art}). Activists organized countless marches, vigils, and marathons through Facebook/Instagram ``events'' features to raise awareness, a practice that grew especially popular during the COVID-19 pandemic. When in-person gatherings such as marches, prayer circles, and walks became impossible, families, movement leaders, and \mmiw groups turned to virtual tributes and online discussions, filling the internet with images and stories of missing and murdered loved ones~\cite{gable-when-2023}.
Families create art representing red-colored skirts, lamps, dresses, flags, beadwork, prints, and organize art events through e-commerce platforms (e.g Etsy). Their goal is not to earn money from selling art, but to symbolize the movement and raise awareness~\cite{raftery-missing-2019} (see \autoref{fig:art}). 

\begin{myquotel}{\actortable{Native Womens Wilderness and Indigenous Women (NWWIW)}{nwwiw}, an advocacy website}    
Red is the official color of the \#MMIW campaign, but it goes deep and has significant value. In various tribes, red is known to be the only color spirits see. It is hoped that by wearing red, we can call back the missing spirits of our women and children so we can lay them to rest.
    \end{myquotel}

\subsubsection*{\action{6} Education and Reconciliation Programs}



%
Tribal programs organize educational sessions for youth, \lgbt, and frontline support workers on financial crime, phishing, child luring and trafficking, and online harassment. {For e.g.}, 
frontline professionals utilize the ``Sexually Exploited Youth risk assessment tool'' to help identify youth at risk~\cite{williams-ontarios-2023}, enabling them {``to identify red flags of trafficking''~\cite{tsuutina_nation_police_service_annual_2022}.
In addition, reconciliation programs strengthen relationships between Tribes and the State, an essential step toward addressing historic and ongoing harms.
Indigenous advocates lead education programs using cultural, strength-based, and trauma-informed approaches that respond directly to community needs~\cite{williams-ontarios-2023}. 
These programs hold the State and law enforcement accountable, by ensuring commitments are enacted through sensitizing frontline workers in the justice system to systemic violence against Native peoples~\cite{national-inquiry-into-missing-and-murdered-indigenous-women-and-girls-reclaming-2019,tsuutina_nation_police_service_annual_2022, tsuutina_nation_police_service_annual_2021}.
E.g., RCMP~\cite{government-of-canada-royal-2021} created ``Lived-Experience Advisory Group'' and ``Circle of Change'', led by Indigenous women, to shape policies on missing persons investigations, new risk assessment tools, and youth engagement strategies. 
%
Finally, as non-Native service providers often fail to provide reliable coverage, tribal-led technology companies have expanded high-speed internet access on reservation lands~\cite{telus_corporation_indigenous_2024}.

\section{RQ3: {Discussion and Recommendations for HCI: \textit{``Technology That Honors Our Stories''}
}}
\label{sec:recommendations}



\epigraph{\itshape ``
We honor and remember always the lives and legacies of love from each missing and murdered Indigenous womxn and girl. 
%
\ldots
We plant seeds of resistance through lives of health and wellness. Certainly, the complexities and challenges are ever present; but looking forward we remember the vision of living our power to the fullest, in safety, while thriving. We are calling for this vision of justice to come forward in the same way we call to our relatives in an effort to ensure they can rest in peace and with memory eternal.''}{\actortable{Charlene Aqpik Apok et al.}{charleneaqpikapok}~\cite{apok-we-2021}}


We witness the calls to action made by Indigenous survivors, families, advocates, and tribal police to support them in their resistance. 
{To answer RQ3, we contextualize their needs by enriching them with prior HCI literature. We invite the HCI scholars to  (a) recognize self-determination of data and sovereignty (\autoref{recommendations:data_sovereignty}), (b) direct action (\autoref{recommendations:tech}), and finally, (c) reconcile with Indigenous onto-epistemologies (\autoref{recommendations:research}). HCI scholars must pursue this generative work within the Indigenous Interconnected Infrastructure of Care (\autoref{fig:closed-loop}, \recref{5} for discussion).}

\subsection{Recognizing Self-Determination of Data and Technological Infrastructure
: \textit{``Healing Through Tech Agency''}}
\label{recommendations:data_sovereignty}



\subsubsection{\recommendation{1} Data Stewardship and Sovereignty}
%
%
Communities echo the United Nations' Declaration on the Rights of Indigenous Peoples\footnote{United Nations Declaration on the Rights of Indigenous Peoples (UNDRIP) sets out rights of Indigenous peoples worldwide, covering areas like culture, identity, language, employment, health, education, land, and natural resources --- affirming their right to self-determination, meaning they can freely determine their political status and pursue economic, social, and cultural development.} to call for socio-political and technological self-determination through their own data policies, {recognizing the digital sovereignty ethos~\cite{ricaurte-data-nodate, lehuede_alternative_2024, kukutai_indigenous_2016, yanchapaxi_indigenous_2025, tsosie_models_2020, carroll_care_2020}}.
%
%
%
%
For example, initiatives from SBI\footnote{Sovereign Bodies Institute (SBI)}, NIWRC\footnote{National Indigenous Women's Resource Center (NIWRC)}, and UIHI\footnote{Urban Indian Health Institute (UIHI)} create missing persons databases~\cite{sovereign-bodies-institute--2021, sovereign-bodies-institute-they-nodate, sovereign-bodies-institute-hidden-nodate, sovereign-bodies-institute-zyya-nodate, lucchesi-sovereign-2024, lucchesi-missing-2018} and health epidemiology data~\cite{urban-indian-health-institute-urban-nodate} respectively, have been exceptionally well implemented. {HCI researchers should ensure that tribal nations and families' voices are meaningfully consulted to guarantee a culturally-sensitive data-stewardship practice: giving weight to their lived experiences as equal to the empirical data, how it is collected, stored, and retained, and sustained access to the data.}

\begin{myquote}{\actortable{Trisha Etringer}{trishaetringer}~\cite{butz-special-2021}}
    We need to have Indigenous data keepers that hold that data, because it's sacred to us. It's not just numbers that you throw around. These are people.
\end{myquote}

Moreover, policies are needed that enable a sovereign governance model through transparency, accountability, and oversight, especially, how the data is shared between investigative agencies across State and International borders~\cite{lucchesi-missing-2018, swanson-they-2020, sovereign-bodies-institute--2021}. 
E.g., SBI provides an instance of such a policy by publicly releasing the schema of their database (see \autoref{tab:database}). SBI invites Native advocacy and investigation agencies to access the data through a contact form requiring information on the need and purpose of data storage, usage, and sharing protocol.
Emphasizing their commitment to protecting the sacredness of the data and respecting families, SBI grants access only to those who can demonstrate the ability to handle the information responsibly and in accordance with Indigenous values.
%
As an example, strengthening privacy, security, and usability in data-sharing processes can improve transparency and accountability. A cryptographically verifiable access-control protocol may enable Tribal Nations to share data with State criminal databases, while being able to approve, deny, and audit data access requests.

\begin{table}[t]
\rowcolors{2}{rowAlt}{rowBase}
\centering\footnotesize
\mycaption{Database Schema}{Data Columns from SBI \mmiw Database~\cite{sovereign-bodies-institute-hidden-nodate}.}
\begin{tabular}{p{0.8\linewidth}}
\toprule
{\textbf{Victim Information}: {Name}, Indigenous name and translation, tribal affiliation(s), birth date, age, parental status, other MMIP cases in family} \\ \midrule
{\textbf{Perpetrator information}: race, gender, relationship to the victim} \\ \midrule
{\textbf{Violence details}: missing or murdered, incident date, violence after death, relevant issues (domestic violence, sexual assault, sex work/trafficking, foster care, police brutality, trans victim, death in custody, unsheltered, residential/boarding school)} \\ \midrule
{\textbf{Police and court response}: reward amount, case classification, conviction status, agencies involved in locating deceased individuals} \\ \midrule
{\textbf{Geographic information}: city, state/province, country, location type (tribal land, rural, urban)} \\ \midrule
\end{tabular}
\label{tab:database}
\end{table}

%
%



\subsubsection{\recommendation{2} Technological Infrastructure Sovereignty}
Equitable economic resource distribution would allow tribes to scale up their network infrastructure and improve cellular coverage.
Ensuring the agency to raise and monitor alerts could help resolve problems with the AMBER alerts (\barrierref{3}). Although, {only 4 US states and 12 tribes (out of 574)}\footnote{{As of December 1 2025, only 4 state-level alerts exist in Washington, California, New Mexico, Colorado and 12 tribal alerting authorities exist in Chickasaw Nation, Cocopah Indian Tribe, Hualapai Tribe, Confederated Tribes of the Chehalis, Navajo Nation, Mandan, Hidatsa, and Arikara Nation, Rincon Band of Luiseno Indians, Eastern Band of Cherokee Indian, Ysleta del Sur Pueblo, Pascua Yaqui, Shawnee Tribe, and Inupiat Community of the Arctic Slope.}} with significant economic power have set up their own alert systems~\cite{the-amber-adovcate-amber-2025, wadsworth_update_nodate}, no such National-level alert exists for missing persons. Recently, in Aug 2024, the US Federal Communications Commission (FCC) unanimously voted to approve the {Child Abduction Emergency (CAE) and Missing and Endangered Persons (MEP) Codes~\cite{fcc-national-2024}. The USA's national alerting framework is built on top of the Integrated Public Alert and Warning
System (IPAWS) which includes radio and TV,  cellphones, and internet-based alerts~\cite{wadsworth_update_nodate}. FCC consulted with \textit{some} tribal nations, which will allow them to broadcast alerts on the state or federal level}. This is a good first step. However, it will take years to implement it since the alerts are transmitted through a ``patchwork of notification systems'' with varying local and state laws~\cite{fcc-national-2024}. Technology researchers could put more pressure on private network providers and the FCC by urging more technical support and policy comments to ensure tribal stewardship of the network alert systems.

\begin{myquote}{\actortable{Nits'il?in (Chief) Joe Alphonse}{joealphonse}~\cite{telus_corporation_indigenous_2024}}
Partnerships between network providers and government at all levels are essential for removing the barriers Indigenous communities face to digital connection and unlocking our increased economic potential.
\end{myquote}

\subsection{Building Technology to Help Families, Advocates, and Tribal Police
: \textit{``Tools That Protect Our Loved Ones''}}
\label{recommendations:tech}

\subsubsection{\recommendation{3} Tools for Effective Support and Advocacy}

HCI researchers and technologists could use their academic positions of power to provide direct action by building tools and technologies for tribal support providers and advocates. For e.g. by centering nuances of cultural data erasure (\barrierref{1}), appropriate race and tribal citizenship markers in the healthcare records can prevent disenfranchising violence against survivors. 
Moreover, tech policy research complementing HCI could add accountability checks in the social media platforms, such as highlights for missing posters, culturally-sensitive content moderation, and sensitive transfer of accounts for traumatized families. 
%
 Prior research~\cite{doyle-assessing-2025, doyle_i_2024, ravn_materialising_2024, lefevre_new_2024, doka-disenfranchised-1989} has provided crucial theories and design interfaces for platforms to acknowledge and be sensitive to the grief and diversity of the users. Such a respectful account transfer would enable families and friends to collect important evidence from the \victim's accounts {that can be used for investigations.}



%

\subsubsection{\recommendation{4} Tools to Improve Law Enforcement's Accountability}

Taking inspiration from the tribal police's efforts to improve communication with families, HCI researchers could do critical work to make police technologies transparent and accountable.
%
For example, shifting the onus away from families, technology tools could enable law enforcement to collect digital evidence in a transparent way, ensuring equal access to families, tribal police, and courts~\cite{wild-new-2024}. Tribes can similarly reconfigure surveillance tech and social media for good.
%
%
Similarly, UIHI~\cite{echo-hawk-step-2024}'s recent survey did a needs assessment with survivors and families to provide specific recommendations for law enforcement agencies' MMIP websites.
The recommendations range from appropriate information for families (victim resources, steps to report, what to expect, how to contact detectives, misconduct complaints), law enforcement (cultural training, cold case investigations, communication with families), to visual representation and coverage about the crisis (families' stories, links to State policies). 
\subsection{Reconciliation with Indigenous Epistemologies in HCI: \textit{``Respecting Our Knowledge, Protecting Our Rights''}}
\label{recommendations:research}

\begin{myquote}{\actortable{Abigail Echo-Hawk}{abigailechohawk}~\cite{wilbur-protect-2021}}
And that goes back to why aren't we in those places? And people like, oh, well, Native people haven't achieved this academic place or they haven't gone to college for this or that, like these systems of inequality, including access to Western education, \ldots
 were meant to continue to participate in the ongoing genocide of Native people.
\end{myquote}
%

\subsubsection{\recommendation{5} Embracing Cultural-Sensitivity and 
{Reconciling with} Indigenous Epistemologies}

\paragraph{Embracing diversity.}
Although united, we recognize the diversity of cultures, languages, ways of being and thinking with different socio-cultural and geographical uniqueness. We extend Noe and Kishenbaum's~\cite{noe-where-2024} invitation for the research community to recognize the specificity of and design solutions led by local Indigenous knowledge systems. 
%
For e.g. “Restorative justice” frameworks reflect the Indigenous values of respect for women
in matrilineal societies “based on a model of healing rather than of punishing.”~\cite{smith-decolonizing-2021}.
Prior HCI research points to design solutions that embrace restorative justice in non-Indigenous contexts~\cite{dibenedetto-forgiving-2023, xiao-sensemaking-2022, doan-design-2025, robalo-cyber-2023, henry-promise-2015, moeke-pickering-understanding-2018, kidd-extra-activism-2019, schoenebeck-youth-2021, musgrave-techno-mediated-2025, caletti-criminalizing-2024}.

\begin{myquotel}{\actortable{Brennan McCullagh}{brennanmcCullagh}~\cite{national-centre-for-truth-and-reconciliation-imagine-2021}}
After Badiuk made racist comments on social media, Grand Chief Derek Nepinak dropped all charges. Instead, Nepinak used restorative justice to resolve the problem. [They]
employed some traditional ceremonial methodology in terms of sitting down and resolving the issue. This had a more positive impact and created a better relationship between Badiuk and Indigenous people.
\end{myquotel}

Echoing prior calls in HCI across the globe, we need to decolonize epistemologies by working with communities through respect, reciprocity, and equitable representation. By recognizing Indigenous ways of being and meaning-making, researchers should strive for an accurate representation of families' sacred stories of loss, resilience, and justice. Researchers should make sure that they do not theorize, fetishize, or appropriate sacred traditional and spiritual practices~\cite{claisse_designing_2024, sultana-witchcraft-2019}. E.g., analyzing missing posters content on the search-\&-rescue groups without families' consent extracts and appropriates their emotional labor and grief.
We challenge the scholars to pause and reflect on the biases and notions they bring in before working with Indigenous communities.

\paragraph{Preventing Cission.} HCI scholars should ensure that their work is generative and does not lead to a cission or compartmentalize the Indigenous Interconnected Infrastructure of Care (\autoref{fig:closed-loop}). 
%
At the same time, respect the integrity and sacredness of Indigenous knowledge systems to avoid ``parachuting'' techno-solutionist approaches that side-step their generational work.
Solutions to only one of the action pillars without acknowledging other connected pillars may end up reinforcing epistemic injustices. 
A cission further erases Indigenous knowledge through separation from Western meaning-making~\cite{mbembe-out-2021, mignolo-darker-2011, anzaldua-borderlands-2021, quijano-coloniality-2007}. Instead, scholars invite {adaptive knowledge that builds} ``folds''~\cite{mbembe-out-2021} or {``reconciliation''~\cite{smith-decolonizing-2021, corntassel-indigenous-2009, yang-weaving-2023, farao-transformative-2024}; co-designing across knowledge systems by folding in diverse ways of meaning-making. 
{For e.g}, engagements with intergenerational trauma and Indigenous healing practices (\actionref{3}) have been siloed off from prior Western-dominant trauma-informed HCI work~\cite{chen-trauma-informed-2022, scott-trauma-informed-2023}, merely adding it as an asterisk~\cite{shotton-beyond-2023}.
We urge HCI to decolonize ``trauma-informed care''~\cite{blackdeer-unsettling-2023, asher_blackdeer_decolonizing_2026}  and recognize historically proven practices of cultural and spiritual healing
~\cite{trejo_mendez_decolonizing_2023, smith2020, claisse_designing_2024, claisse_keeping_2023, sheh-hamidulfuad-collaborative-2024, smith-what-2021}.

\begin{myquote}{\actortable{Tegan Swanson}{teganswanson}~\cite{swanson-they-2020}}
Refuse to silo advocacy efforts -- that division only serves to foster white supremacist, settler-colonialist, and capitalist power structures. Ask your local leadership in politics, finance, business, and education to honor the treaties, respect Indigenous sovereignty, and advocate for the end of violent environmental exploitation.
\end{myquote}

\citeauthor{escobar-designs-2018}~\cite{escobar-designs-2018} invites us to consider technology design as a space of autonomy that carries spirit, memory, and ceremony, and re-think the technologies of domination.  
Through a descriptive research positionality and methodology, we have provided a methodological contribution to HCI that bridges Western and Indigenous ways of knowing and being (\autoref{sec:positionality} and \autoref{sec:methods}). 

\paragraph{Intersectionality and reciprocity.}
Driven by racialized and hegemonic algorithms~\cite{alkhatib-live-2021, benjamin-race-2019, noble-algorithms-2018} and content moderation systems~\cite{shahid-decolonizing-2023}, missing posters and advocacy posts may not be shown to many users or in the required geographic spaces, creating pockets of isolated ``death spaces in darkness''~\cite{islekel-absent-2017, romeo-towards-2021, mbembe-necropolitics-2016}.
Therefore, more work is needed to identify the challenges faced by families and advocates on these platforms and their effectiveness for investigation and \mmiw-related advocacy. 
Recent trends in digital safety research have proposed solutions through an academic lens. Despite noble intentions, such lenses overlook communities' refusal and the intersectionality of oppressive structural~\cite{patricia-hill-collins-black-1990, crenshaw-critical-1995, lorde-masters-1979, morgan-describing-1996}. Moreover, scholars oftentimes limit their contributions through myopic and siloed solutions without doing the work outside academic spaces.
As a consequence, such solutions to support those who are situated at the margins of Western meaning-making turn harmful, ineffective, or often reinforce oppression. {For example, HCI research may have a limited definition of sustainable, ethical, or private LLMs. However, as we discussed earlier, BigTech AI mimics extractive and colonial logics, creating dangerous conditions for the communities by enabling \mmiw crisis (\autoref{sec:limitations}). To combat extraction, e.g., Māori technologists have created locally sustainable models that respect and preserve Indigenous languages~\cite{harper_how_2022, boyanton_faces_2025}. Following prior work on critical AI~\cite{tacheva_ai_2023, pendse-treatment-2022, smith_designing_2024}}
We invite HCI researchers to make these intersectional connections visible by honoring the community's lived experiences and voices without appropriation. 
We conclude our work by sharing a poem.

\begin{myquotel}{\actortable{Abigail Echo-Hawk}{abigailechohawk}~\cite{apok-we-2021}}
\centering\em
Indigenous is not a survival story \\it is a genealogy \\an ancestral story of Matriarchs\\ with bright eyes \\long hair\\ fiery strength\\ and gentle words\\ tripping over colonial tongues\\ the settlers language can't translate\\ the words\\it was never meant for their ears
\end{myquotel}

\bibliographystyle{ACM-Reference-Format}
\bibliography{bib}

\section*{Appendix}
\label{appendix}

\subsection{Prompts for LLM}
\begin{figure}[t]
    \centering
    \begin{cutegreenbox}{}
        \begin{lstlisting}
DOMAIN_CATEGORIES = {
    1: "News",
    2: "Blog article",
    3: "Non Profit Organization",
    4: "E-commerce platform",
    5: "Government",
    6: "International Organization",
    7: "Law enforcement",
    8: "Education",
    9: "Social Media",
    10: "Unknown"
}
        \end{lstlisting}
        
        Given the content, categorize the URL into one or more categories from the list: {DOMAIN\_CATEGORIES}. Provide the answer in JSON format with 'categories' (list of numbers) and 'justification'. Do not add any additional information or notes.
        \end{cutegreenbox}
\mycaption{LLM Prompt 1}{For categorization of domains through DC-LLM}
\Description{The image features a code snippet set inside a rounded rectangle with a green border. The background within the rectangle is a light mint green, contrasting with the black text. The code is written in a structured fashion, using curly braces to define a dictionary labeled "DOMAIN\_CATEGORIES." The dictionary includes ten numbered categories: "News," "Blog article," "Non Profit Organization," "E-commerce platform," "Government," "International Organization," "Law enforcement," "Education," "Social Media," and "Unknown." Below the code, the text instructs to categorize a URL using the provided categories and to format the answer in JSON with 'categories' and 'justification'. This instructional text appears beneath the rectangle. At the bottom, it’s labeled as "Fig. 3. LLM Prompt 1 – For categorization of domains through DC-LLM."}
    \label{fig:prompt1}
\end{figure}

\begin{figure}[t]
    \centering
    \begin{cutegreenbox}{}
    \begin{lstlisting}
TECH_CATEGORIES = {
    1: "social media platforms",
    2: "smart-home or Internet of Things (IoT) devices",
    3: "mobile or phone applications",
    4: "software databases for storing information on missing persons",
    5: "search engines to find information",
    6: "software or hardware for general-purpose computing",
    7: "cloud computing for storing and accessing data",
    8: "privacy for protecting personal information",
    9: "security for protecting information from unauthorized access",
    11: "podcast or films",
    12: "AMBER alerts or other network-based broadcast alerts",
    13: "digital photo or sharing photos for a missing poster",
    14: "tools for reporting missing persons",
    15: "Data Sovereignty: having control over Indigenous data",
    16: "DNA databases to find missing persons"
    17: "Other: any other technology not listed",
}
    \end{lstlisting}
    
    Given the document, identify if it mentions one or more of the following technology categories: {TECH\_CATEGORIES}. Provide only the answer in de-serializable JSON format: ``categories``: [list of numbers], ``justification``: ``brief explanation``, ``direct\_quotes``: [direct text quotes if applicable]. If no technology use is mentioned, say `No'. Do not summarize or explain the rest of the content. If no technology is mentioned, say 'I don’t know'.
    \end{cutegreenbox}
    \mycaption{LLM Prompt 2}{For categorizing web pages with technology categories through CC-LLM.}
    \Description{The image is a screenshot of a document with a list of technology categories enclosed in a green rectangle. The list, labeled as "TECH\_CATEGORIES," consists of 17 numbered items, each describing a different technology category, such as "social media platforms" and "smart-home or Internet of Things (IoT) devices." Below the list is a prompt instructing the reader to identify if any of these categories are mentioned in a given document, and to provide an answer in a specific JSON format. The background is white, and the text is primarily in a monospace font.}
    \label{fig:prompt2}
\end{figure}

\subsection{Table of \mmiw Advocates}
{
  \rowcolors{2}{rowAlt}{rowBase}
  \tabfontsize
  \renewcommand{\arraystretch}{1.4}
  \begin{longtable}{>{\centering\arraybackslash}
  m{.02\textwidth}|m{0.18\textwidth}|m{0.8\textwidth}
  }
    \mycaption{Native Advocates behind the \mmiw movement}{The families, advocates, and tribal police are fighting back to reclaim the narrative behind their stolen relatives.}
    \Description{The image displays a structured table titled "Native Advocates behind the MMIR movement." It outlines individuals and their roles in advocating for this movement, which focuses on raising awareness of missing and murdered Indigenous relatives. The table is divided into three columns: the first column is numbered, the second lists the names of advocates, and the third provides detailed descriptions of their contributions. There are twelve entries in total. The descriptions include personal backgrounds, professional roles, and specific efforts made by each advocate to support the MMIR movement. Some entries have reference numbers for further reading, which are indicated in brackets.}
    \label{tab:actors}
    
    \\
    \toprule
    \rowcolor{headerWarm}\color{white} \textbf{\#} & \color{white}\textbf{Advocate} & \color{white}\textbf{Role in advocating for the \mmiw movement} \\
    \endfirsthead

    \multicolumn{3}{c}{{\bfseries \tablename\ \thetable{} -- continued from previous page}} \\
    \toprule
    \rowcolor{headerWarm}\color{white} \textbf{\#} & \color{white}\textbf{Advocate} & \color{white}\textbf{Role in advocating for the \mmiw movement} \\
    \endhead

    \arrayrulecolor{accentRule}\midrule \multicolumn{3}{r}{{Continued on next page}} \\
    \endfoot

    \arrayrulecolor{accentRule}\bottomrule
    \endlastfoot

    \rowlabel{jordanmarie} & Jordan Marie Brings Three White Horses Daniel & ``Jordan Marie Brings Three White Horses Daniel, a Kul Wicasa Oyate/Lower Brule Sioux MMIW activist in the documentary short No More Stolen Sisters: Real America with Jorge Ramos.''~\cite{swanson-they-2020} \\
    \rowlabel{juliawoock} & Julia Annette Woock & ``Julia Annette Woock is a Latina born in Tijuana, Mexico, raised in a binational borderlands community by a single mother. 
    Woock is the Editor-in-Chief of America’s \#1-ranked community college newspaper, The Southwestern College Sun, where she covers local politics, immigration and indigenous civil rights.''~\cite{voice_of_san_diego_our_nodate} \\
    \rowlabel{maggiecywink} & Maggie Cywink & Maggie Cywink's sister Sonya Nadine in 1994.
    In 2004, Cywink shared her story with Amnesty International in their groundbreaking report  titled ``Stolen Sisters: A Human Rights Response to Discrimination and Violence Against Indigenous Women in Canada.'' Cywink has become a strong advocate for the movement since then~\cite{brown-indigenous-2018}. \\
    \rowlabel{lisabrunner} & Lisa Brunner & Lisa Brunner [is] co-director of Indigenous Women’s Human Rights Collective and professor and cultural coordinator at White Earth Tribal and Community College in Mahnomen, Minnesota. Brunner, who is also an Anishinaabe member of White Earth Nation in Minnesota, told News21 that she has survived numerous sexual assaults by non-Native and Native American men alike, which drove her to advocate for the past 20 years on behalf of other victimized Native American women.''~\cite{bleir-murdered-2018} \\
    \rowlabel{lelamailman} & Lela Mailman & Lela Mailman lost her eighteen year old daughter Melissa. ``She went to the police in Farmington to report her missing, but was dismissed.`` In the following years, Mailman became a strong activists at marches, protests, and prayer gatherings.~\cite{monroe-is-2024} \\
    \rowlabel{annitalucchesi} & Annita Lucchesi & Annita Hetoevehotohke'e Lucchesi, a Southern Cheyenne cartographer, researcher, and advocate for Indigenous cartography, geography, and earth sciences; Indigenous data sovereignty and research methodologies; and violence against Indigenous peoples. As the director of Sovereign Bodies Institute (SBI), she built and maintains the \mmiw database.~\cite{lucchesi-home-nodate} \\
    \rowlabel{abigailechohawk} & Abigail Echo-Hawk & ``Abigail Echo-Hawk, Pawnee Nation of Oklahoma, is currently the Executive Vice President at Seattle Indian Health Board and the Director of Urban Indian Health Institute.''~\cite{echo-hawk-step-2024} \\
    \rowlabel{marykathryn} & Mary Kathryn & ``Mary Kathryn Nagle is an enrolled citizen of the Cherokee Nation of Oklahoma, a playwright and partner at Pipestem Law and the executive director of the Yale Indigenous Performing Arts Program.''\cite{wilbur-protect-2021} \\
    \rowlabel{matikawilbur} & Matika Wilbur & Matika Wilbur is ``a visual storyteller from the Swinomish and Tulalip peoples of coastal Washington, for the past five years has been traveling and photographing Indian Country in pursuit of one goal: To Change the Way We See Native America.''~\cite{wilbur-protect-2021} \\
    \rowlabel{kellyhayes} & Kelly Hayes & Kelly Hayes is ``a Menominee author, organizer, movement educator and photographer. She is the host of Truthout‘s podcast Movement Memos, and the creator of Organizing My Thoughts, a weekly newsletter about politics and justice work.''~\cite{hayes-indigenous-2022} \\
    \rowlabel{megsinger} & Meg Singer & \changed{Meg Singer is ``the Indigenous Justice Program manager with the American Civil Liberty Union’s Montana chapter in Missoula.''~\cite{bleir-murdered-2018}} \\
    \rowlabel{patinapark} & Patina Park & Patina Park is the ``Executive Director of the Minnesota Indian Women's Resource Center.''~\cite{swanson-they-2020} \\
    \rowlabel{nativehope} & Native Hope & Native Hope ``has made a profound impact through its mission to empower Native voices, preserve cultural heritage, and create positive change for Indigenous communities.''~\cite{native-hope-missing-2024}~\cite{native-hope-missing-2024} \\
    \rowlabel{tsuutina} & Tsuut’ina Nation Police Service & ``In consultation with Tsuut’ina First Nation elders, the Tsuut’ina Nation Police Service designs and delivers a comprehensive cultural training program for Alberta law enforcement agencies and other representatives in the justice system that will challenge and dispel stereotypes of Indigenous peoples and address unconscious biases.''~\cite{tsuutina_nation_police_service_annual_2022} \\
    \rowlabel{jodivoiceyellowfish} & Jodi Voice Yellowfish &  ``Jodi Voice Yellowfish is founder and chair of the Missing or Murdered Indigenous Women Texas Rematriate, a Dallas-based organization that helps Indigenous families search for and bring home missing and murdered relatives, to support and offer healing processes to the missing and murdered and their families, and to advocate for social change.''~\cite{yellowfish-missing-2023}\\
    \rowlabel{fawndouglas} & Fawn Douglas & ``Fawn Douglas is an indigenous American artist, activist and registered member of the Las Vegas Paiute Tribe.''~\cite{mueller-nevada-2020}\\
    \rowlabel{marilenejames} & Marilene James & ``Marilene wrote a Facebook post asking people to share information with her about Yazzie's disappearance. She made a list of all the tips and provided it to the police.''~\cite{monroe-is-2024} \\
    \rowlabel{deborahshipman} & Deborah Maytubee Shipman & ``Deborah Maytubee Shipman of Portland, founded the Missing and Murdered Indigenous Women USA page on Facebook, adds two: \#MMIWUSA and \#JUSTICEFORYAKAMA.''~\cite{ayer-first-2018} \\
        \rowlabel{staceyschreiberschinko} & Stacey Schreiber Schinko & ``Stacey Schreiber Schinko, whose children are related to Decorah, was relieved by the conviction and 25-year prison sentence Decorah's partner ultimately received. She knows all too well that many never find their missing family member or never see a conviction.''~\cite{volpenhein-rally-2022} \\
        \rowlabel{tawnasanchez} & Rep. Tawna Sanchez & In 2019, Oregon State Rep. Tawna Sanchez, sponsored a bill 
        focused on ``increasing and improving the reporting, investigation, and response to incidents involving Missing and Murdered Native American Women''~\cite{reyna-secrecy-2024} \\
        \rowlabel{charleneaqpikapok} & Charlene Aqpik Apok & ``Charlene Aqpik Apok (Iñupiaq),  Malia Villegas (Native Village of Afognak), Abigail Echo-Hawk (Pawnee Nation of Oklahoma), Jody Juneby Potts (Han Gwich’in from Eagle Village, Alaska), and  Kelsey Ciugun Wallace ( Yup'ik, Yaaruin Creative LLC) created a report for the Data for Indigenous Justice (DIJ). They say ``This report is a reclamation of our stories that we have always had and maintained. This ancestral knowledge of data that we put forward is for our families and communities to self-determine our pathways to justice.''~\cite{apok-we-2021} \\
        \rowlabel{malindaharris} & Malinda Harris Limberhand & ``Malinda hadn’t heard from her 21-year-old daughter, Hanna Harris, since she’d left to watch fireworks the previous night. Malinda babied her ``Hanna Bear'' or ``Hanna Banana,'' but her youngest daughter was now a mother herself. Her son, Jeremiah, was 10 months old, and wasn’t taking his bottle. He was hungry, and Malinda was worried. It wasn’t like Hanna not to come home to breastfeed him.''~\cite{mabie-mothers-2022}\\
         \rowlabel{hunter} & Hunter Old Elk & ``Hunter Old Elk (Crow \& Yakama) of the Plains Indian Museum at the Buffalo Bill Center of the West, grew up on the Crow Indian Reservation in Southeastern Montana. Old Elk uses museum engagement through object curation, exhibition development, social media, and education to explore the complexities of historic and contemporary Indigenous culture.''~\cite{elk-protect-2021}\\
         \rowlabel{montemills} & Monte Mills & ``Monte Mills is a professor and co-director of Indian law clinic at the University of Montana.''~\cite{mabie-mothers-2022}\\
         \rowlabel{cheryl} & Cheryl Bennett & ``Cheryl Bennett is an Arizona State University professor ``researched the race of perpetrators and the use of racist slurs during sexual assaults targeting indigenous women. She believes that they should be considered in most cases to be hate crimes.''~\cite{bleir-murdered-2018}\\
         \rowlabel{teganswanson} & Tegan Swanson & ``Teegan Swanson is a Systems Change Coordinator at End Domestic Abuse Wisconsin member, Missing and Murdered Indigenous Women (MMIW) Task Force of Wisconsin''~\cite{swanson-they-2020}\\
         \rowlabel{paulacastro} & Paula Castro & ``Paula Castro is the mother of Henny Scott --- ``Henny Scott was 14 years old, a high school freshman on the Northern Cheyenne Indian reservation in Lame Deer, Montana, when she went missing after a house party in late December 2018'' \cite{horton-families-2023}\\
         \rowlabel{carolyndeFord} & Carolyn DeFord & ``Carolyn DeFord, a Puyallup tribal member Leona, who lost her mother LeClair Kinsey in 1999. ``She is now a member of Washington’s recently created Missing and Murdered Indigenous Women and People Task Force''. \cite{turner-we-2022}\\
         \rowlabel{reneegralewica} & Renee Gralewicz & ``Dr. Renee Gralewicz, Brothertown Indian Nation Peacemaker, retired Professor of Anthropology in the University of Wisconsin Systems, and co-chair of the Wisconsin MMIW Task Force Legal/Policy subcommittee'' \cite{swanson-they-2020}\\
         \rowlabel{desismallrodriguez} & Desi Small-Rodriguez Lonebear & ``Desi Small-Rodriguez Lonebear (Northern Cheyenne \& Chicana) is a dual PhD candidate in sociology at the University of Arizona and demography at the University of Waikato in New Zealand. She is an incoming assistant professor in the departments of sociology and American Indian studies at UCLA. Her research examines the intersection of Indigenous erasure and inequality, including health equity for Indigenous Peoples.''\cite{carroll_indigenous_2020}\\
         \rowlabel{sheilanorth} & Sheila North Wilson & ``Sheila North Wilson, the former Grand Chief of Manitoba Keewatinowi Okimakanak Inc., who coined the hashtag \#MMIW while working for the Assembly of Manitoba Chiefs in 2012.''~\cite{moeke-pickering-understanding-2018}\\
         \rowlabel{melindaharris} & Melinda Harris Limberhand & ``Melinda Harris Limberhand is a a member of the Northern Cheyenne Tribe, was twenty-one when she disappeared on July 4, 2013, in Lame Deer, Montana. The previous night she had gone to meet some friends. Like many Native women who vanish inexplicably, Harris was a mother, the devoted single parent of a ten-month-old son. The ``National Day of Awareness for Missing and Murdered Native Women and Girls'' honors the memory of Hanna Harris and countless missing \victims.''~\cite{gable-when-2023}\\
         \rowlabel{sasha-chelsea} & Sasha Reid & ``Sasha Reid is the former University of Calgary instructor and current law student''~\cite{culbert-bcs-2023} \\
         \rowlabel{morningstargali} & Morning Star Gali & ``Morning Star is a member of the Ajumawi band of the Pit River Nation. She is a lifelong Indigenous activist and the project director of Restoring Justice for Indigenous Peoples. Morning Star is also a tribal water organizer for Save California Salmon. She supports Indigenous families who have been impacted by the crisis of missing and murdered Indigenous relatives.''~\cite{hayes-indigenous-2022}\\
         \rowlabel{cutcharislingbaldy} & Cutcha Risling Baldy & ``Cutcha Risling Baldy is an assistant professor and department chair of Native American Studies at Humboldt State University.''~\cite{mueller-nevada-2020} \\
         \rowlabel{alleenbrown} & Alleen Brown & ``Alleen Brown is a New York-based reporter, focused on environmental justice issues. Her work has been published by The Intercept, The Nation, In These Times, YES! Magazine, and various Twin Cities publications.''~\cite{brown-indigenous-2018}\\
         \rowlabel{reneegralewicz} & Renee Gralewicz  & {Dr. Renee Gralewicz, Brothertown Indian Nation Peacemaker, retired Professor of Anthropology in the University of Wisconsin Systems, and co-chair of the Wisconsin MMIW Task Force Legal/Policy subcommittee} \\
         \rowlabel{jamieday} & Jamie Day & ``Jamie Day works full time as an evidential medium and spiritual development teacher.~\cite{day-missing-2022} \\
         \rowlabel{andrealemkerochon} & Andrea "Andry" Lemke-Rochon & Andrea "Andry" Lemke-Rochon is a member of Wisconsin's MMIW Task Force~\cite{swanson-they-2020}\\
         \rowlabel{peggyflanagan} & Lt. Gov. Peggy Flanagan & Minnesota Lt. Gov. Peggy Flanagan is a member of the White Earth Band of Ojibwe \cite{nelson-native-2022}\\
         \rowlabel{razellebenally-matthewgalkin} & Razelle Benally \& Matthew Galkin & Filmmakers Razelle Benally and Matthew Galkin spent more than two years working on the documentary focuses on Montana cases of missing, murdered Indigenous girls. Series details the murder of three girls found dead in Big Horn County and the lack of law enforcement response.~\cite{paige-showtime-2023} \\
         \rowlabel{sheyashe-maggie-ahlisha} & Sheyahshe Littledave, Maggie Jackson \& Ahli-sha "Osh" Stephens & Sheyahshe Littledave, Maggie Jackson and Ahli-sha "Osh" Stephens launched the “We are Resilient” podcast in 2021. The true-crime podcast shines a light on missing Indigenous women with a community perspective. All three are members of the Eastern Band of Cherokee. ~\cite{knoepp-murdered-2024}\\
         \rowlabel{kandimossett} & Kandi Mossett & Kandi Mossett is a member of the MHA Nation and the director of the Native Energy and Climate Campaign of the Indigenous Environmental Network ~\cite{bleir-murdered-2018}\\
         \rowlabel{dollyalfred} & Dolly Alfred & Dolly Alfred is a Wet’suwet’en language teacher who is friends with Gracie and Florence and joined us at Gracie’s house, believes the spirits of MMIWG are restless.~\cite{morin-stench-2021}\\
         \rowlabel{niwrc} & National Indigenous Women’s Resource Center (NIWRC) & ``NIWRC made the MMIW Toolkit for Understanding and Responding to Missing and Murdered Indigenous Women for Families and Communities is designed to assist families, communities, and advocacy organizations in understanding and responding to a case of a missing or murdered Native woman. While there is no one-size-fits-all approach to developing a community response, these resources provide a starting point and outline important information and resources available.''~\cite{niwrc-mmiw-2024}\\
         \rowlabel{nwwiw} & Native Women's Wilderness and Indigenous Women (NWWIW) & Native Women's Wilderness was created to bring Native women together to share our stories, support each other, and learn from one another as we endeavor to explore and celebrate the wilderness and our native lands.~\cite{native-womens-wilderness-mmiw-2025}\\
         \rowlabel{ttp} & Treaty Three Police & Treaty Three Police is a self-administered Policing entity under the First Nations Policing Program in serving First Nations in the greater Treaty \#3 region.~\cite{treaty-three-police-service-annual-2023, treaty-three-police-service-annual-2024}\\
         \rowlabel{kenoramakwapatrol} & Kenora Makwa Patrol & Kenora Makwa Patrol ``project will provide an opportunity for people to share their own stories and engage with local service providers and law enforcement The project aims to address hate motivated speech in social media and counteract it with true stories and an understanding of the effects of systemic and structural racism.''~\cite{williams-ontarios-2023}\\
         \rowlabel{joealphonse} & Nits'il?in (Chief) Joe Alphonse & ``Joe Alphonse has served as head of the Tl'etinqox Government for 16 years and has been tribal chair of the Tŝilhqot'in National Government for nearly as long.''~\cite{kurjata_chief_2025}\\
         \rowlabel{trishaetringer} & Trisha Etringer & ``Trisha Etringer, Muriel Walker's daughter, said Indigenous people need to be the ones controlling their own data. She cited the MMIWG2 Database, which Annita Lucchesi, a Cheyenne descendent and executive director of the nonprofit Sovereign Bodies Institute, created to log cases of missing and murdered Indigenous women, girls and those who are two-spirit -- a term used in some Native American cultures to describe gender-variant individuals in their communities.''~\cite{butz-special-2021}\\
         \rowlabel{abbyabinanti} & Abby Abinanti & ``Abby Abinanti is the chief judge of the Yurok Tribe and the first Native American woman to be admitted to the California State Bar''~\cite{ortiz-why-2020}\\
         \rowlabel{isabellemeawasige} & Isabelle Meawasige & ``Isabelle Meawasige is a bear clan woman from Serpent River, whose roots are Ojibway and Algonquin.''~\cite{auger-force-2023}\\
         \rowlabel{lupelopez} & Lupe Lopez & Lupe Lopez is a counselor and workshop leader~\cite{raftery-missing-2019}\\
         \rowlabel{emmahall} & Emma Hall & Emma Hall covers Sacramento County for The Sacramento Bee. Hall graduated from Sacramento State and Diablo Valley College. She is Blackfeet and Cherokee.~\cite{hall-missing-2024}\\
         \rowlabel{mikebalczer} & Mike Balczer & Mike Balczer is the father of a 18-year-old daughter, Jessica Patrick. ~\cite{morin-stench-2021}\\
         \rowlabel{amanda} & Amanda & Amanda is a news reporter at ListVerse~\cite{amanda-10-2015}\\
         \rowlabel{tamaracolaque} & Tamara Colaque & Tamara shares her journey to know herself through her mother’s MMIW story. An aspect that is often forgotten when someone goes missing or is murdered is the impact on the family—especially the children. The wound is deep and the answers often intangible. ~\cite{colaque-part-2023}\\
         \rowlabel{melisemarubbio} & M. Elise Marubbio & M. Elise Marubbio examines the sacrificial role of what she terms the "Celluloid Maiden"—a young Native woman who allies herself with a white male hero and dies as a result of that choice. Marubbio intertwines theories of colonization, gender, race, and film studies to ground her study in sociohistorical context all in an attempt to define what it means to be an American.~\cite{marubbio-killing-2006}\\
         \rowlabel{ericaficklin} & Erica Ficklin & Erica Ficklin is a proud member of the Tlingit and Oglala Lakota tribes. She is currently in the Combined Clinical and Counseling Psychology program at Utah State University and mentored by Dr. Melissa Tehee.~\cite{ficklin-fighting-2022} \\
         \rowlabel{theamberadvocate} & The Amber Advocate & The mission of the AMBER Alert Training and Technical Assistance Program (AATTAP) is to safely recover missing, endangered, or abducted children through the coordinated efforts of law enforcement, media, transportation, and other partners by using training and technology to enhance response capacities and capabilities and increase public participation.~\cite{the-amber-adovcate-amber-2025}\\
         \rowlabel{luellabrien} & Luella Brien & Luella Brien is a journalist based in Hardin who grew up on the Crow Reservation~\cite{paige-showtime-2023} \\
         \rowlabel{monagable} & Mona Gable & Mona Gable is a journalist in Los Angeles who covers gender, science, and travel. Her work has appeared in The New York Times, Outside, The Atlantic, Vogue, Los Angeles magazine, BBC, the Los Angeles Times and many other publications. Her story in Los Angeles magazine about sexual assault at Occidental College was named one of the Best Longreads of 2015.~\cite{gable-when-2023}\\
         \rowlabel{brennanmcCullagh} & Brennan McCullagh & Brennan McCullagh is a Grade 11 student at St. John’s Ravenscourt  School,  Winnipeg, Manitoba  ~\cite{national-centre-for-truth-and-reconciliation-imagine-2021}\\
         \rowlabel{debhaaland} & \changed{Deborah Haaland} & Secretary of the Interior Deb Haaland established the Missing and Murdered Unit (MMU) within the Bureau of Indian Affairs (BIA) to investigate \mmiw cases in Indian Country.~\cite{native-news-online-missing-2021}\\
         \rowlabel{cherylhorn} & \changed{Cheryl Horn} & 
         Cheryl Horn, Magazu Nahzi (Standing in the Rain), an Assiniboine from Ft Belknap, MT. She is a Member for the Montana Missing Indigenous Persons Task Force. 
\end{longtable}
}




\end{document}